\documentclass[12pt]{elsarticle}                                          
\usepackage{axodraw4j}
\usepackage{hyperref}
\usepackage{graphicx}                                                         
\usepackage{bbold}
\usepackage{a41}                                                                
\usepackage{xcolor}                     
\usepackage{slashed}
\usepackage[rflt]{floatflt}
\usepackage{float}
\usepackage{hyperref}
\hypersetup{colorlinks=true, 
linkcolor=blue, filecolor=blue, 
urlcolor=mygreen}
\usepackage{breakurl} 
\usepackage{times}     %
\usepackage{array}
\usepackage[english]{babel}
\usepackage[latin1]{inputenc}
\usepackage[T1]{fontenc}
\usepackage{ae}
\usepackage{url}
\usepackage{amsmath, amsthm, amssymb}
\usepackage{slashed}

\usepackage{rotating}
\usepackage{graphicx}
\usepackage{comment}
\newcounter{mmacnt}
\def\restartmma{\setcounter{mmacnt}{0}}
\restartmma \catcode`|=\active
\def|#1|{\mathrm{#1}}
\catcode`|=12
\newenvironment{mma}{
\par\smallskip
\catcode`|=\active
\parskip=0pt\parindent=0pt 
\small
\def\In##1\\{%
\def\linebreak{\hfill\break\null\qquad}%
\refstepcounter{mmacnt}
\hangindent=2.5em\hangafter=0
\leavevmode
\llap{\tiny\sffamily In[\arabic{mmacnt}]:=\kern.5em}%
\mathversion{bold}\footnotesiVe$
\displaystyle##1$\normalsiVe
\mathversion{normal}\par
 }%
\def\Print##1\\{%
\def\linebreak{\hfill\break}%
\hangindent=2.5em\hangafter=0
\leavevmode ##1\par}%
\def\Out##1\\{%
\def\linebreak{$\hfill\break\null\hfill$}%
\kern\abovedisplayskip\par
\hangindent=2.5em\hangafter=0
\leavevmode
\llap{\tiny\sffamily Out[\arabic{mmacnt}]=\kern.5em}
\footnotesiVe$\displaystyle##1$
\normalsiVe\hfill\null\par
\kern\belowdisplayskip
}%
\def\Warning##1##2\\{%
\def\linebreak{\hfill\break}%
\hangindent=2.5em\hangafter=0
\leavevmode
{\scriptsiVe##1 : ##2}\par}%
}{%
\par\smallskip
}

\usepackage{color}
\newenvironment{fshaded}{%
\MakeFramed {\FrameRestore}
}%
{\endMakeFramed}

\makeatletter
\def\ps@pprintTitle{%
\let\@oddhead\@empty
\let\@evenhead\@empty
\def\@oddfoot{\reset@font\hfil\thepage\hfil}
\let\@evenfoot\@oddfoot
}
\makeatother
\newcommand{\n}{\nonumber}

\usepackage{tcolorbox}
\usepackage{slashed}
\usepackage{comment}

\begin{document}  
\begin{frontmatter}
\title{\Large 
\textbf{ One-loop expressions for $h\rightarrow l\bar{l}\gamma$ in Higgs extensions of  the Standard Model}}
\author[1,2]{L. T. Hue}
\ead{lethohue@vlu.edu.vn}
\author[3,4]{Dzung Tri Tran}
\author[5]{Thanh Huy Nguyen}
\author[3,4]{Khiem Hong Phan}
\ead{phanhongkhiem@duytan.edu.vn}
\address[1]{ Subatomic Physics Research Group, Science and Technology Advanced Institute, Van Lang University, Ho Chi Minh City 70000, Vietnam
}
\address[2]{Faculty of Applied Technology, School of Technology, Van Lang University, Ho Chi Minh City 70000, Vietnam
}
\address[3]{\it Institute of Fundamental 
and Applied Sciences, Duy Tan University, 
Ho Chi Minh City $700000$, Vietnam}
\address[4]{Faculty of Natural Sciences, 
Duy Tan University, Da Nang City $550000$, 
Vietnam}
\address[5]{\it 
University of Science Ho Chi Minh City, 
$227$ Nguyen Van Cu, District $5$, 
Ho Chi Minh City, Vietnam}
\pagestyle{myheadings}
\markright{}
\begin{abstract} 
A systematic study of one-loop contributions to the decay channels $h\rightarrow l\bar{l}\gamma$ with $l=\nu_{e,\mu, \tau}, e, \mu$, performed in Higgs extended versions of the Standard Model, is presented in the 't Hooft-Veltman gauge.	Analytic formulas  for one-loop form factors are expressed in terms of  the logarithm and di-logarithmic functions. As a result, these form factors can be reduced to those relating to the loop-induced decay processes $h\rightarrow \gamma\gamma, Z\gamma$,  confirming not only previous results using different approaches but also close relations between the three kinds of the loop-induced Higgs decay rates.  For phenomenological study, we focus on the two observables, namely the enhancement factors defined as ratios of the decay rates calculated between the Higgs extended versions and the standard model, and  the forward-backward 	asymmetries of fermions, which can be used  to  search for  Higgs extensions of the SM.   We show that direct effects	of mixing between neutral Higgs bosons and indirect contributions of charged Higg boson exchanges can be probed at future colliders.  
\end{abstract}
\begin{keyword} 
{\it
Higgs phenomenology, 
Physics beyond the 
Standard Models,  
One-loop Feynman integrals, 
Analytic methods 
for Quantum Field Theory, 
Dimensional regularization, 
Future colliders.}
\end{keyword}
\end{frontmatter}
\allowdisplaybreaks
\section{Introduction}
Future colliders such as the High-Luminosity Large Hadron Collider  (HL-LHC) \cite{Liss:2013hbb,CMS:2013xfa} and  International Linear Collider (LC) \cite{Baer:2013cma}  focus greatly on the precise measurements for the properties  of the Standard Model-like (SM-like) Higgs boson. The measurements provide an important information to answer the nature of  the electroweak spontaneous symmetry  breaking (EWSB)$-$in other words,  to verify the structure of scalar  potential.   It is well-known that the Higgs potential of the SM is  the simplest  such that a scalar  doublet is only taken into account for EWSB. In some Higgs extended version of  the SM (called as HESM hereafter), the respective  scalar potential  is   enlarged by including  new scalar particles such as neutral, singly  and doubly charged Higgs bosons, etc. They could contribute the decay channels of the SM-like Higgs boson and these effects could be tested at  future colliders. Recently, the loop-induced decays of the SM-like Higgs  boson ($h$) into $\gamma\gamma$~\cite{CMS:2014fzn, ATLAS:2015egz} and $Z\gamma$ have been probed at the LHC~\cite{D0:2008swt,  CMS:2013rmy, ATLAS:2017zdf,ATLAS:2020qcv}.  
Together with $h\rightarrow Z\gamma, \gamma\gamma$, the decay processes $h\rightarrow f\bar{f}\gamma$ with $l=\nu_{e,\mu, \tau}, e, \mu$ also play a key role for testing the SM \cite{Chen:2012ju, Gainer:2011aa, Korchin:2014kha} and constraining parameters of many models beyond the SM (BSMs) \cite{ Korchin:2014kha}. Morerecently, the decay channels $h\rightarrow f\bar{f}\gamma$ have been greatly paid attention at the LHC~\cite{CMS:2015tzs,  CMS:2017dyb,CMS:2018myz, ATLAS:2021wwb}.

From the theoretical viewpoint, one-loop corrections to $h\rightarrow f\bar{f}\gamma$ 
are important for matching high-precision data at future  colliders. In the SM, many calculations for these one-loop corrections  were reported in Refs. \cite{Abbasabadi:1995rc, Djouadi:1996ws, Abbasabadi:2000pb, Dicus:2013ycd, Sun:2013rqa, Passarino:2013nka, Dicus:2013lta, Kachanovich:2020xyg}. One-loop formulas for 
$h\rightarrow f\bar{f}\gamma$ in Two Higgs Doublet Models (THDM) were 
computed in Refs. \cite{Li:1998rp, Sasaki:2017fvk}. In our previous works~\cite{Phan:2021xwc,VanOn:2021myp}, we have performed the evaluations for one-loop
contributions to $h\rightarrow l\bar{l}\gamma$ in general  BSM frameworks. The computations
have done in the unitarity gauge.  It is well-known that the results for one-loop contributions to $h\rightarrow \gamma\gamma, Z\gamma$ in  unitarity gauge
may face to  large numerical cancellations~\cite{Marciano:2011gm, 
Phan:2020xih}, because the  higher-rank tensor one-loop integrals appear from Feynman diagrams containing gauge boson exchanges in the loops. The same  problem may occur in the  computations for one-loop contributions to the decays $h\rightarrow l\bar{l}\gamma$. In particular, the issue may 
appear in boundary of the phase space in which low-values of the invariant mass of lepton-pair, lepton-photon and  the low-energy of photon is applied, and the case that the photon may go nearly parallel to the lepton momenta. To cure these problems, several solutions are proposed such as evaluating the decay processes  in the general space-time with dimension $d$  \cite{Phan:2021xwc,VanOn:2021myp, Phan:2020xih}, or considering the calculations in the 't Hooft-Feynman (HF) gauge \cite{Phan:2021pcc}.

Motivation from the above issues and confirming the consistent results in different gauges $-$ checking the gauge invariance of the results, gaining the stability of the numerical  results in full kinematic regions of  the mentioned decay channels, we study systematically one-loop corrections to the decay channels $h\rightarrow l\bar{l}\gamma$ with $l=\nu_{e,\mu, \tau}, e, \mu$
in the HF gauge. In the limit of this paper, we consider the decay processes  $h\rightarrow l\bar{l}\gamma$ with $l=\nu_{e,\mu, \tau}, e, \mu$ within HESM. In further detail, one-loop analytical formulas for the form factors are written in terms of scalar one-loop integrals which are then 
expressed  with reference to the logarithm and di-logarithmic functions.  One-loop form factors in this work can be also reduced to the  corresponding ones for the decay channels $h\rightarrow 
\gamma\gamma, Z\gamma$ and these results confirm the previous results which are available in the SM as well as HESM. In phenomenological results, we show the enhancement factor in the Higgs extension for the SM models. Last but not least, the forward-backward (FB)  asymmetries of fermions are also studied  in this paper. Other beyond the SM frameworks such as the left-right models (LR) constructed from the  $SU(2)_L\times SU(2)_R\times U(1)_Y$~\cite{Pati:1974yy,
Mohapatra:1974gc, Senjanovic:1975rk},  the 3-3-1 models  ($SU(3)_L\times U(1)_X$)~\cite{Singer:1980sw, Valle:1983dk, Pisano:1991ee, Frampton:1992wt, Diaz:2004fs,Fonseca:2016tbn, Foot:1994ym},  the $3$-$4$-$1$ models ($SU(4)_L\times U(1)_X$)~\cite{Foot:1994ym, Sanchez:2004uf, Ponce:2006vw, Riazuddin:2008yx, Jaramillo:2011qu, Long:2016lmj}, etc,  will be  addressed in our future works.

The layout of  our work is as follows. In  section \ref{sec_MDs}, we review three specific HESMs investigated in detail in our work, namely the  Inert Doublet  (IDM),  THDM,  and Triplet-Higgs (THM)  models. Section \ref{sec_OL} shows the detailed evaluations for one-loop contributions to the decay amplitudes $h\rightarrow l\bar{l}\gamma$. Phenomenological results for all mentioned HESMs  are analyzed in section \ref{sec_pheno}. Conclusions and outlook are devoted in section \ref{sec_con}. Four appendices show more detailed expressions of the relevant couplings used in our calculations,  precise expressions of scalar one-loop integrals in terms of di-logarithm functions,  the finite and consistent results of the form factors being independent with the ultraviolet divergent part $1/\epsilon$ and the renormalization scale $\mu^2$, and the basis integrals relating to previous calculations.

\section{ \label{sec_MDs} Higgs extended versions of 
	the Standard Model}           
This section will review  specific models corresponding to the HESMs we are interested in this work. The simplest case is the SM extension adding  a singlet neutral Higgs boson.  As a result, 
the respective analytical expressions for the decay rates 	$h\rightarrow 	l\bar{l}\gamma$  are  the same as those predicted by the SM, except 	an overall factor relating	to the mixing angle between 	neutral Higgs bosons.  For this reason, we will pay attention to the more interesting models, namely the IDM, THDM, and the real Triplet Higgs Model (THM), which consist of charged Higgs bosons giving new one-loop contributions to the decay amplitudes under consideration.  
\subsection{The IDM}
The IDM is constructed by adding into the SM an  inert scalar 
$SU(2)_L$ doublet, which provide stable particles playing roles as dark matter  candidates \cite{Borah:2012pu, 
Gustafsson:2012aj,Arhrib:2012ia,
Klasen:2013btp,Krawczyk:2013jta,
Arhrib:2014pva,Chakrabarty:2015yia,
Ilnicka:2015jba,Datta:2016nfz,
Kalinowski:2018ylg,Dercks:2018wch,
Chiang:2012qz,Benbrik:2022bol}. The Higgs
potential corresponding to the renormalizable and gauge invariant theory  is:
\begin{eqnarray}
\label{VIDM} 
\mathcal{V}(\Phi_1,\Phi_2) 
&=& \mu_1^2 |\Phi_1|^2 
+ \mu_2^2 |\Phi_2|^2  
+ \lambda_1 |\Phi_1|^4
+ \lambda_2 |\Phi_2|^4 
+  \lambda_3 |\Phi_1|^2 
|\Phi_2|^2\nonumber \\
&&
+ \lambda_4 |\Phi_1^\dagger 
\Phi_2|^2 + \frac{\lambda_5}{2} 
\left\{ (\Phi_1^\dagger \Phi_2)^2 
+ {\rm h.c} \right\},
\end{eqnarray}
where an additional unbroken global $Z_2$-symmetry  is imposed, namely $\Phi_1\leftrightarrow -\Phi_1$ is odd,  while  $\Phi_2\leftrightarrow \Phi_2$ and all SM particles  are  even. Because  $Z_2$ is maintained  after the EWSB, all $\Phi_2$ components   develop the zero of vacuum expectation  values (VEV), leading  to the following expansions around VEV of  the Higgs components: 
\begin{eqnarray}
\Phi_1 = \left (\begin{array}{c}
G^\pm \\
\frac{1}{\sqrt{2}}(v + h + i G^0) \\
\end{array} \right)
, \qquad
\Phi_2 = \left( \begin{array}{c}
H^\pm\\ 
\frac{1}{\sqrt{2}}(H + i A^0) \\ 
\end{array} \right),
\end{eqnarray}
where $G^0,\; G^{\pm}$ are Nambu-Goldstone bosons absorbed by the massive gauge bosons
$Z$ and $W^{\pm}$, respectively. In addition, there is no mixing of the two neutral components  $h$ and $H$. After  EWSB, the matching condition with the SM  results in $v = 246$ GeV as the electroweak scale.  The IDM consist of   three neutral physical states, in which $h$ is identified as the SM-like Higgs boson observed at the LHC, while  $H$ and $A^0$ are  the two CP-even and -odd ones predicted by the IDM.  This model also predicts a pair of singly charged Higgs bosons   $H^\pm$. All Higgs boson masses are functions of  the parameters $\mu_1, \mu_2, \lambda_1, \cdots, \lambda_5$, namely: 
\begin{eqnarray}
\label{MassesIDM}
m_h^2 &=& - 2 \mu_1^2
=2 \lambda_1 v^2,\\
M_{H}^2 &=& \mu_2^2 + 
\frac{  v^2}{2}\left(\lambda_3 
+ \lambda_4 + \lambda_5\right),\\ 
M_{A^0}^2 &=& \mu_2^2 
+ \frac{v^2}{2}\left(\lambda_3 
+ \lambda_4 - \lambda_5\right),
\\
M_{H^{\pm}}^2 &=& 
\mu_2^2 + \frac{v^2}{2} \lambda_3.
\end{eqnarray}
Due to the
unbroken $Z_2$-symmetry, all "inert" Higgs bosons $H,A^0$ and $H^{\pm}$ are  odd under $Z_2$, therefore they do not interact  with quarks and leptons. As the result, the   lightest neutral Higgs boson 
($H$ or $A^0$) is  a dark matter candidate. The set of scanning parameters chosen in our analysis is   
\begin{eqnarray}
\label{eq_PIDM}
 \mathcal{P}_{\rm IDM} =
\{\mu_2^2, \lambda_2^2, m_h^2, 
M_H^2, M_{A^0}^2, M^2_{H^{\pm}}\}.
\end{eqnarray}
The Yukawa Lagrangian of this model is exactly the same as that of the SM, namely 
\begin{eqnarray}
\label{YukawaIDM}
{\mathcal{L}}_{\rm Yukawa} 
= -\sum_{f=u,d,l} 
g^{\rm IDM}_{hff} \; h \bar{f} f  
+ \cdots,
\end{eqnarray}
where $g^{\rm IDM}_{hff} = g^{\rm SM}_{hff} =m_f/v$, and $f$ is a SM lepton with mass $m_f$.  All related couplings to the  decay channels $h\rightarrow l\bar{l}\gamma$ are listed in Table~\ref{IDM-coupling},
\begin{table}[ht]
\centering
{\begin{tabular}
{l@{\hspace{2cm}}
l@{\hspace{2cm}}l }
\hline \hline
\textbf{Vertices} &\textbf{Notations} 
& \textbf{Couplings}\\
\hline \hline \\
$hH^{\pm}H^{\mp}$  
& $g^{\textrm{IDM} }_{hH^{\pm}H^{\mp}}$
& $-i\; \frac{2(M_{H^{\pm}}^2 -\mu_2^2)}{v}$ \\
\hline  \\
$Z_{\mu}H^{\pm}(p^{+})H^{\mp}(p^{-})$  
& $g^{\textrm{IDM} }_{ZH^{\pm}H^{\mp}}$
& $\frac{M_Z}{v}\; 
c_{2 W}\; 
(p^{+}-p^{-})_{\mu}$ \\
\hline  \\
$A_{\mu} H^{\pm}(p^{+})H^{\mp}(p^{-})$  
& $g^{\textrm{IDM} 
}_{A H^{\pm}H^{\mp}}$
& $\frac{M_Z}{v} 
\;  s_{2W}\; 
(p^{+}-p^{-})_{\mu}$ \\
\hline  \\
$hf\bar{f}$  
& $g^{\textrm{IDM} }_{hf\bar{f}}$
& $i m_f/v$ \\
\hline\hline  
\end{tabular}}
\caption{
\label{IDM-coupling}
All couplings giving one-loop contribution to 
the decay amplitudes 
$H\rightarrow l\bar{l} \gamma$ 
in the IDM. Here $A_{\mu}$ is the photon field,  $p^\pm$ is the incoming momentum of $H^\pm$, 
$s_W (c_W)$ is sine (and cosine) 
of the Weinberg's angle, respectively.}
\end{table}
see a detailed derivation in 
\ref{app_coupling}.
\subsection{THDM}%
The second HESM considered in this work is the THDM, in which a  new complex  Higgs doublet with the hypercharge $Y = 1/2$ is added into the SM, see  Ref.\cite{Branco:2011iw} for a detailed theoretical and phenomenological review.  The  renormalizable and gauge invariant Higgs potential is 
\begin{eqnarray}
\label{V2HDM}
\mathcal{V}
(\Phi_1,\Phi_2) &=&  m_{11}^2\Phi_1^\dagger \Phi_1+m_{22}^2\Phi_2^\dagger \Phi_2-
\Big[
m_{12}^2\Phi_1^\dagger \Phi_2
+{\rm h.c.}
\Big] 
+ \frac{\lambda_1}{2}(\Phi_1^\dagger \Phi_1)^2
+\frac{\lambda_2}{2}(\Phi_2^\dagger \Phi_2)^2 \nonumber \\
&&+ \lambda_3(\Phi_1^\dagger \Phi_1)(\Phi_2^\dagger \Phi_2)+\lambda_4(\Phi_1^\dagger \Phi_2)(\Phi_2^\dagger \Phi_1) 
+\frac{1}{2}[\lambda_5~(\Phi_1^\dagger \Phi_2)^2 +~{\rm h.c.}]. 
\end{eqnarray}
For the EWSB, two scalar doublets 
can be written as 
\begin{eqnarray}
\Phi_1 =
\begin{bmatrix}
\phi_1^+ \\
(v_1+\rho_1+i\eta_1)/\sqrt{2} 
\end{bmatrix}
\quad 
{\rm and}
\quad 
\Phi_2 =
\begin{bmatrix}
\phi_2^+ \\
(v_2+\rho_2+i\eta_2)/\sqrt{2} 
\end{bmatrix},
\label{representa-htm}
\end{eqnarray}
where $v=\sqrt{v_1^2+v_2^2}=246$ GeV from the matching condition with the SM. After EWSB, the THDM  consist of two CP-even Higgs bosons $h$ and $H$,  a CP-odd $A^0$, and a pair of singly charged ones $H^\pm$. One of them, namely $h$ is identified with  the SM-like Higgs boson  discovered at LHC. 

The  mass and flavor base of all Higgs bosons relate to each other by the following rotations 
\begin{eqnarray}
\begin{pmatrix}
\phi_1^{\pm}\\
 \phi_2^{\pm}
\end{pmatrix}
&=&
\begin{pmatrix}
c_{\beta } & 
-s_{\beta} \\
s_{\beta}
& c_{\beta}
\end{pmatrix}
\begin{pmatrix}
G^{\pm}\\
H^{\pm}
\end{pmatrix},
\\ \begin{pmatrix}
\rho_1\\
 \rho_2
\end{pmatrix}
&=&
\begin{pmatrix}
c_{\alpha} &  -s_{\alpha} \\
s_{\alpha}
& c_{\alpha}
\end{pmatrix}
\begin{pmatrix}
h\\
H
\end{pmatrix},
\end{eqnarray}
and 
\begin{eqnarray}
\begin{pmatrix}
\eta_1\\
 \eta_2
\end{pmatrix}
=
\begin{pmatrix}
c_{\beta} & 
-s_{\beta} \\
s_{\beta}
& c_{\beta}
\end{pmatrix}
\begin{pmatrix}
G^{0}\\
A^0
\end{pmatrix}.
\end{eqnarray}
where the mixing $\alpha$ between two neutral Higgs
is taken into account and
$\beta$ is the mixing angle 
defined as $t_{\beta} \equiv \tan \beta= v_2/v_1$. 
All Higgs boson masses 
are determined as follows: 
\begin{align}
M_{H^{\pm}}^{2}&=\mu^{2}-
\frac{1}{2}(\lambda_{4}+\lambda_{5})v^{2},
\\ M_{A^0}^{2}&=\mu^{2}-\lambda_{5}v^{2},
\\m_{h}^{2} &= M_{11}^{2}
s_{\beta-\alpha}^{2}+M_{22}^{2}c_{\beta-\alpha}^{2}
+M_{12}^{2}s_{2(\beta-\alpha)},
\\M_{H}^{2} &= M_{11}^{2}c_{\beta-\alpha}^{2}
+M_{22}^{2}s_{\beta-\alpha}^{2}
-M_{12}^{2}s_{2(\beta-\alpha)}, 
\end{align}
where $\mu^{2}=m_{12}^{2}/(s_{\beta}c_{\beta})$, $s_{2x}\equiv 2 s_xc_x$, $c_{2x}\equiv c_x^2-s_x^2$ ($x=\beta-\alpha$, $\beta$), 
\begin{eqnarray}
M_{11}^{2}&=&
(\lambda_{1}c_{\beta}^{4}
+\lambda_{2}s_{\beta}^{4})v^{2}
+\frac{v^{2}}{2}(\lambda_{3}+\lambda_{4}
+\lambda_{5})s_{2\beta}^{2}, \\
M_{22}^{2}&=&\mu^{2} 
+ \frac{v^{2}}{4}
\Big[
\lambda_{1}+\lambda_{2}
-2(\lambda_{3}+\lambda_{4}+\lambda_{5})
\Big]
s_{2\beta}^{2}, \\
M_{12}^{2} &=&
-\frac{v^{2}}{2}
\Big[
\lambda_{1}c_{\beta}^{2}
-\lambda_{2}s_{\beta}^{2}
-(\lambda_{3}+\lambda_{4}
+\lambda_{5}) c_{2\beta}
\Big]
s_{2\beta}.
\end{eqnarray}
All couplings involving the decay processes under consideration were presented  
in detail in \ref{app_coupling}, where the couplings between Higgs and gauge bosons are listed in Table \ref{THDM-coupling}. 
\begin{table}[ht]
\centering
{\begin{tabular}{l@{\hspace{1cm}}
l@{\hspace{1cm}}l }
\hline \hline
\textbf{Vertices} &\textbf{Notations} 
& \textbf{Couplings}\\
\hline \hline \\
$hW_{\mu}W_{\nu}$  
& $g^{\rm THDM}_{hWW}$
& $ -i \dfrac{2 M_W^2 }{v}
\; s_{\beta-\alpha}\; 
g_{\mu\nu}$ \\
\hline  \\
$hZ_{\mu}Z_{\nu}$  
& $g^{\rm THDM}_{hZZ}$
& $-i 
\dfrac{2M_Z^2 }{v}
\; s_{\beta-\alpha}\; 
g_{\mu\nu}$ \\
\hline  \\
$hH^{\pm}H^{\mp}$
& $
g^{\rm THDM}_{hH^{\pm}H^{\mp}}
$ 
& 
$ -\dfrac{i}{v}
\Big[
(2\mu^2 - 2M_{H^{\pm}}^2 - m_h^2)
s_{\beta-\alpha}
$
$+ 2 \cot (2\beta)
(\mu^2 -m_{h}^2)
c_{\beta-\alpha}$
\Big]
\\
\hline  \\
$Z_{\mu}H^{\pm}(p^{+})H^{\mp}(p^{-})$  
& $g^{\rm THDM}_{ZH^{\pm}H^{\mp}}$
& $
\dfrac{M_Z}{v} \; c_{2W} 
(p^+- p^-)_{\mu}$ \\
\hline  \\
$A_{\mu} H^{\pm}(p^{+})H^{\mp}(p^{-})$
& $g^{\rm THDM}_{A H^{\pm}H^{\mp}}$
& 
$
\dfrac{M_Z  
}{v}\; s_{2W}
(p^+- p^-)_{\mu} 
$ 
\\
\hline
\hline  
\end{tabular}}
\caption{
\label{THDM-coupling}
All the couplings involving 
the decay processes
$h\rightarrow l\bar{l} \gamma$ 
in the THDM. $A_{\mu}$ is photon field. }
\end{table}

The Yukawa Lagrangian 
written in terms of the mass eigenstates is \cite{Branco:2011iw} 
\begin{eqnarray}
	\label{YukawaTHDM}
	{\mathcal{L}}_{\rm Yukawa} = 
	-\sum_{f=u,d,l} 
	\left(g^{\rm THDM}_{hff} 
	\bar{f} f h + 
	g^{\rm THDM}_{Hff}  \bar{f} f H 
	- i g^{\rm THDM}_{A^0ff}  
	\bar{f} \gamma_5 f A^0 \right) + \cdots,
\end{eqnarray}
where particular formulas for different types of the THDM were shown in \ref{app_coupling}, see the final results  presented in Table~\ref{YukawaTHDM}. 
\begin{table}[ht]
\centering
\begin{tabular}{l@{\hspace{2cm}}l
@{\hspace{2cm}}l@{\hspace{2cm}}l}
\hline\hline
Type&$g^{\rm THDM}_{huu}$ 
&$g^{\rm THDM}_{hdd}$ 
&$g^{\rm THDM}_{hll}$\\  \hline \hline\\
I&$\dfrac{m_u}{\sqrt{2}v}\dfrac{c_\alpha}{s_\beta}$
&$
\dfrac{m_d}{\sqrt{2}v}\dfrac{c_\alpha}{s_\beta}
$ 
&$\dfrac{m_l}{\sqrt{2}v}\dfrac{c_\alpha}{s_\beta}
$\\ \hline\\
II&$\dfrac{m_u}{\sqrt{2}v}\dfrac{c_\alpha}{s_\beta}$ 
&$-\dfrac{m_d}{\sqrt{2}v}\dfrac{s_\alpha}{c_\beta}$ 
&$-\dfrac{m_l}{\sqrt{2}v}\dfrac{s_\alpha}{c_\beta}
$\\ \hline\\
X&$\dfrac{m_u}{\sqrt{2}v}\dfrac{c_\alpha}{s_\beta}$ 
&$\dfrac{m_d}{\sqrt{2}v}\dfrac{c_\alpha}{s_\beta}$ 
&$-\dfrac{m_l}{\sqrt{2}v}\dfrac{s_\alpha}{c_\beta}$ 
\\ \hline\\
Y&$\dfrac{m_u}{\sqrt{2}v}\dfrac{c_\alpha}{s_\beta}$ 
&$-\dfrac{m_d}{\sqrt{2}v}\dfrac{s_\alpha}{c_\beta}$ 
&$\dfrac{m_l}{\sqrt{2}v}\dfrac{c_\alpha}{s_\beta}$ \\ 
\hline\hline
\end{tabular}
\caption{\label{YukawaTHDM} The Yukawa 
couplings in THDMs with type I,II, X, and Y 
respectively.}
\end{table}
In the limit of 
$s_{\beta-\alpha} \rightarrow 1$, 
called as $h^{\rm SM}$-scenario, 
one has 
\begin{eqnarray}
	\label{eq_sba1}
g^{\rm THDM}_{hH^{\pm}H^{\mp}}
&\rightarrow& -i \;
\dfrac{(2\mu^2 - 2 M_{H^{\pm}}^2 - m_h^2)}{v}
\; s_{\beta-\alpha}. 
\end{eqnarray}

In the case of THDM, Finally, the set $	\mathcal{P}_{\rm THDM} $ of scanning parameters used for our numerical investigation is chosen as follows 
\begin{eqnarray}
	\mathcal{P}_{\rm THDM} =
	\{m_h^2, M_H^2, M_{A^0}^2, 
	M^2_{H^{\pm}}, m_{12}^2, 
	t_{\beta}, s_{\beta-\alpha} \}.
\end{eqnarray}
\subsection{THM}%
We finally mention THM, which is the SM adding only one additional real Higgs 
triplet, denoted as  $\Delta$ with 
$Y_{\Delta}=2$ \cite{Chun:2012jw,Chen:2013dh, Arhrib:2011uy, Arhrib:2011vc,Akeroyd:2012ms,Akeroyd:2011zza,Akeroyd:2011ir,Aoki:2011pz,Kanemura:2012rs,Chabab:2014ara,Han:2015hba,Chabab:2015nel,Ghosh:2017pxl, Ashanujjaman:2021txz,Zhou:2022mlz}.
The renormalizable gauge invariant  Higgs potential is 
\begin{eqnarray}
\mathcal{V}(\Phi, \Delta) 
&=& -m_{\Phi}^2{\Phi^{\dagger}\Phi}
+\frac{\lambda}{4}(\Phi^{\dagger}\Phi)^2 
+ M_{\Delta}^2 \textrm{Tr}
(\Delta^{\dagger} \Delta)
+[\mu(\Phi^T{i}\sigma^2\Delta^{\dagger}\Phi)
+{\rm h.c.}]\nonumber\\
&&+
\lambda_1(\Phi^{\dagger}\Phi) 
\textrm{Tr}(\Delta^{\dagger}{\Delta})+\lambda_2( \textrm{Tr}\Delta^{\dagger}{\Delta})^2
+\lambda_3 \textrm{Tr}(\Delta^{\dagger}{\Delta})^2+ 
\lambda_4{\Phi^\dagger\Delta\Delta^{\dagger}\Phi},
\end{eqnarray}
where $\sigma^2$ is Pauli matrix, and all Higgs self couplins $\lambda_i$ ($i=\overline{i,4}$) 
are real. For the EWSB,  two Higgs multiplets are parameterized as follows: 
\begin{eqnarray}
\Delta &=\begin{bmatrix}
{\delta^+ \over \sqrt{2}} &  \delta^{++} \\
\frac{1}{\sqrt{2} } 
(v_{\Delta} +\eta_{\Delta} 
+i \chi_{\Delta})
& -{\delta^+ \over \sqrt{2}}
\end{bmatrix}
\quad 
{\rm and}
\quad 
\Phi=
\begin{bmatrix}
\phi^+ \\
\frac{1}{\sqrt{2} } 
(v_{\Phi} +\eta_{\Phi} 
+i \chi_{\Phi})
\end{bmatrix},
\label{representa-htm}
\end{eqnarray}
where $v_{\Phi}$ and $v_{\Delta}$ are correspondingly the VEVs of the two neutral Higgs components. The electroweak scale is 
$v= \sqrt{v_{\Phi}^2+2v_{\Delta}^2}
= 246$ GeV when matching with the SM.

After EWSB, the physical Higgs spectrum of the  THM  consist of  two pairs of  
charged Higgs bosons, namely doubly $H^{\pm\pm}$ and singly
 $H^\pm$, 
a neutral CP-odd $A^0$, and two CP-even 
$H$ and $h$ being identified with the SM-like Higgs boson.  The relations between two  mass and flavor base are 
\begin{eqnarray}
\begin{pmatrix}
\phi^{\pm}\\
 \delta^{\pm}
\end{pmatrix}
&=&
\begin{pmatrix}
c_{\beta^{\pm}} & 
-s_{\beta^{\pm}} \\
s_{\beta^{\pm}}
& c_{\beta^{\pm}}
\end{pmatrix}
\begin{pmatrix}
G^{\pm}\\
H^{\pm}
\end{pmatrix},
\\\begin{pmatrix}
\eta_{\Phi}\\
 \eta_{\Delta}
\end{pmatrix}
&=&
\begin{pmatrix}
c_{\alpha} &  -s_{\alpha} \\
s_{\alpha}
& c_{\alpha}
\end{pmatrix}
\begin{pmatrix}
h\\
H
\end{pmatrix},
\end{eqnarray}
and 
\begin{eqnarray}
\begin{pmatrix}
\chi_{\Phi}\\
 \chi_{\Delta}
\end{pmatrix}
=
\begin{pmatrix}
c_{\beta^{0}} & 
-s_{\beta^{0}} \\
s_{\beta^{0}}
& c_{\beta^{0}}
\end{pmatrix}
\begin{pmatrix}
G^{0}\\
A^0
\end{pmatrix},
\end{eqnarray}
where $t_{\beta^{\pm}} = \frac{\sqrt{2} v_{\Delta}}{v_{\Phi}}$, $t_{\beta^0} = \sqrt{2} t_{\beta^{\pm}}$ and the mixing angles $\alpha$ between two neutral Higgs is taken into account. 
Their masses  are functions of  Higgs self couplings and $\mu$ as follows
\begin{eqnarray}
&& M_{H^{\pm\pm}}^2  =  \frac{\sqrt{2}\mu{v_{\Phi}^2}-\lambda_4v_{\Phi}^2v_\Delta-2\lambda_3v_\Delta^3}{2v_\Delta},
\\
&& M_{H^{\pm}}^2 =  \frac{(v_{\Phi}^2+2v_\Delta^2)\,[2\sqrt{2}\mu-\lambda_4v_\Delta]}{4v_\Delta},\\
&& M_{A^0}^2 =  \frac{\mu(v_{\Phi}^2+4v_\Delta^2)}{\sqrt{2}v_\Delta},
\\
&& M_H^2 = \frac{1}{2}\Big\{\lambda v_{\Phi}^2 s_\alpha^2 + c_\alpha^2 \Big[\sqrt{2}\mu\frac{v_{\Phi}^2}{v_\Delta}\big(1+4\frac{v_\Delta}{v_{\Phi}} t_{\alpha}\big) +4v_\Delta^2\big((\lambda_2+\lambda_3) - (\lambda_1+\lambda_4)\frac{v_{\Phi}}{v_\Delta} t_{\alpha}\big) \Big]\Big\},
\\
&& m_h^2 = \frac{1}{2}
\Big\{
\lambda v_{\Phi}^2 c_\alpha^2 + s_\alpha^2 \Big[ \sqrt{2}\mu\frac{v_{\Phi}^2}{v_\Delta}\big(1-4\frac{v_\Delta}{v_{\Phi} t_{\alpha}}\big) +4v_\Delta^2\big( (\lambda_1+\lambda_4) \frac{v_{\Phi}}{v_\Delta t_{\alpha}} + 
(\lambda_2+\lambda_3)
\big)\Big]
\Big\}.
\end{eqnarray}
The Yukawa Lagrangian is written in terms of  the mass eigenstates as follows 
\begin{eqnarray}
\label{YukawaTHM}
{\mathcal{L}}_{\rm Yukawa} = 
{\mathcal{L}}^{\rm SM}_{\rm Yukawa}
- L^T y_{\nu}C (i \sigma^2 \Delta) L 
+ \textrm{h.c},
\end{eqnarray}
where $L=(L_e,L_{\mu},L_{\tau})$ consists of three left-handed lepton doublets, 
$y_{\nu}$ is the $3\times3$ Yukawa 
coupling matrix  generating the neutrino masses, and  $C$ is the charge 
conjugation operator. Expanding 
the Yukawa Lagrangian and the Higgs potential in the mass basis of all particles, we derived all couplings that give one-loop contributions to the decay amplitudes $h\to f\bar{f} \gamma$. The respective vertices  are presented in Table~\ref{THM-coupling}, see a detailed derivation given in \ref{app_coupling}. 
\begin{table}[h!]
\centering
{\begin{tabular}
{l@{\hspace{1cm}}l@{\hspace{1cm}}l }
\hline \hline
\textbf{Vertices} &\textbf{Notations} 
& \textbf{Couplings}\\
\hline \hline \\
$hW_{\mu}W_{\nu}$  
& $g^{\rm THM}_{hWW}$
& 
$
-i \frac{2M_W^2}{v} ( c_{\alpha}\; c_{\beta^{\pm} }  
+ \sqrt{2} s_{\alpha} \; s_{\beta^{\pm} } )
g_{\mu\nu}
$
\\
\hline  \\
$hZ_{\mu}Z_{\nu}$  
& $g^{\rm THM}_{hZZ}$
&
$
-i \frac{2M_Z^2}{\sqrt{v^2+2v_{\Delta}^2 }}
(c_{\beta^0} c_{\alpha} 
+ 2 s_{\beta^0} s_{\alpha} )
g_{\mu\nu}
$
\\  \hline  \\
$hH^{\pm}H^{\mp}$
& $g^{\rm THM}_{hH^{\pm}H^{\mp}}$
&
$
-i\dfrac{c_{\alpha}}{v_{\Phi}}
\Big[ 
2 M_{H^{\pm}}^2 \dfrac{v_{\Phi}^2}{v^2}
+ 2m_h^2 \dfrac{v_{\Delta}^2}{v^2}
\Big]
$
\\
&&
$ 
-i\dfrac{s_{\alpha} 
}{v_{\Delta}}
\Big[ 
4 M_{H^{\pm}}^2 \dfrac{v_{\Delta}^2}{v^2}
+ m_h^2 \dfrac{v_{\Phi}^2}{v^2}
- M_{A^0}^2 \dfrac{v^2}{v^2+2 v_{\Delta}^2}
\Big]
$
\\
\hline  \\
$hH^{\pm\pm}H^{\mp\mp}$  
& $g^{\rm THM}_{hH^{\pm\pm}H^{\mp\mp}}$
&
$
-i\dfrac{2 v_{\Phi} \; c_{\alpha}}
{v_{\Phi}^2 + 4v_{\Delta}^2}
\Big[ 
2 M_{H^{\pm}}^2 
\Big(1+ \dfrac{2 v_{\Delta}^2}{v^2} \Big)
- M_{A^0}^2
\Big]
$
\\
&&
$
-i\dfrac{s_{\alpha}}{v_{\Delta}}
\Big[ 
2 M_{H^{\pm\pm}}^2 
- 4 M_{H^{\pm}}^2 \dfrac{v_{\Phi}^2}{v^2}
+ m_h^2 
$
\\
&& 
\hspace{2.5cm}
$
+ M_{A^0}^2 
\Big(1-\dfrac{4v^2_{\Delta}}
{v^2+2 v_{\Delta}^2} \Big)
\Big]
$
\\
\hline  \\
$Z_{\mu}H^{\pm}H^{\mp}$  
& $g^{\rm THM}_{ZH^{\pm}H^{\mp}}$
&
$
\frac{M_Z}{ \sqrt{v^2+2v_{\Delta}^2} }
(c^2_{W}-s_W^2-c^2_{\beta^{\pm}})
(p^+-p^-)_{\mu}
$
\\
\hline  \\
$Z_{\mu}H^{\pm\pm}H^{\mp\mp}$  
& $g^{\rm THM}_{ZH^{\pm\pm}H^{\mp\mp}}$
&
$
\frac{2M_Z}{  \sqrt{v^2+2v_{\Delta}^2} 
}(c^2_{W}-s_W^2)
(p^{++}-p^{--})_{\mu}
$
\\  \hline  \\
$A_{\mu} H^{\pm}H^{\mp}$  
& $g^{\rm THM}_{A H^{\pm}H^{\mp}}$
&
$ e (p^+-p^-)_{\mu}
$
\\   \hline  \\
$A_{\mu} H^{\pm\pm}H^{\mp\mp}$  
& $g^{\rm THM}_{A H^{\pm\pm}H^{\mp\mp}}$
&
$ (2 e) (p^{++}-p^{--})_{\mu}  $
\\
\hline  \\
$hf\bar{f}$  
& $g^{\rm THM}_{hf\bar{f}}$
&$
i\dfrac{m_f}{v} \dfrac{c_{\alpha}}{c_{\beta^{\pm} }}
$
\\\hline \\
$h\nu_l\bar{\nu}_l$  
& $g^{\rm THM}_{h\nu_l\bar{\nu}_l}$
&$
i\dfrac{m_{\nu_l}}{v_{\Delta}}
s_{\alpha} 
$
\\
\hline\hline  
\end{tabular}}
\caption{ The couplings involving 
the decay processes $h\rightarrow l\bar{l} \gamma$ 
in the THM. $A_{\mu}$ is the photon. \label{THM-coupling}}
\end{table}

It is noted that in the limit 
of $\frac{v_{\Delta} }{v} 
\rightarrow 0$, we have
\begin{eqnarray}
	\label{aprTHM}
	\dfrac{s_{\alpha}}{v_{\Delta}}
	\simeq \dfrac{v}{(m_h^2- M_{A^0}^2)c_{\alpha}}
	\left( 
	\lambda_1 + \dfrac{2M_{A^0}^2-4 M_{H^{\pm}}^2 }{v^2}
	\right)+ \mathcal{O}\left(\frac{v_{\Delta}^2}{v^2}\right).
\end{eqnarray}
All above couplings can be obtained appropriately by using the relation in (\ref{aprTHM}) in this limit. 

In the  THM, all parameters used in our analysis are 
\begin{eqnarray}
	\mathcal{P}_{\rm THM} =
	\{m_h^2, M_H^2, M_{A^0}^2, 
	M^2_{H^{\pm}},M^2_{H^{\pm\pm}},
	t_{\beta^{\pm}}, 
	s_{\alpha} \}.
\end{eqnarray}

\section{ \label{sec_OL} One-loop expressions 
for  $h\rightarrow l\bar{l}\gamma$ in HESMs }             
\begin{figure}[ht]
\centering
\includegraphics[width=7cm, height=5cm]
{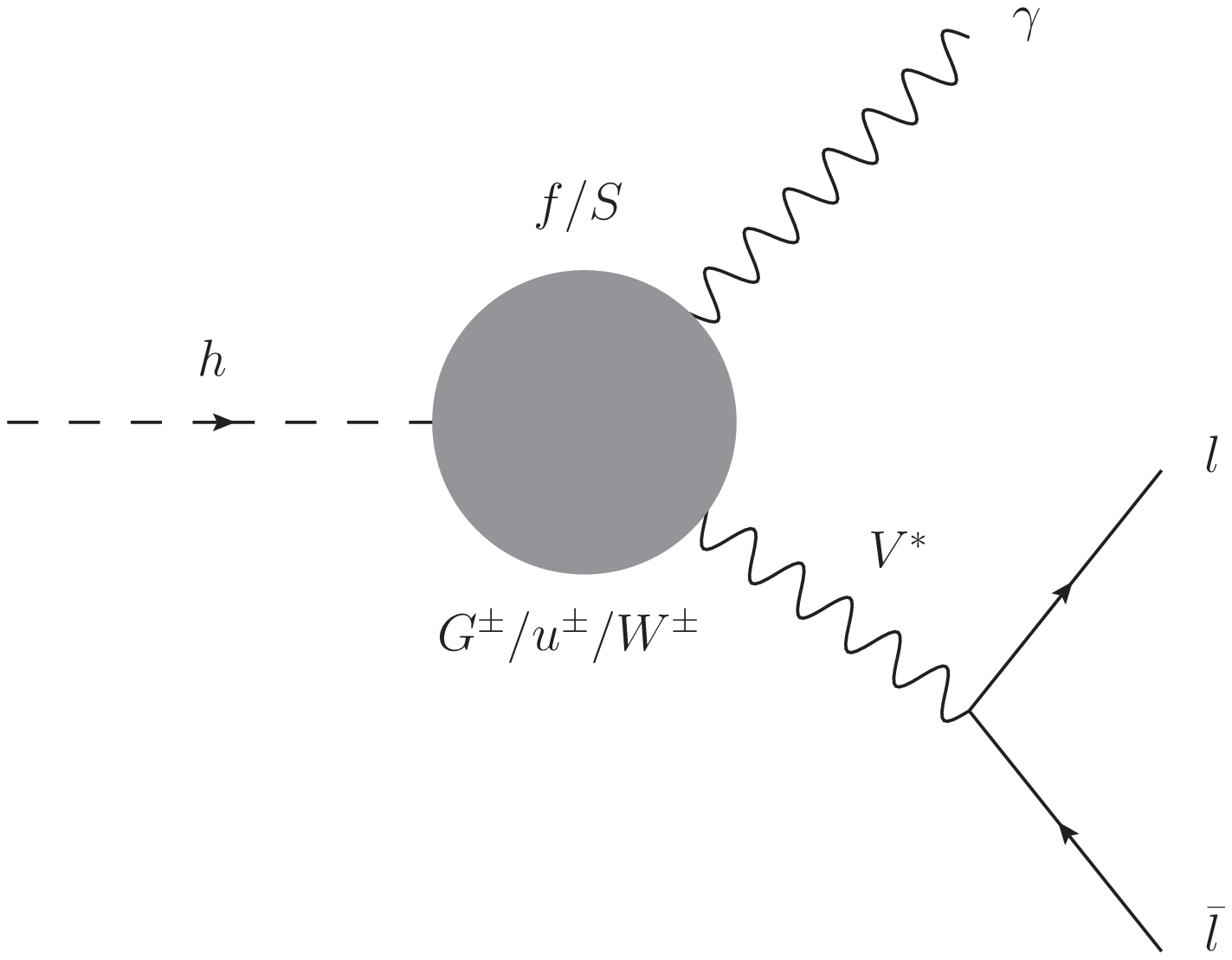} \caption{\label{feynFFgPoleV0} One-loop Feynman diagrams for  the decay $h\rightarrow l \bar{l} \gamma$ belonging to group 1. Here  $S$, $G^{\pm}$, and $u^{\pm }$  are Higgs boson, Nambu-Goldstone bosons,  and 
Ghost, respectively. }
\end{figure}

\begin{figure}[ht]
\centering
\includegraphics[width=12cm, height=5cm]
{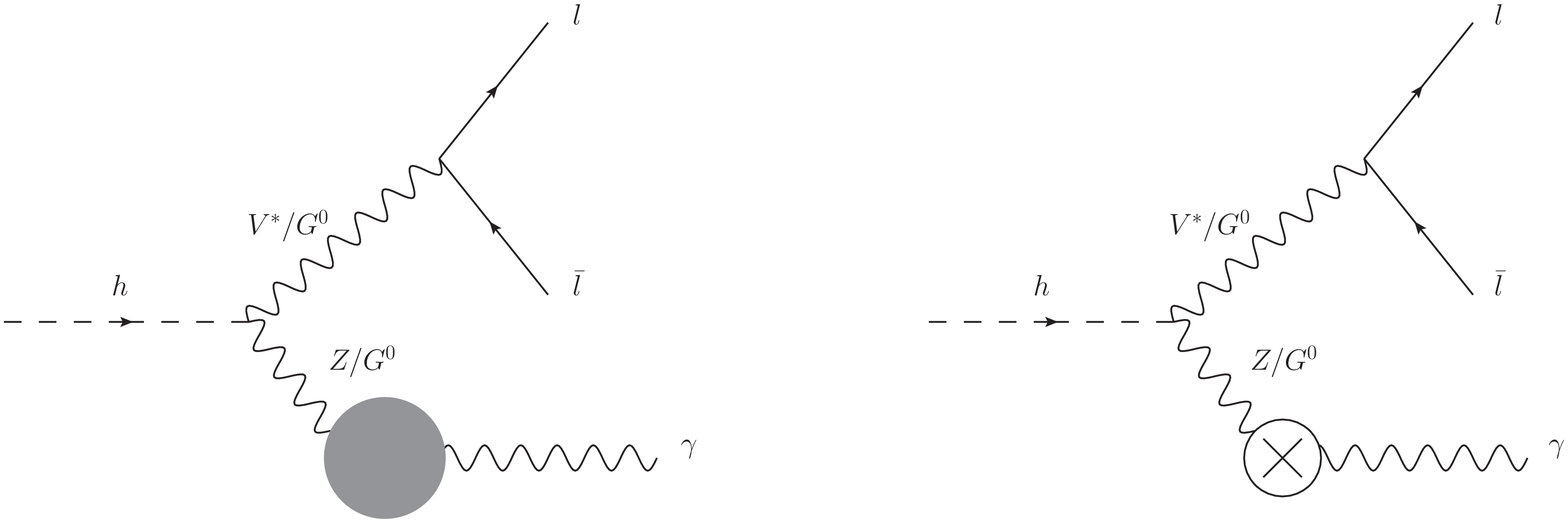}
\caption{\label{feynFFg-VpoleZRO} One-loop Feynman diagrams for the decay  $h\rightarrow l \bar{l} \gamma$ in group 1. $G^{0}$ is the	Nambu-Goldstone boson. 
}
\end{figure}

\begin{figure}[ht]
\centering
\includegraphics[width=12cm, height=5cm]
{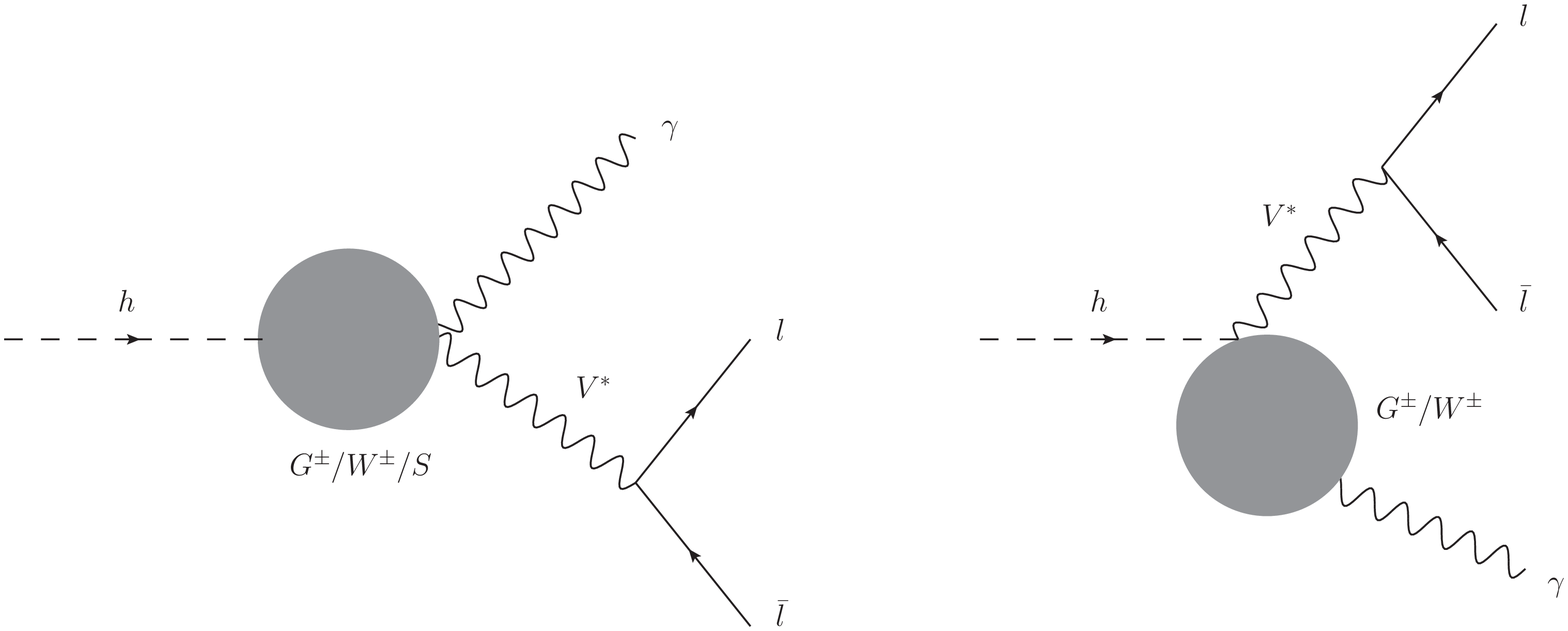}
\caption{\label{feynFFgV0Pole} 
One-loop Feynman diagrams for
the decay 
$h\rightarrow l 
\bar{l} \gamma$ in group 1.}
\end{figure}

\begin{figure}[ht]
\centering
\includegraphics[width=12cm, height=7cm]
{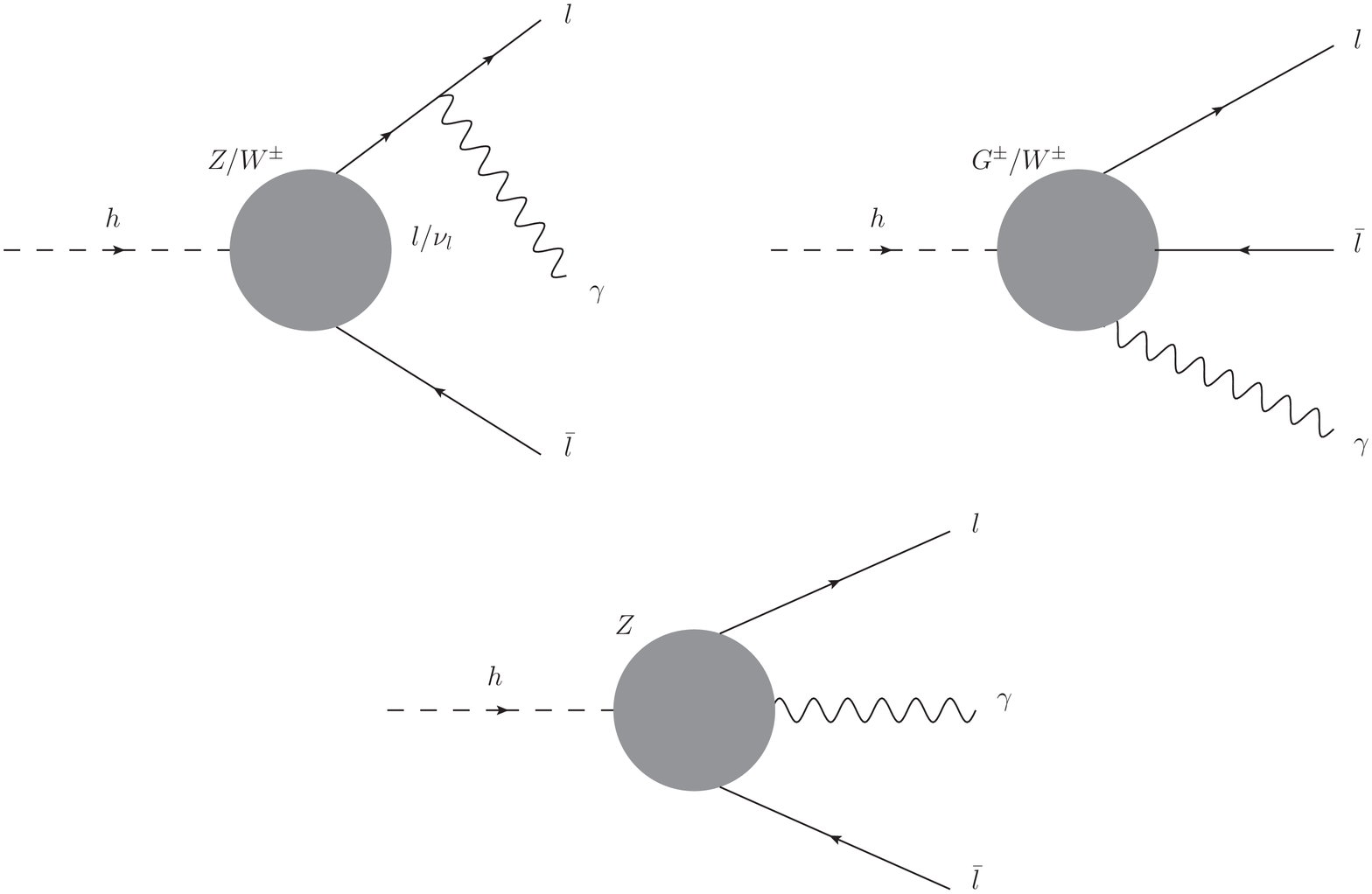}
\caption{\label{feynFFgNonPole} 
One-loop Feynman diagrams for
the decays  
$h\rightarrow l 
\bar{l} \gamma$ 
classifying into group 2. 
}
\end{figure}

In order to generalize one-loop expressions for $h\rightarrow l(q_1)\bar{l}(q_2)\gamma(q_3)$ in  the HESMs considered in this work, we define the following notations. The common notation $g^{(\textrm{NP})}_{\textrm{vertex}}$ denotes all vertices appearing in the HESMs with NP $\equiv$ IDM, THDM and THM.  These couplings were listed in Tables~\ref{IDM-coupling},
\ref{THDM-coupling}, \ref{YukawaTHDM}, and \ref{THM-coupling} for particular HESM.  
Noting that we use $f$ for all internal fermions exchanging in the loop
and $l$ is for external fermions in our compuatations. 
For this calculation, we choose to work in the on-shell renormalization scheme, hence  no
diagrams giving one-loop corrections  to external legs. In the HF gauge, all one-loop Feynman diagrams can be separated into  two following groups. Group 1 includes all 
$V^*$-pole diagrams $h\to  V^*\gamma\to l\bar{l}\gamma$, where the $V^*$ exchange may be neutral or charged gauge $Z^*, \gamma^*$, or Higgs  $S^*$  bosons, as  described in  Figs.~\ref{feynFFgPoleV0},  \ref{feynFFg-VpoleZRO}, and \ref{feynFFgV0Pole}, where  particle exchangings in loops may be 
fermions, gauge bosons, Nambu-Goldstone,  Higgs bosons or Ghost particles. 
 A  $V^*$-pole diagram is realized as it always consists of a virtual boson $V^*$ decaying into two final lepton states through the vertex $V^*l\bar{l}$.   It is well-known that the sum of all  diagrams given in Fig.~\ref{feynFFg-VpoleZRO}  vanishes for the final on-shell external photon \cite{VanOn:2021myp}. 
 Furthermore, we are going to collect only form factors relating with the parts proportional to $q^{\mu}q^{\nu}_3$ appearing in Eqs.~(\ref{ampV0POLE}, \ref{boxAmp}).  Therefore,  all diagrams in  Fig.~\ref{feynFFgV0Pole} are ignored in this work, because their contributions  are always proportional to  $g^{\mu\nu}$. 
Group 2 consists of all remaining Feynman 
diagrams giving none $V^*$-pole contributions to the decay amplitudes $h\to l\bar{l} \gamma$ 
(Non-pole contributions
hereafter). 
This group contains  several one-loop 
box diagrams, as given in  Fig.~\ref{feynFFgNonPole}.

Our calculations are performed as follows. 
We firstly make a model file in {\tt FeynArt}~\cite{Hahn:2000kx}, then use {\tt FormCalc}~\cite{Hahn:1998yk} and  {\tt FeynCalc}~\cite{Mertig:1990an} to generate automatically  
one-loop amplitudes. Related one-loop form factors  are collected in terms of Passarino-Veltman (PV-) functions before transformed in terms of  scalar one-loop integrals~\cite{Hahn:1998yk,Denner:2005nn}. We finally use {\tt Package-X} \cite{Patel:2015tea}  
to perform the $\epsilon$-expansions for scalar one-loop integrals with respect to logarithmic and di-logarithm functions. 
\subsection{$V^*$-pole contributions}
 As mentioned 
in the above arguments, it is enough to consider the contributions from diagrams given in Fig.~\ref{feynFFgPoleV0}.  In principle,  
$V^*$-pole diagrams may appear with all $V^*=Z^*, \gamma^*$, and neutral Higgs bosons.  Since the couplings of  $S^*$ to leptons are proportional to $m_l/v$ in most of the HESMs,  the later cases
give much smaller contributions than the former ones. As a result, we only concern $Z^*, \gamma^*$-poles in the current work. Now, 
the one-loop amplitude can be decomposed into the following Lorentz  structures: 
\begin{equation}
\label{ampV0POLE}
\mathcal{A}_{V^*\text{-pole}}
=
	\sum\limits_{V^*=\{\gamma^*, Z^*\} }
	F^{V^*\textrm{-pole}}_{\gamma V^*}
	\Big[
	q^\mu q_3^\nu 
	-
	(q \cdot q_3) g^{\mu \nu}
	\Big]
	\Big[
	\bar{u}(q_1)
	\gamma_\nu
	\Big(
	\sum\limits_{j=\{L,R\}}
	g_{V^*l \bar{l}}^j \; P_j 
	\Big)
	v(q_2)
	\Big]
	\varepsilon^*_\mu (q_3),
\end{equation}
where 
$q= q_1+q_2$, 
$\varepsilon^*_\mu (q_3)$ is the  final
photon polarization, and 
\begin{eqnarray}
\label{gV*ll}
g_{V^* l \bar{l}}^j
&=&
\begin{cases}
- e Q_l 
&\text{if $V^* \equiv \gamma^*$},
\vspace{0.25cm}
\\
\dfrac{e}{s_W c_W}
\Big(
I_{3,l}^j - Q_l s_W^2
\Big) 
&\text{if $V^* \equiv Z^*$}
\end{cases}     
\end{eqnarray}
is the coupling of the 
vertex $V^* \cdot 
l \cdot \bar{l}$  with $j = \{L,R\}$, $P_{L,R} =
\frac{1\mp \gamma_5}{2}$. In addition,  $Q_l =-1$,  $I_{3,l}^L=1/2,I_{3,l}^R=0$ for $l\equiv e,\mu$; and  $Q_l =0$,  $I_{3,l}^L=-1/2,I_{3,l}^R=0$ for  $l\equiv \nu_{e,\mu,\tau}$.  Each one-loop form factor $F^{V^*\textrm{-pole}}_{\gamma V^*}$
is decomposed into three parts as follows 
\begin{eqnarray}
 F^{V^*\textrm{-pole}}_{\gamma V^*}
 =
\dfrac{e^2 g}{(4\pi)^2 M_W}   
\dfrac{1}{q_{12} - 
M_V^2 + i M_V \Gamma_V}
\Big[
F^{(f)}_{\gamma V^*}
+F^{(W)}_{\gamma V^*}
+F^{(S)}_{\gamma V^*} 
\Big],
\end{eqnarray}
where $M_V, \; \Gamma_V$ are mass and 
decay width of  the gauge  boson $V^*$, and 
$q_{12} = (q_1+q_2)^2$. Here, we include all contributions 
from fermion exchanges $f$ in the loops, namely  
\begin{eqnarray}
F^{(f)}_{\gamma V^*}
&=&
\sum\limits_{f} 
\dfrac{ N_f^C
m_f\;v
}{(2 \pi \alpha) (m_h^2-q_{12})^2}
\; g_{hf\bar{f}}^{\textrm{NP} }
g_{\gamma f\bar{f}}^L 
(g_{V^*f\bar{f}}^L+g_{V^*f\bar{f}}^R)
\times
\\
&& \times
\Bigg\{ 
2 q_{12} 
\Big[
B_{0}(m_h^2,m_f^2,m_f^2)
- B_{0}(q_{12},m_f^2,m_f^2)
\Big]
\n \\
&& \hspace{0cm}
+(m_h^2-q_{12}) 
\Big[
2
+
(q_{12}-m_h^2+4 m_f^2)
C_{0}(0,q_{12},m_h^2,m_f^2,m_f^2,m_f^2)
\Big]
\Bigg\}.
\nonumber
\end{eqnarray}
Here $g_{\gamma f\bar{f}}^{L, R}$ and  $g_{V^*f\bar{f}}^{L, R}$ are the couplings of  photon 
and $V^*$ to fermions, respectively.

Regarding the $W$ exchange in the loop diagrams,  we know that  there are also involved Nambu-Goldstone  bosons, and Ghost exchanges in the loop within the HF  gauge. In order to find out a representation which are valid for 
both $V^* =Z^*$ and $\gamma^*$, one 
adds the parameters 
$\delta_{V^*W}$ and 
$\delta_{V^*G}$ to the couplings 
of $V^*$ with W bosons, Goldstone bosons
and Ghost particles as in 
Table~\ref{VGcouplings}.
\begin{table}[ht]
\begin{center}
{\begin{tabular}
{l@{\hspace{2cm}}l}
\hline \hline
\textbf{Vertices}  
& \textbf{Couplings}\\
\hline \hline \\
$g_{V^*WW}$ & $\delta_{V^* W}\;e$ 
\\ \hline \\
$g_{V^*WG}$ & $\delta_{V^*G}\;(-eM_W)$
\\ \hline \\
$g_{V^*u^{\pm}u^{\mp}}$ 
& $\mp \delta_{V^*W}\;e$
\\ \hline \\
$g_{V^*G^{\pm}G^{\mp}}$ & 
$\frac{e}{2}
 \left(
 \delta_{V^*W}-\delta_{V^*G}
 \right)
 $
 \\ \hline \hline
\end{tabular}}
\end{center}
\caption{\label{VGcouplings}
Generalizations of the couplings
of $V^* =Z^*$ and $\gamma^*$ to
$W$, $G^\pm$, and the Ghost 
particle $G$ in the HF gauge.}
\end{table}
Subsequently, one-loop 
form factors for this contribution
are given in the compact form as 
\begin{eqnarray}
F^{(W)}_{\gamma V^*}
&=&
\dfrac{g_{hWW}^{\textrm{NP}} \; v}
{2 M_W^2\; (m_h^2-q_{12})^2}
\Bigg\{
2 M_W^2 (m_h^2-q_{12}) 
\Big[
(\delta_{V^*G}-5 \delta_{V^*W}) 
(m_h^2-2 M_W^2)
\\
&&
\hspace{3.5cm}
-2 q_{12} 
(\delta_{V^*G}-3 \delta_{V^*W})
\Big]
C_{0}(0,q_{12},m_h^2,M_W^2,M_W^2,M_W^2)
\n \\
&&
\hspace{2cm}
-\Big[
\delta_{V^*G} (m_h^2+2 M_W^2)
-\delta_{V^*W} (m_h^2+10 M_W^2)
\Big] 
\times
\nonumber\\
&& 
\hspace{2cm}
\times
\Big[
q_{12} 
\Big(
B_{0}(m_h^2,M_W^2,M_W^2)
- B_{0}(q_{12},M_W^2,M_W^2)
\Big)
+(m_h^2-q_{12})
\Big]
\Bigg\}.
\nonumber 
\end{eqnarray}
The mentioned parameters are taken for $Z^*$-pole and $\gamma^*$-pole
as follows:
\begin{eqnarray}
\label{coeVll}
\delta_{V^*W}
\, ; \,
\delta_{V^*G}
&=&
\begin{cases}
-1
\, ; \,
1
&\text{if $V^* \equiv \gamma^*$},
\vspace{0.25cm}
\\
\dfrac{c_W}{s_W}
\, ; \,
\dfrac{s_W}{c_W}
&\text{if $V^* \equiv Z^*$}.
\end{cases}
\end{eqnarray}
It is worth to mention that we have a full set of gauge invariance Feynman diagrams for each $V^*$-pole contribution in  the HF gauge. Hence, we can derive the above results to many of well-known  expressions for $h\rightarrow \gamma\gamma, Z \gamma$
by considering the final state as  $V^* \rightarrow Z, \gamma$, respectively. For this purpose, the PV-functions $B_0$ and $C_0$ appearing in the above one-loop form factors can expressed via the 
basic $I_k$ functions (for $k = 1, 2$)~ \cite{Djouadi:2005gi} as follows:
\begin{align}
C_{0}(0,q_{12},m_h^2,M_i^2,M_i^2,M_i^2)
&=
- \dfrac{I_2 (\tau_i, \lambda_i)}{M_i^2},
\\
B_{0}(m_h^2,M_i^2,M_i^2)
- B_{0}(q_{12},M_i^2,M_i^2)
&=
\dfrac{m_h^2 - q_{12}}{2 M_i^2 q_{12}}
\Big\{
2 M_i^2
\Big[
2 I_2 (\tau_i, \lambda_i)
- 1
\Big]
+ (q_{12} - m_h^2)
I_1 (\tau_i, \lambda_i) 
\Big\},
\nonumber\\
\end{align}
where $\tau_{i} = 4M_i / m_h^2$ 
and $\lambda_{i} = 4 M_i^2  / q_{12}$.
The reductions for  
$h\rightarrow \gamma\gamma, Z\gamma$
with on-shell $Z$ and photon in 
final states are presented in the 
following paragraphs. 
\begin{itemize}
\item \underline{
$\gamma^*$-pole contributions:}

In case of $\gamma^*$-pole contributions,
taking a limit of $M_{V^*}$ tends to zero 
as well as 
the corresponding couplings in 
Eq.~(\ref{gV*ll})
and selecting the appropriated
parameters in (\ref{coeVll}), one-loop 
form factors $F^{(f,W)}_{\gamma \gamma^*}$
are casted into the form of 
\begin{eqnarray}
F^{(f,W)}_{\gamma \gamma^*}
= F^{(f)}_{\gamma \gamma^*} 
+ F^{(W)}_{\gamma \gamma^*}
&=&
\sum \limits_f
N^C_f
Q_f^2
F^{(1/2)}_{\gamma \gamma^*} (\tau_f, \lambda_f)
+
F^{(1)}_{\gamma \gamma^*} (\tau_W, \lambda_W).
\end{eqnarray}
Where form factors 
$F^{(1)}_{\gamma \gamma^*}$,
$F^{(1/2)}_{\gamma \gamma^*}$
are taken into
account in the above equation
as follows:
\begin{eqnarray}
F^{(1)}_{\gamma \gamma^*} (\tau_W, \lambda_W)
&=&
16 \, I_2 (\tau_W, \lambda_W)
-
\big(
4/\tau_W
+ 6
\big)
I_1 (\tau_W, \lambda_W),
\\
F^{(1/2)}_{\gamma \gamma^*} 
(\tau_f, \lambda_f)
&=&
4 I_1 (\tau_f, \lambda_f)
-
4 I_2 (\tau_f, \lambda_f).
\end{eqnarray}
For the decay $h \rightarrow \gamma \gamma$ corresponding to the limit of $q_{12} \rightarrow 0$, 
we  derived once again the SM results  shown
in Ref. \cite{Marciano:2011gm}. 
\item \underline{$Z^*$-pole 
contributions:}

Similarly, in the $Z^*$-pole 
contributions with  
$M_{V^*} = M_Z, $ and $\Gamma_{V^*} 
= \Gamma_Z$, the respective one-loop form factor 
$F^{(f,W)}_{\gamma Z^*}$ is:
\begin{eqnarray}
F^{(f,W)}_{\gamma Z^*}
=F^{(f)}_{\gamma Z^*}
+
F^{(W)}_{\gamma Z^*}
&=&
\sum \limits_f
N^C_f
\dfrac{Q_f v_f}{s_W c_W}
F^{(1/2)}_{\gamma Z^*} (\tau_f, \lambda_f)
+
F^{(1)}_{\gamma Z^*} (\tau_W, \lambda_W),
\end{eqnarray}
where $v_f = 2 I_3^f - 4 Q_f s_W^2$, and  
\begin{align}
F^{(1)}_{\gamma Z^*} (\tau_W, \lambda_W)
&=
\dfrac{c_W}{s_W}
\Big\{
\Big[
\Big(
\dfrac{2}{\tau_W}
+ 1
\Big)
\dfrac{s_W^2}{c_W^2}
-
\Big(
\dfrac{2}{\tau_W} + 5
\Big)
\Big]
I_1 (\tau_W, \lambda_W)
+
4
\Big(
3-\dfrac{s_W^2}{c_W^2}
\Big)
I_2 (\tau_W, \lambda_W)
\Big\},
\nonumber
\\
F^{(1/2)}_{\gamma Z^*} (\tau_f, \lambda_f)
&=
I_1 (\tau_f, \lambda_f)
-
I_2 (\tau_f, \lambda_f).
\end{align}
The decay process $h \rightarrow Z \gamma$  corresponds to  
$q_{12} =M_Z^2$, leading to the  SM results presented in Refs.  
 \cite{Djouadi:1996yq, Hue:2017cph}.
\end{itemize}
We turn our attention to the case of a Higgs boson $S$ exchanging in the loop of a diagram, which gives the one-loop form factor  $F^{(S)}_{\gamma V^*}$  determined as follows 
\begin{eqnarray}
F^{(S)}_{\gamma V^*}
&=&
\dfrac{4  M_W}{g (m_h^2-q_{12})^2}
\sum \limits_S
Q_S^2 
\, g^{(NP)}_{hSS} 
\, g^{(NP)}_{ASS} 
\, g^{(NP)}_{VSS} 
\Bigg\{ 
q_{12} 
\Big[
B_{0}(m_h^2,M_S^2,M_S^2)
- B_{0}(q_{12},M_S^2,M_S^2)
\Big]
+
\n \\
&&\hspace{0.2cm}
+(m_h^2-q_{12}) 
\Big[
2 M_S^2 \,
C_{0}(0,q_{12},m_h^2,M_S^2,M_S^2,M_S^2)
+1
\Big]
\Bigg\},
\end{eqnarray}
where $Q_S$ is the electric charge of  $S$. Any neutral Higgs bosons $S$ with $Q_S=0$ do not contribute to $F^{(S)}_{\gamma V^*}$. The decay processes $h \rightarrow V \gamma$ have amplitudes with one-loop form factors derived from $F^{(S)}_{\gamma V}$ as follows  
\begin{eqnarray}
F^{(S)}_{\gamma \gamma}
&=&
\dfrac{4  M_W}{g m_h^2}
\sum \limits_S
Q_S^2 
\, g^{(NP)}_{hSS} 
\, \Big[g^{(NP)}_{ASS}
\Big]^2
\,
\Big[ 
2 M_S^2 \,
C_{0}(0,0,m_h^2,M_S^2,M_S^2,M_S^2)
+1
\Big],
\\
F^{(S)}_{\gamma Z}
&=&
\dfrac{4  M_W}{g (m_h^2-M_Z^2)^2}
\sum \limits_S
Q_S^2 
\, g^{(NP)}_{hSS} 
\, g^{(NP)}_{ASS} 
\, g^{(NP)}_{ZSS} 
\,
\Bigg\{ 
M_Z^2
\Big[
B_{0}(m_h^2,M_S^2,M_S^2)
- B_{0}(M_Z^2,M_S^2,M_S^2)
\Big]
\n \\
&&\hspace{3.5cm}
+(m_h^2-M_Z^2) 
\Big[
2 M_S^2 \,
C_{0}(0,M_Z^2,m_h^2,M_S^2,M_S^2,M_S^2)
+1
\Big]
\Bigg\}.
\end{eqnarray}
The above form factors can be written via the well-known $I_k$ functions for $k = 1, 2$ \cite{Djouadi:2005gi}, namely:
\begin{eqnarray}
F^{(S)}_{\gamma \gamma}
&=&
\dfrac{M_W}{g}
\sum \limits_S
Q_S^2 
\, 
\dfrac{g^{(NP)}_{hSS} 
\, [g^{(NP)}_{ASS} ]^2}{M_S^2}
\,
\tau_S
\Big[
1-\tau_S f(\tau_S)
\Big],
\\
F^{(S)}_{\gamma Z}
&=&
-
\dfrac{M_W}{2 g}
\sum \limits_S
Q_S^2 
\, 
\dfrac{g^{(NP)}_{hSS} 
\, g^{(NP)}_{ASS} 
\, g^{(NP)}_{ZSS} }{M_S^2}
\,
\tau_S
I_1 (\tau_S, \rho_S),
\end{eqnarray}
where $\tau_S = 4 M_S^2 / m_h^2$,
$\rho_S = 4 M_S^2 / M_Z^2$, and  all analytic expressions 
of the basis functions $I_1 (x, y), I_2 (x, y)$ 
and $f(z)$ are shown given in \ref{app_Ik}.

We also give several interesting 
comments on our results as follows. 
First, summing of all 
$\gamma^*$-contributions, we can derive 
the results for $h\rightarrow \gamma \gamma$
in many EHSM, for examples, the 
IDM~\cite{Arhrib:2012ia}, THM~\cite{Akeroyd:2012ms}, and a class
of EHSM~\cite{Chiang:2012qz}. For 
$h\rightarrow Z \gamma$
in many of EHSM~\cite{Chiang:2012qz, Benbrik:2022bol,Chen:2013dh,Chabab:2014ara}, we can select all the contributions from $Z^*$-pole. 
Our formulas  confirm again the previous results for many 
EHSM. 
\subsection{Non-pole $V$-contributions}
The non-pole contributions come from the Feynman diagrams in group $2$. There are many kinds of 
Feynman diagrams which are 
involed to vector $W, Z$ bosons, 
their Goldstone bosons and the 
cases of scalar Higgs exchanging, 
mixing of 
vector boson and scalar Higgs 
in the loop diagrams. The later 
give much smaller contributions than 
the former due to the appearance 
of the couplings $S\bar{l}l$ proportional to $m_l$. 
Therefore, we ignore all contributions from diagrams having this kind of couplings. 
The remaining ones can be divided into two parts. The first (second)  part includes $Z$ ($W$)  and its Goldstone boson  in the loop diagrams, 
respectively. One-loop amplitude 
is given
\begin{equation}
\label{boxAmp}
\mathcal{A}_{\text{Non-pole, V}} 
= \sum\limits_{V=\{Z,W, S\}}
\sum\limits_{k=1}^2 
\Big\{
[q_3^{\mu}q_k^{\nu} 
- g^{\mu\nu}q_3\cdot q_k]
\bar{u}(q_1)
\Big(
\sum\limits_{j=L,R}
F_{k,j}^{\text{Non-pole, V}}
\gamma_{\mu}P_j 
\Big)
v(q_2) 
\Big\}
\varepsilon^{*}_{\nu}(q_3).
\end{equation}

As we have pointed out in above arguments that
the contributions of $S$ in the loop diagrams
can be obmited. It is enough to consider 
$Z, W$ exchanging in the loop 
diagrams. In the first contribution, we 
concern one-loop diagrams with $Z$ boson
internal lines. One loop form factors 
are written in terms of 
scalar PV-functions as follows:
\begin{eqnarray}
F^{\text{Non-pole, Z}}_{1,L}
&=&\dfrac{\alpha \; v}{\pi }
\dfrac{g_{hZZ}^{\textrm{NP} } 
\; (g_{Z l \bar{l}}^L)^2 
}{M_Z \; s_{2W}}
\Big[
D_{2}
+D_{12}
+D_{23} 
\Big]
(0,q_{13},m_h^2,q_{23},
0,0,m_l^2,m_l^2,M_Z^2,M_Z^2) ,
\nonumber\\
&&\\
F^{\text{Non-pole, Z}}_{1,R}
&=&
F^{\text{Non-pole, Z}}_{1,L}
\Big|
g_{Z l \bar{l}}^L
\rightarrow
g_{Z l \bar{l}}^R,
\\
F^{\text{Non-pole, Z}}_{2,L}
&=&
F^{\text{Non-pole, Z}}_{1,L}
\Big|
q_{13}
\leftrightarrow
q_{23}, 
\\
F^{\text{Non-pole, Z}}_{2,R}
&=&
F^{\text{Non-pole, Z}}_{2,L}
\Big|
g_{Z l \bar{l}}^L
\rightarrow
g_{Z l \bar{l}}^R,
\end{eqnarray}
where  $q_{13}= (q_1+q_3)^2, 
q_{23} = (q_2+q_3)^2$.  Changing into the forms of scalar integrals, we have
\begin{eqnarray}
F^{\text{Non-pole, Z}}_{1,L}
&=&
\dfrac{\alpha \; v\;}
{2\pi} 
\dfrac{g_{hZZ}^{\textrm{NP}}
\;
(g_{Z l \bar{l}}^L)^2
}{M_Z s_{2W} \; 
q_{12}\; q_{13}^2 \;  
(q_{12}+q_{13})}
\; 
\dfrac{q_{23}}{q_{23}-m_l^2}
\times
\\
&&
\times
\Bigg\{
2 q_{12} q_{13} 
\Big[
B_{0}(m_h^2,M_Z^2,M_Z^2)
- 
B_{0}(q_{23},0,M_Z^2)
\Big]
\n \\
&&
-\Big[
M_Z^2 
\big(q_{12}^2
+2 q_{12} q_{13}
-q_{13}^2\big)
+
\big(q_{13}^2
-q_{12}^2 
\big)
q_{13}
\Big] 
C_{0}(0,m_h^2,q_{23},0,M_Z^2,M_Z^2)
\n \\
&&
- q_{23} (M_Z^2-q_{13}) 
(q_{12}+q_{13})  
C_{0}(0,0,q_{23},0,0,M_Z^2)
\n \\
&&
+(m_h^2-q_{13}) 
(M_Z^2-q_{13}) (q_{12}+q_{13}) 
C_{0}(q_{13},m_h^2,0,0,M_Z^2,M_Z^2)
\n \\
&&
+q_{13} (q_{13}-M_Z^2) (q_{12}+q_{13}) 
C_{0}(0,q_{13},0,0,0,M_Z^2) 
\n \\
&&
-(M_Z^2-q_{13}) 
(q_{12}+q_{13}) 
\Big[
m_h^2 (M_Z^2-q_{13})
-M_Z^2 q_{12}
+q_{13} (q_{12}+q_{13})
\Big]
\times
\n \\
&&\hspace{5.2cm} \times 
D_{0}(0,q_{13},m_h^2,
q_{23},0,0,0,0,M_Z^2,M_Z^2) 
\Bigg\}.  
\n 
\end{eqnarray}
We next arrive at the non-pole      
$W$-contributions described in Fig. \ref{feynFFgNonPole}. 
The one-loop amplitude is the same
 as that in \eqref{boxAmp} with the following form
factors:
\begin{eqnarray}
F^{\text{Non-pole, W}}_{1,L}
&=&- 
\dfrac{\alpha^2\; v 
\;
g_{hWW}^{\textrm{NP} }}{2\;M_W s_W^3}
\Bigg\{
\Big[
D_{1}
+D_{13} 
\Big]
(0,q_{12},0,q_{23},0,m_h^2,
0,M_W^2,M_W^2,M_W^2) 
\\
&&\hspace{2cm}
+
\Big[
D_{2}
-D_{23}
-D_{33} 
\Big]
(0,q_{12},0,q_{13},
0,m_h^2,0,M_W^2,M_W^2,M_W^2) 
\Bigg\},  \n \\
&& \n \\
F^{\text{Non-pole, W}}_{1,R}
&=& F^{\text{Non-pole, W}}_{2,R}=0, 
\\
F^{\text{Non-pole, W}}_{2,L}
&=&
F^{\text{Non-pole, W}}_{1,L}
\Big|
q_{13}
\leftrightarrow
q_{23}. 
\end{eqnarray}
 $F^{\text{Non-pole, W}}_{k,R} = 0$  with  $k = 1, 2$ because the $W$ boson only couples to left-handed neutrinos. The expression of non-pole $W$-boson contributions in terms of one-loop scalar integrals are: 
\begin{eqnarray}
F^{\text{Non-pole, W}}_{1,L}
&=&
\dfrac{
\alpha^2\; v\;
g_{hWW}^{\textrm{NP} }
}
{2\; M_W \; s_W^3 q_{12} 
(q_{12}+q_{13}) q_{13}^2}
\dfrac{q_{23}}{q_{23}-m_l^2}
\times
\\
&&\hspace{-0.0cm} \times
\Bigg\{
- 2 q_{12} q_{13}  
\Big[
B_{0}(m_h^2,M_W^2,M_W^2)
- 
B_{0}(q_{23},0,M_W^2)
\Big]
\n \\
&& 
-  
(q_{12}^2+2 q_{12} 
q_{13}-q_{13}^2) 
(-M_W^2+q_{12}+q_{13}) 
C_{0}(0,m_h^2,q_{23},0,M_W^2,M_W^2)
\n \\
&& 
+ q_{12}^2 (q_{12}+q_{13})  
C_{0}(0,q_{12},0,0,M_W^2,M_W^2)
\n \\
&& 
+ q_{23}
(q_{12}+q_{13}) 
(-M_W^2+q_{12}+q_{13}) 
C_{0}(0,0,q_{23},0,M_W^2,M_W^2) 
\n \\
&& 
- q_{13} 
(M_W^2-q_{13}) 
(q_{12}+q_{13}) 
C_{0}(0,0,q_{13},0,M_W^2,M_W^2) 
\n \\
&& 
+ (m_h^2-q_{13}) 
(q_{13}-M_W^2) 
(q_{12}+q_{13}) 
C_{0}(0,m_h^2,q_{13},0,M_W^2,M_W^2) 
\n \\
&& 
- 
(m_h^2-q_{12}) 
(q_{12}+q_{13})  
(-2 M_W^2+q_{12}+2 q_{13}) 
C_{0}(q_{12},0, m_h^2, M_W^2,M_W^2,M_W^2) 
\n \\
&& 
- (q_{12}+q_{13})  
\Big[
q_{13} M_W^2 
(M_W^2+q_{12}-q_{13})
\n \\
&& 
+ (M_W^2-q_{12}) 
(M_W^2-q_{12}-q_{13})
\Big] 
D_{0}(0,q_{12},0,q_{23},0,
m_h^2,0,M_W^2,M_W^2,M_W^2) 
\n \\
&& 
- (M_W^2-q_{13}) 
(q_{12}+q_{13}) 
\Big[
m_h^2 M_W^2
+q_{12} (q_{13}-M_W^2)
\Big] 
\times
\n \\
&&\hspace{5cm} \times
D_{0}(0,q_{12},0,q_{13},
0,m_h^2,0,M_W^2,M_W^2,M_W^2) 
\Bigg\}. 
\n
\end{eqnarray}
The analytic expressions of the PV-fucntion $C_0$, $D_0$
 are then written in terms of the logarithm and the
di-logarithmic functions as given in \ref{app_loga}.

The total one-loop 
form factors are computed as follows:
\begin{eqnarray}
\label{FLR}
 F_{k,L/R} &=& 
 \sum\limits_{V^*=\{\gamma^*, Z^*\}}
 g^{L/R}_{V^*l\bar{l}}
 \cdot 
 F^{V^*\textrm{-pole}}_{\gamma V^*}
 + 
 \sum\limits_{V=\{Z, W\}}
 F_{k,L/R}^{\text{Non-pole, $V$} }
\end{eqnarray}
for $k=1,2$. 

We first check our  calculations by testing the independent of the  $UV$-divergent, $IR$-divergent, 
$\mu^2$-independent of the form factors.  We also cross-check  the decay rates to our previous work and other references if they are  available \cite{Kachanovich:2020xyg, VanOn:2021myp,Phan:2020xih}. For these decay processes, we have no $IR$-divergent. Since we 
don't have photon exchange in the loop diagrams. For  $UV$-divergence, $\mu^2$-independence checking analytically,  we refer \ref{eq_UV} for more details. Numerical checks our work with other
papers and our previous works will be shown in the next section. We confirm the correctness of the calculations.

Having the correctness form factors for the decay processes,  the decay rate 
is given by \cite{Kachanovich:2020xyg}:   
\begin{eqnarray}
\label{decayrate}
 \dfrac{d\Gamma}{d q_{12}q_{13}} 
 = \dfrac{q_{12}}{512 \pi^3 m_h^3}
 \Big[
 q_{13}^2(|F_{1,R}|^2 
 + |F_{2,R}|^2)
 + q_{23}^2(|F_{1,L}|^2 
 + |F_{2,L}|^2)
 \Big]. 
\end{eqnarray}
Taking the above integrand over $\left(m_{ll}^{\rm cut}\right)^2\leq q_{12} \leq m_h^2$ and $0\leq q_{13} \leq m_h^2 - q_{12}$, one gets the  total decay rates. For our later discussions, we   define the standard formula for the enhancement factor of the decay rates as follows: 
\begin{eqnarray}
 R_{\textrm{NP}}(\mathcal{P}_{\rm NP}) =
 \dfrac{\Gamma_{h\rightarrow l
 \bar{l}\gamma}^{\textrm{NP}} (\mathcal{P}_{\rm NP}) }
 {\Gamma_{h\rightarrow l
 \bar{l}\gamma}^{\textrm{SM} } },
\end{eqnarray}
where $\Gamma_{h\rightarrow l \bar{l}\gamma}^{\textrm{NP}} (\mathcal{P}_{\rm NP})$  (and $\Gamma_{h\rightarrow l  \bar{l}\gamma}^{\textrm{SM} }$ ) is the decay rate of the EHSM and SM, respectively. 

We also interested in the fermion  FB asymmetry in $h\rightarrow l\bar{l}\gamma$, which is defined through the following integrated FB asymmetry: 
\begin{eqnarray}\label{AFB2}
\mathcal{A}_{\rm FB}(\mathcal{P}_{\rm NP})
=\frac{\int\limits_{E_{\gamma}^{\rm cut} }
^{m_h/2}\int_0^1 \frac{{d\Gamma}}
{dE_{\gamma}{d\cos\theta_l}} d{\cos\theta_l }dE_{\gamma} 
-\int_{E_{\gamma}^{\rm cut} }
^{m_h/2}
\int_{-1}^0 \frac{{d\Gamma}}{
dE_{\gamma}{d\cos\theta_l}} 
d{\cos\theta_l} dE_{\gamma} }
{
\int_{E_{\gamma}^{\rm cut} }
^{m_h/2} \int_0^1 \frac{{d\Gamma}}
{dE_{\gamma}{d\cos\theta_l}}
d{\cos\theta_l }dE_{\gamma} 
+ 
\int_{E_{\gamma}^{\rm cut} }
^{m_h/2}
\int_{-1}^0 \frac{{d\Gamma}}{
dE_{\gamma}{d\cos\theta_l}} 
d{\cos\theta_l} dE_{\gamma}
},
\end{eqnarray}
 where $E_{\gamma}$, $\cos\theta_l$ are the energy of photon and cosine angle between two momentum directions of photon and lepton in the rest frame of the SM-like Higgs boson. The enegy cut for photon is $E_{\gamma}^{\rm cut} \geq 4 m_l^2$. We refer \ref{app_Phase} for the details of  kinematic and phase space  of $1\rightarrow 3$ in the rest frame of the SM-like Higgs boson. 
\section{ \label{sec_pheno} Phenomenological results}
For numerical investigation, we will use the following input parameters: $M_Z = 91.1876$ GeV, 
$\Gamma_Z  = 2.4952$ GeV,  $M_W = 80.379$ GeV, $\Gamma_W  = 2.085$ GeV, $m_h =125.1$ GeV, $\Gamma_h =4.07\cdot 10^{-3}$ GeV.  For lepton masses we take $m_e =0.00052$ GeV,
$m_{\mu}=0.10566$ GeV and $m_{\tau} = 1.77686$ GeV. All quark masses are  $m_u= 0.00216$ GeV $m_d= 0.0048$ GeV, $m_c=1.27$ GeV, $m_s = 0.93$ GeV, 
$m_t= 173.0$ GeV, and $m_b= 4.18$ GeV. We are working in the so-called $G_{\mu}$-scheme, where  the Fermi constant is an input parameter with  $G_{\mu}=1.16638\cdot 10^{-5}$ GeV$^{-2}$. The  electroweak constant is then   calculated appropriately as follows:
\begin{eqnarray}
\alpha = \sqrt{2}/\pi G_{\mu}
M_W^2(1-M_W^2/M_Z^2)
= 1/132.184.
\end{eqnarray}
The inputs for scanning parameters will be presented corresponding to  particular  HESM models. 

We first cross-check numerically the results obtained in this work with our previous paper \cite{VanOn:2021myp} and other references \cite{Kachanovich:2020xyg}.  We take the THDM (type I or X)  as an illustration. In the first column of Table \ref{decaySM},  the singly charged Higgs mass  changes from $400$ GeV to $1000$ GeV.
\begin{table}[ht]
	\begin{center}
		\begin{tabular}{l@{\hspace{2cm}}l@{\hspace{2cm}}l}  
			\hline \hline 
			$M_{H^{\pm}}$ [GeV]
			& $\Gamma$ [KeV] & Ref.~\cite{VanOn:2021myp}  
			\\ \hline \hline
			$400$ & $6.157263708843869$ & $6.157263708844675$ 
			\\ \hline 
			$600$ & $1.097388878039453$ & $1.097388878039259$ 
			\\  \hline 
			$800$ & $0.304842878283716$ & $0.304842878276348$ 
			\\  \hline 
			$1000$ & $0.105534527108802$ & $0.105534527102357$
			\\  \hline\hline
		\end{tabular}
		\caption{\label{decaySM}
			$t_{\beta} = 10$, $\alpha = 0.4\pi$
			and $k_{\rm cut} = 0.1$. 
			Decay rate in KeV.}
	\end{center}
\end{table}
 In this test, we take $t_{\beta}=10$, $\alpha = 0.4 \pi$, apply the cut of the invariant mass of lepton-pair as $m_{ll}^{\rm cut} = k_{\rm cut} m_h$ with  $k_{\rm cut}=0.1$.  The second and third columns present the two numerical results evaluated in this work and Ref. \cite{VanOn:2021myp}, respectively. We find that the two results are in good agreement (more than $13$ digits),  confirming that two results are consistent when they are calculated in different gauges. 
Therefore, the test relies on gauge  invariance of the processes. A cross-checking the integrated FB asymmetries $\mathcal{A}_{\text{FB}}^{(l)}$ for $l = e, \mu$ with the ones in~\cite{Kachanovich:2020xyg} in the SM framework is presented in Table~\ref{AFBSM}. 
\begin{table}[ht]
	\begin{center}
		\begin{tabular}{l@{\hspace{2cm}}c
				@{\hspace{1cm}}c@{\hspace{1cm}}}  
			\hline \hline 
			$\mathcal{A}_{\text{FB}}^{(l)}$ 
			&\text{This work}  
			& Ref.~\cite{Kachanovich:2020xyg}
			\\ \hline
			$\mathcal{A}_{\text{FB}}^{(e)}$   
			& $0.3673$ 
			& $0.366$  
			\\ \hline 
			$\mathcal{A}_{\text{FB}}^{(\mu)}$   
			& $0.2843$ 
			& $0.280$  
			\\ \hline   
			\hline      
		\end{tabular}
		\caption{\label{AFBSM} Cross-checking the integrated 
			FB asymmetries 
			$\mathcal{A}_{\text{FB}}^{(l)}$ 
			for $l = e, \mu$ over full 
			phase-space with \cite{Kachanovich:2020xyg}
			in the SM framework.}
	\end{center}
\end{table}
In this test we use the cuts 
$m_{ll}^{\rm cut} = 0.1 m_h$ and $E_{\gamma}^{\rm cut} = 5$ GeV. We find that our results are good agreement with the corresponding ones in~\cite{Kachanovich:2020xyg}. 

Before going to discuss phenomenological results for HESM, we note that for the study of the enhancement factors,  the cuts of $m_{ll}^{\rm cut} = 0.1 m_h$ are applied. Furthermore, once we consider the FB asymmetries of fermions, we apply a futher cut for external photon as $E_{\gamma}^{\rm cut} = 5$ GeV.
\subsection{The IDM}     
For our analysis, we first review the current constraints on the scanning set of  parameters given in Eq. \eqref{eq_PIDM}.  The experimental constraints including Electroweak Precision Tests (EWPT) of the IDM, Dark matter search  at the LHC,  and LEP data were given in Refs. \cite{Borah:2012pu, Gustafsson:2012aj,Arhrib:2012ia, Klasen:2013btp}. The  loop-induced decays were also studied in the IDM framework, for example $h\rightarrow \gamma\gamma$ \cite{Krawczyk:2013jta, Chiang:2012qz,Benbrik:2022bol}, and  $h\rightarrow Z \gamma$  \cite{Chiang:2012qz,Benbrik:2022bol}. Benchmarks for searching
signals of the IDM at future colliders  were discussed in Ref. ~\cite{Arhrib:2014pva,Chakrabarty:2015yia, Ilnicka:2015jba,Datta:2016nfz,Kalinowski:2018ylg,Dercks:2018wch}.  The theoretical and experimental constraints on these parameters were also studied in the above papers, where  the most important  theoretical constraints must ensure the tree-level unitarity, vacuum stability, perturbativity regime, giving strict constraints on the Higgs self couplings 
$\lambda_i$ for $i=1,2, \cdots, 5$ and $\mu_2$. Taking into account all   constraints mentioned in the above works, it is reasonable to choose the ranges of  parameters as follows: $5$ GeV $\leq M_H \leq 150$ GeV,  $70$ GeV $\leq M_{H^{\pm}},  M_{A^0}\leq 1000$ GeV,  $|\mu_2| \leq 500$ GeV, and $0 \leq \lambda_2 \leq 8 \pi$.

\begin{figure}[ht]
	\centering
	\includegraphics[width=12cm, 
	height=7cm]{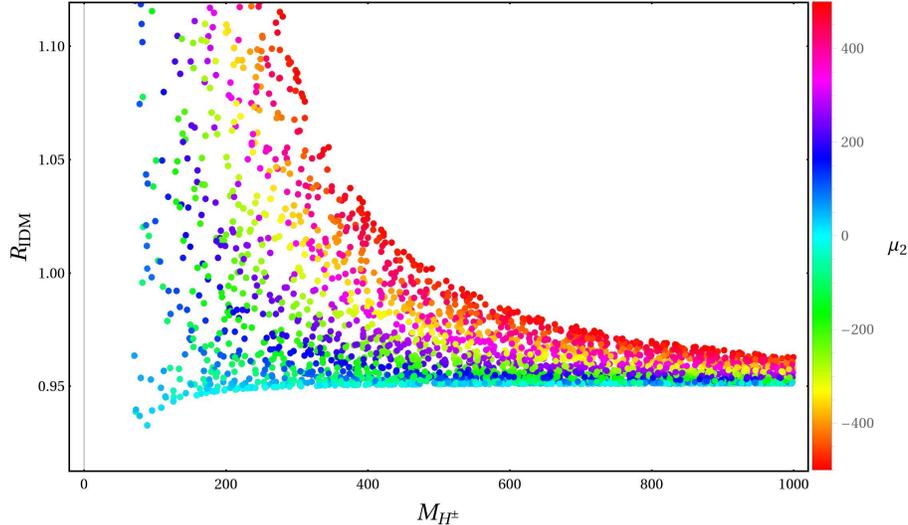}
	\caption{\label{RIDM} 
		The enhancement factor as function 
		of charged Higgs mass $M_{H^{\pm}}$ 
		and $\mu_2$ is shown.
	}
\end{figure}

Fig.~\ref{RIDM} shows a scatter plot of the enhancement factor as a function of the charged Higgs mass $M_{H^{\pm}}$ and $\mu_2$. 
There exist  large deviations of the enhancement factor $R_{\mathrm{IDM}}$ from $1$, which corresponds to the SM limit,  for  $70$ GeV $\leq M_{H^{\pm}} 
\leq 500$ GeV, but in the whole scanning range  $-500 \leq \mu_2 \leq 500$. Morever, $R_{\mathrm{IDM}}$ depends strongly   on $\mu_2$ parameter for$M_{H^{\pm}} \leq 500$ GeV. It is interesting to observe that $R_{\mathrm{IDM}}$ tends to  $\sim 0.95$ with  heavy values of singly charged Higgs  mass, implying that the factor 
is not sensitive with $\mu_2$. In the region of $M_{H^{\pm}}\geq 500$ GeV, the contributions of charged Higgs exchange ($\sim 5\%$) are smaller than other ones. With high-luminosity  of  future colliders, the IDM effect  could be tested. The plot provides a key role to probe indirect effect of the charged Higgs boson at future colliders.

In Table~\ref{AFBIDM}, the FB asymmetries of fermions are presented as a functions of  $M_{H^{\pm}}$ and $\mu_2$. We change the values of $M_{H^\pm}$, $\mu_2$ in the first column, by accessing all possible values in the scanning ranges. Consequently, the second and third columns show numerical values of the FB asymmetries of electron and muon, respectively. It can be seen that these  FB asymmetries  get contribution mainly from the  interference of the $Z^*$-pole diagrams with the remaining ones.  In this model, only more diagrams with singly charged Higgs exchange appear. However, their contributions are  smaller than those  observed in the enhancement factor plot,  resulting that  we get nearly same results of FB asymmetries of fermions as those predicted by the SM.
\begin{table}[ht]
\begin{center}
\begin{tabular}{l@{\hspace{2cm}}
c@{\hspace{2cm}}c}  
\hline \hline 
\\
$\big(M_{H^\pm}, \mu_2 \big)$ [GeV]
&$\big( R^{(e)}_{\text{IDM}} 
\,\,, \,\, \mathcal{A}_{\text{FB}}^{(e)} \big)$  
&$\big( R^{(\mu)}_{\text{IDM}} 
\,\,,\,\, \mathcal{A}_{\text{FB}}^{(\mu)} \big)$ 
\\
\\ \hline
\\
$\big(200 , \pm 200 \big)$  
& $(1.0109 \,\,,\,\, 0.3689)$ 
& $(1.0126 \,\,,\,\, 0.2860)$  \\
\\ \hline \\
$\big( 400 , \pm 100 \big)$  
& $(0.9491 \,\,,\,\, 0.3617)$ 
& $(0.9579 \,\,,\,\, 0.2758)$  \\
\\ \hline \\
$\big( 400 , \pm 200 \big)$   
& $(0.9592 \,\,,\,\, 0.3639)$ 
& $(0.9662 \,\,,\,\, 0.2779)$  \\
\\ \hline \\
$\big( 400 , \pm 400 \big)$  
& $(1.0057 \,\,,\,\, 0.3690)$ 
& $(1.0036 \,\,,\,\, 0.2859)$  \\
\\ \hline \\
$\big( 800 , \pm 100 \big)$   
& $(0.9472 \,\,,\,\, 0.3613)$ 
& $(0.9564 \,\,,\,\, 0.2754)$  \\
\\ \hline \\
$\big( 800 , \pm 200 \big)$   
& $(0.9497 \,\,,\,\, 0.3618)$ 
& $(0.9584 \,\,,\,\, 0.2759)$  \\
\\ \hline \\
$\big( 800 , \pm 400 \big)$  
& $(0.9696 \,\,,\,\, 0.3613)$ 
& $(0.9666 \,\,,\,\, 0.2779)$  \\
\\ \hline \\
$\big( 1000 , \pm 100 \big)$   
& $(0.9470 \,\,,\,\, 0.3613)$ 
& $(0.9562 \,\,,\,\, 0.2753)$  \\
\\ \hline \\
$\big( 1000 , \pm 200 \big)$   
& $(0.9485 \,\,,\,\, 0.3616)$ 
& $(0.9575 \,\,,\,\, 0.2757)$  \\
\\ \hline   
\hline      
\end{tabular}
\caption{\label{AFBIDM} The enhancement factor $R^{(l)}_{\text{IDM}}$ and integrated FB asymmetries $\mathcal{A}_{\text{FB}}^{(l)}$ with $l = e, \mu$ in the  IDM framework. The masses are in GeV. The SM values are $\mathcal{A}_{\text{FB}}^{(e)}
=0.3673$ and  $\mathcal{A}_{\text{FB}}^{(\mu)} =0.2843$.}
\end{center}
\end{table}
\subsection{THDM}       
Following the same strategies  for phenomenological studies in many of BSMs, one first summaries the theoretical  and experimental constraints on the scanning parameters in THDM. Theoretical constraints subject to the tree-level unitarity, vacuum stability, perturbativity regime were shown 
in Ref. \cite{Nie:1998yn, Kanemura:1999xf,Akeroyd:2000wc, Ginzburg:2005dt,Kanemura:2015ska},
giving  strict bounds on $\lambda_i$ for $i=1,2, \cdots, 5$ and $m_{11}, m_{12}, m_{22}$. The experimental  constraints are obtained from the EWPT of THDM, implying the LEP data~\cite{Bian:2016awe, Xie:2018yiv}. Furthermore, direct and indirect searching for scalar masses in THDM were reported  from  LEP, Tevaron and the LHC~\cite{Kanemura:2011sj}. Implications for loop-induced decays of $h\rightarrow \gamma\gamma$ and $h\rightarrow Z\gamma$ were presented in Refs. \cite{Krawczyk:2013jta, Chiang:2012qz,Benbrik:2022bol}. Combining all the constraints, we scan consistently the input parameters  for THDM as follows.  In the type-I and X: we take $126$ GeV $\leq M_H \leq 1000$ GeV, $60$ GeV $\leq M_{A^0} \leq 1000$ GeV and $80$ GeV $\leq M_{H^{\pm} }  \leq 1000$ GeV. In the Type-II and  Y: we scan logically the physical parameters as  
$500$ GeV $\leq M_H \leq 1000$ GeV, $500$ GeV $\leq M_{A^0} \leq 1000$ GeV and 
$580$ GeV $\leq M_{H^{\pm} }  \leq 1500$ GeV. In both types, one takes $2 \leq t_{\beta} \leq 20$,  $0.95\leq s_{\beta-\alpha} \leq 1$ and $m_{12}^2 =M_H^2 s_\beta c_\beta$.

\begin{figure}[ht]
	\centering
	$
	\begin{array}{cc}
		\includegraphics[width=8.5cm, height=6cm]
		{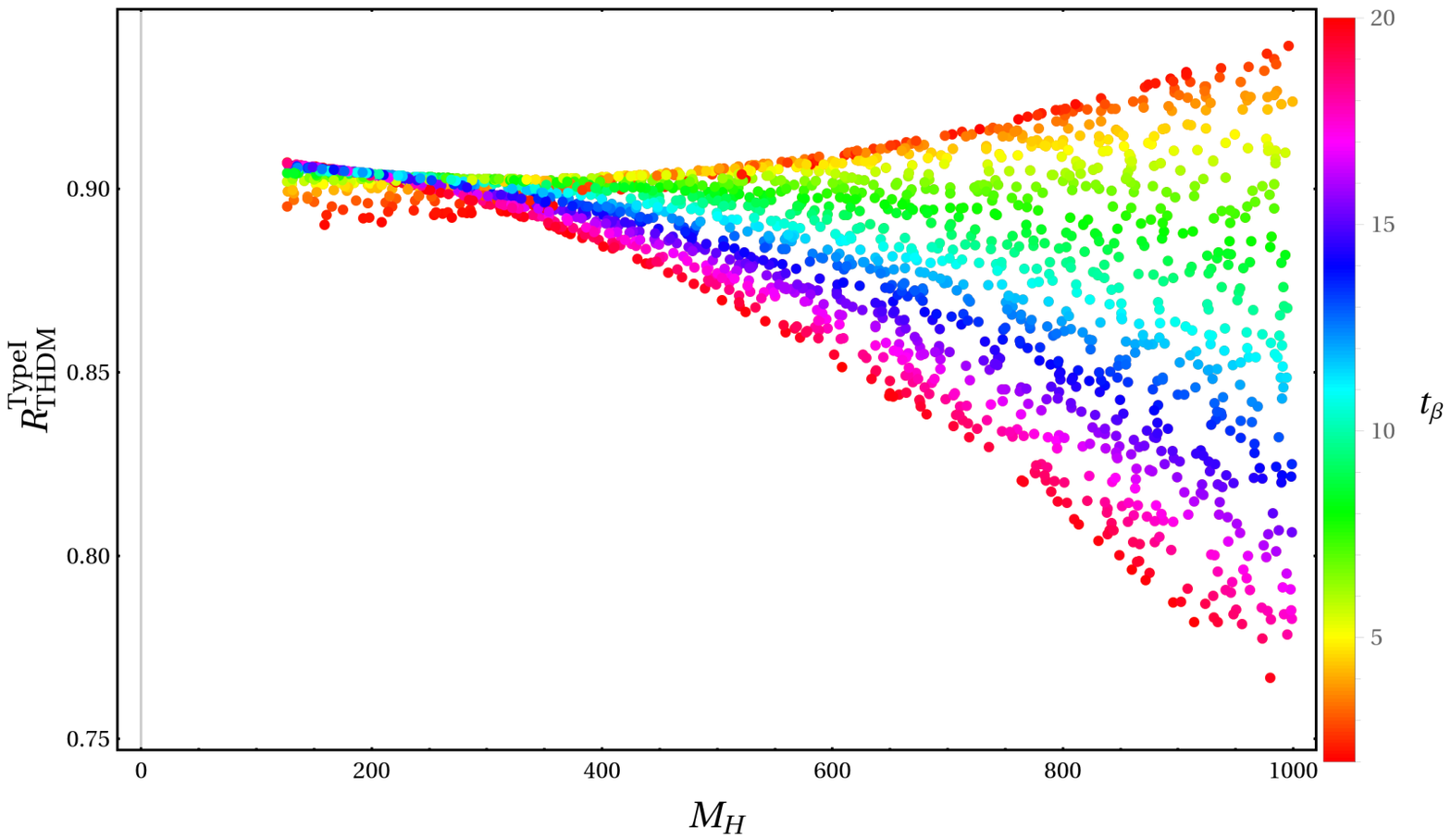}
		& 
		\includegraphics[width=8.5cm, height=6cm]
		{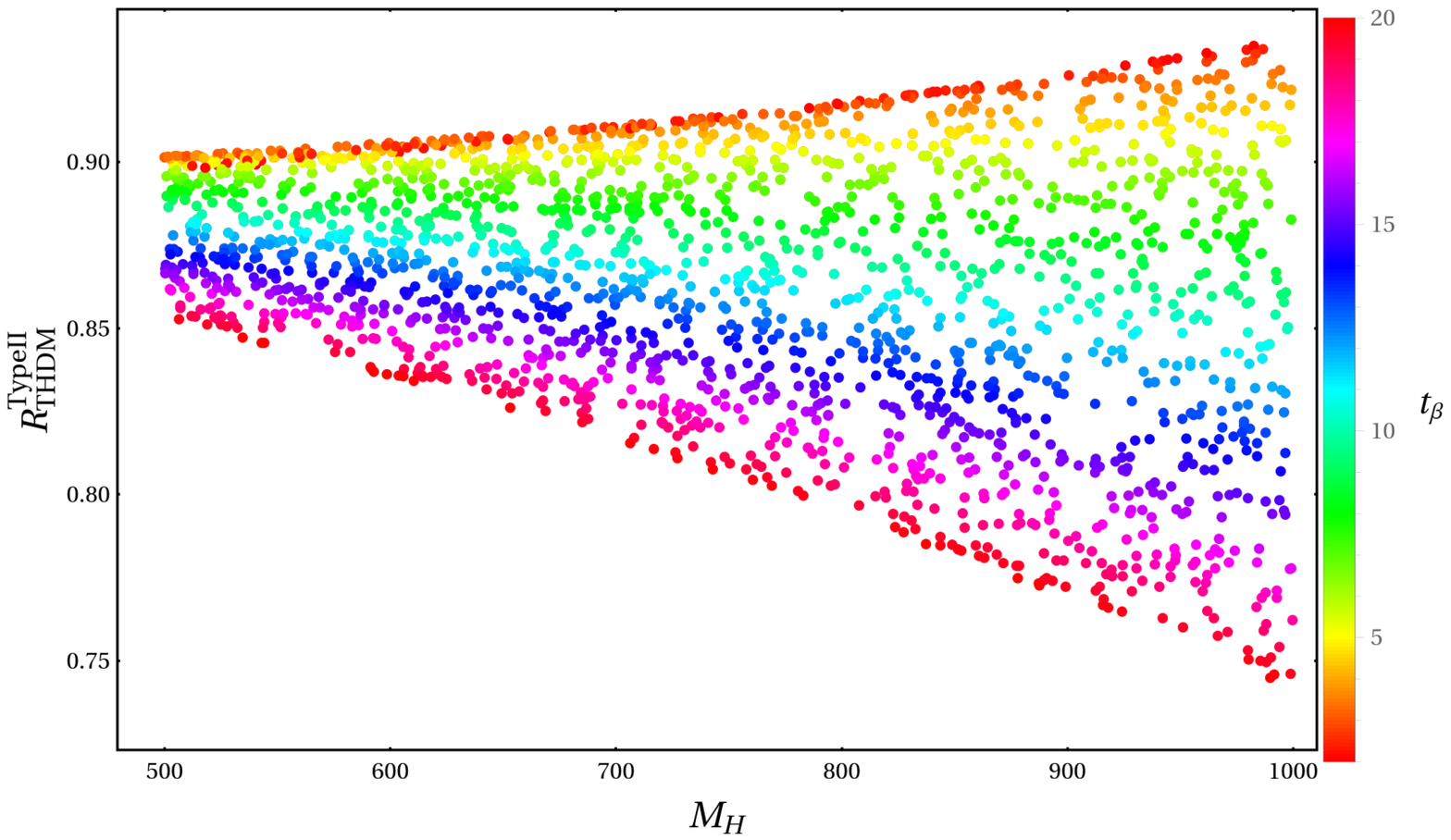}
		\\
		&
		\\
		\includegraphics[width=8.5cm, height=6cm]
		{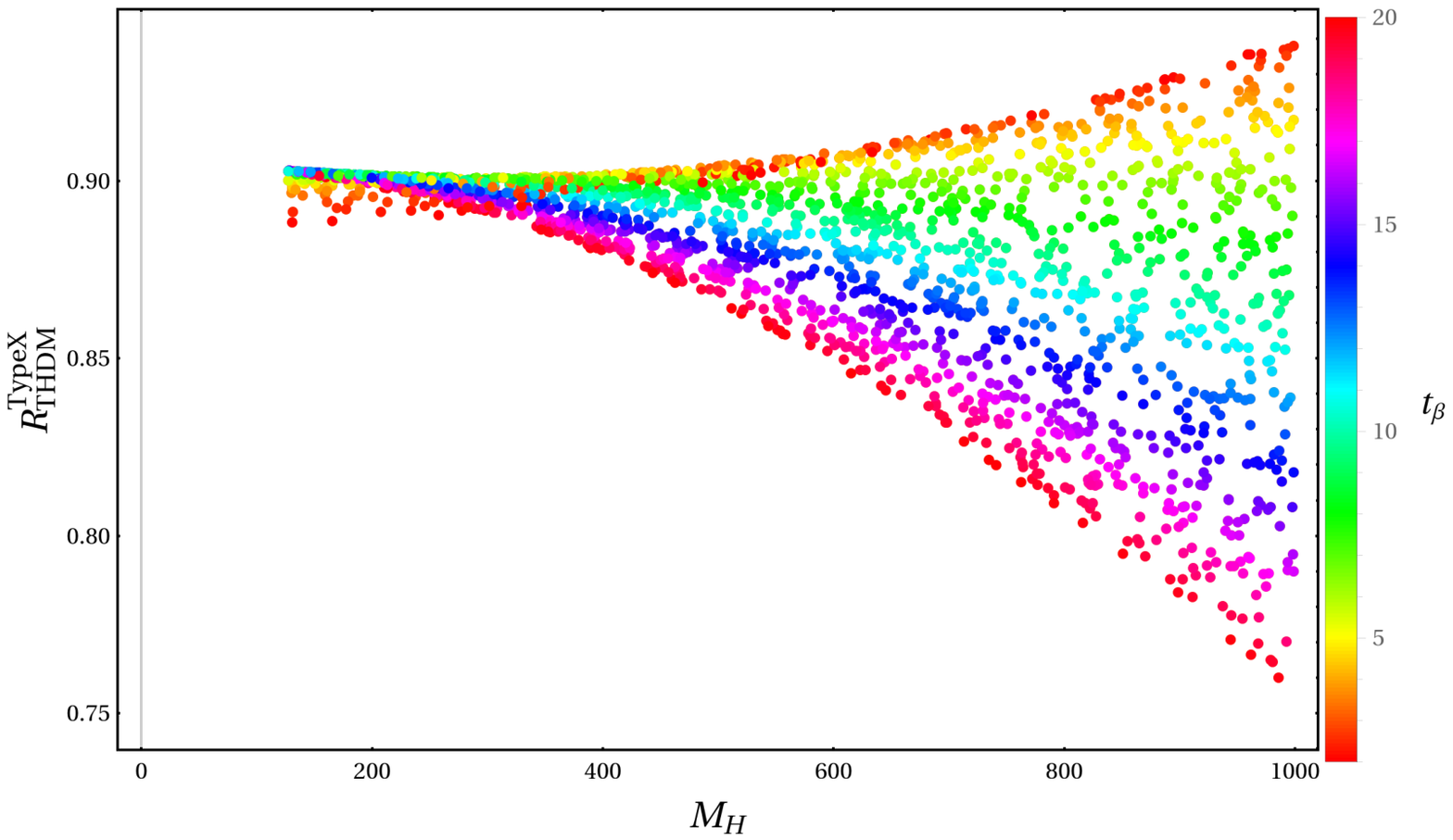}
		& 
		\includegraphics[width=8.5cm, height=6cm]
		{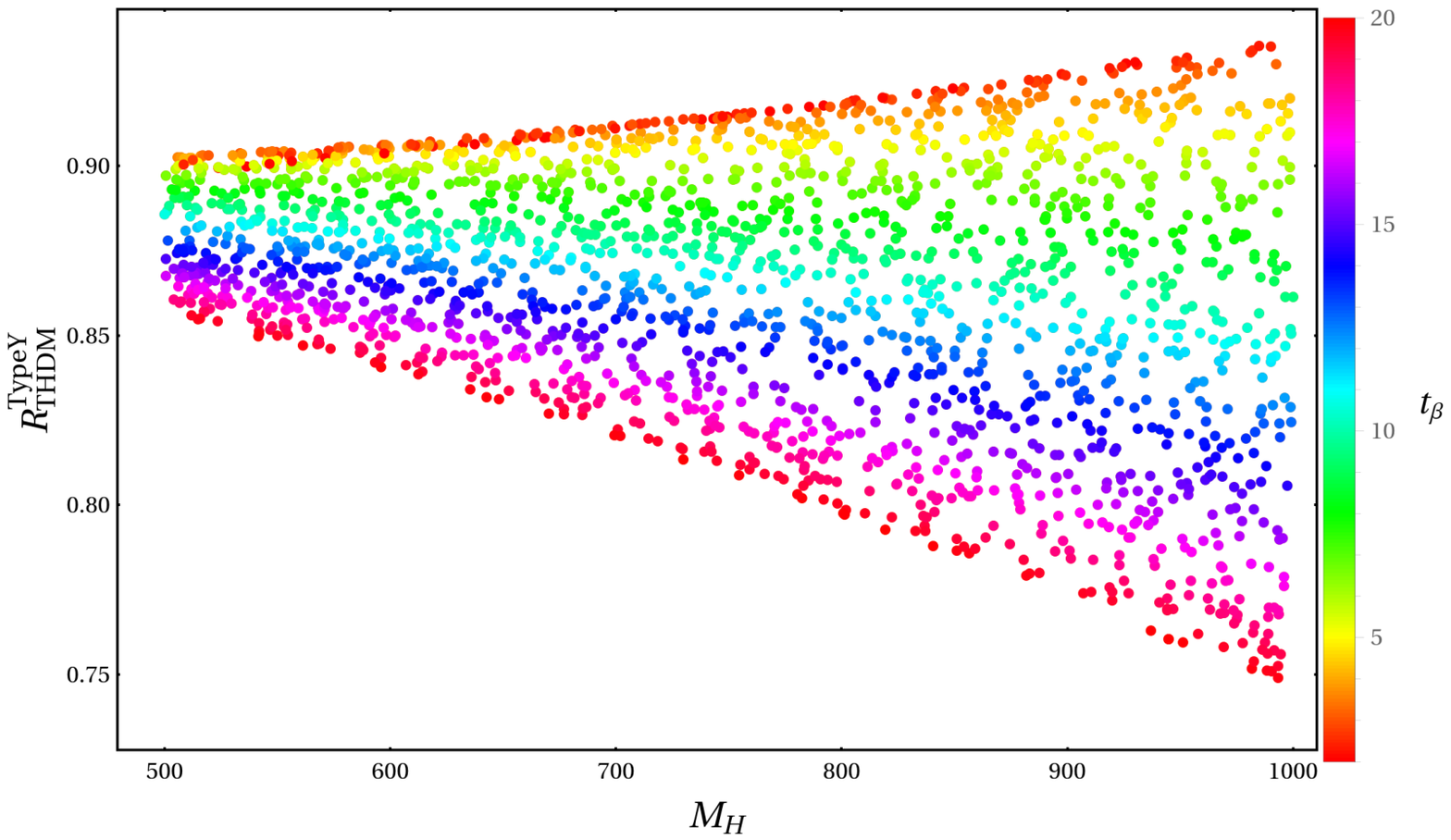}
	\end{array}$
	\caption{\label{RTHDMbeta} The enhancement 
		factors $R_{\mathrm{THDM}}$ as functions of $t_{\beta}$ and $M_H$
		for four types of the THDM.}
\end{figure}

In Fig.~\ref{RTHDMbeta}, we analysis the enhancement factors as functions  of $t_{\beta}$ and neutral Higgs mass $M_H$ for four types of THDM, fixing   $s_{\beta-\alpha} = 0.95$, and  $M_{H^{\pm}} = 800$ GeV.  
The scanning ranges of  $t_{\beta}$ and $M_H$  are  $2\leq t_{\beta}\leq 20$ and  $126$ GeV $\leq M_H \leq 1000$ GeV for  types I and X,  and $500$ GeV  $\leq M_H \leq 1000$ GeV for  types II and Y. The two models  types I and  X predict that the enhancement factors tend to $0.9$ when $M_{H}\leq 400$ GeV. Beyond this region, namely with  $M_{H}\geq 400$ GeV, $R_{\mathrm{THDM}}$ are sensitive with both $t_{\beta}$ and $m_H$, which give  large deviations of the decay rates between all four THDM types and the SM.  It is stressed that the enhancement factors in the two THDM types I and X are slightly different from those of types II and Y, which results from the properties of Yukawa couplings giving smaller contributions than those obtained from the $W$ boson exchange.

\begin{figure}[ht]
	\centering
	$
	\begin{array}{cc}
		\includegraphics[width=8.5cm, height=6cm]
		{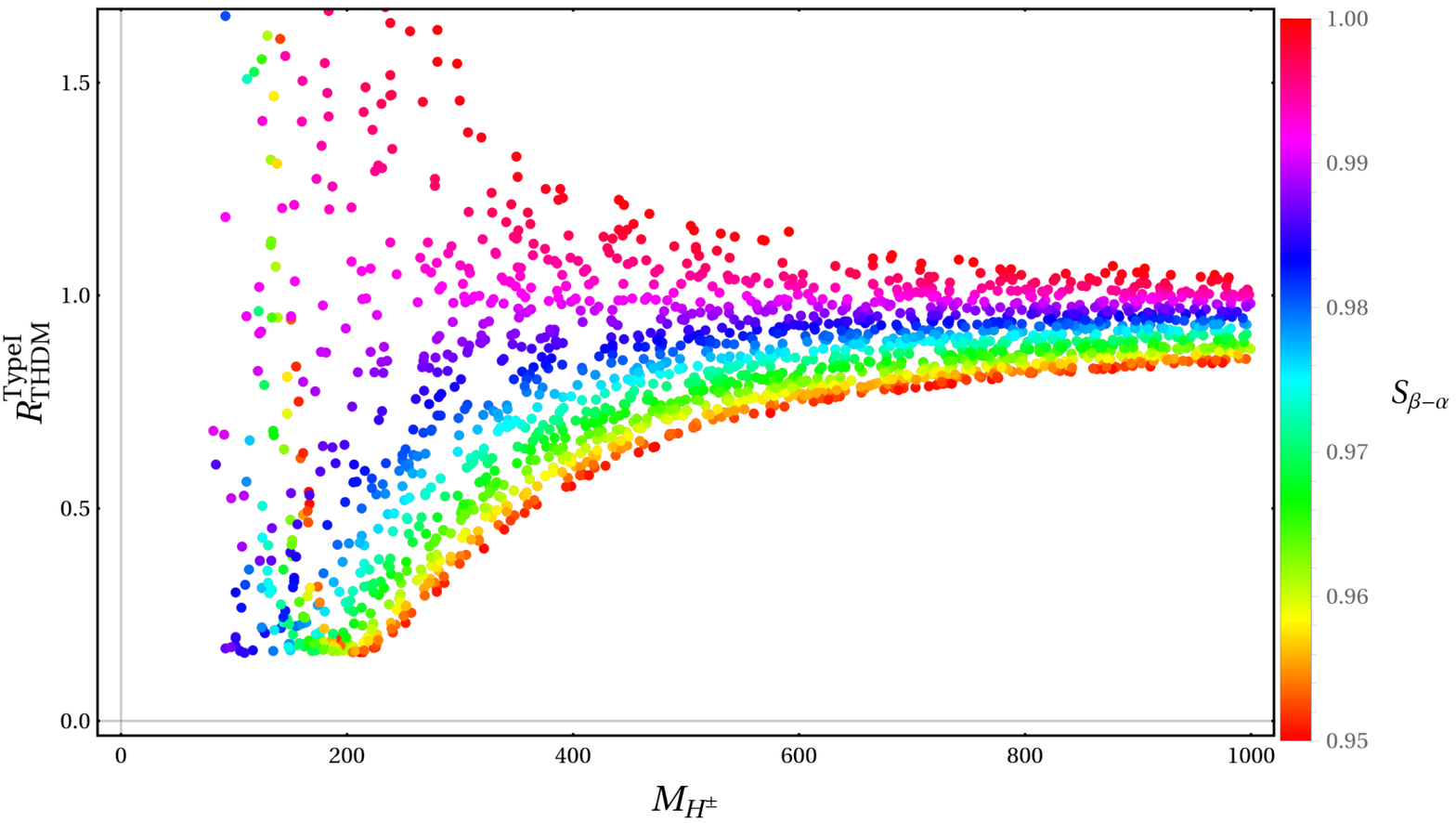}
		& 
		\includegraphics[width=8.5cm, height=6cm]
		{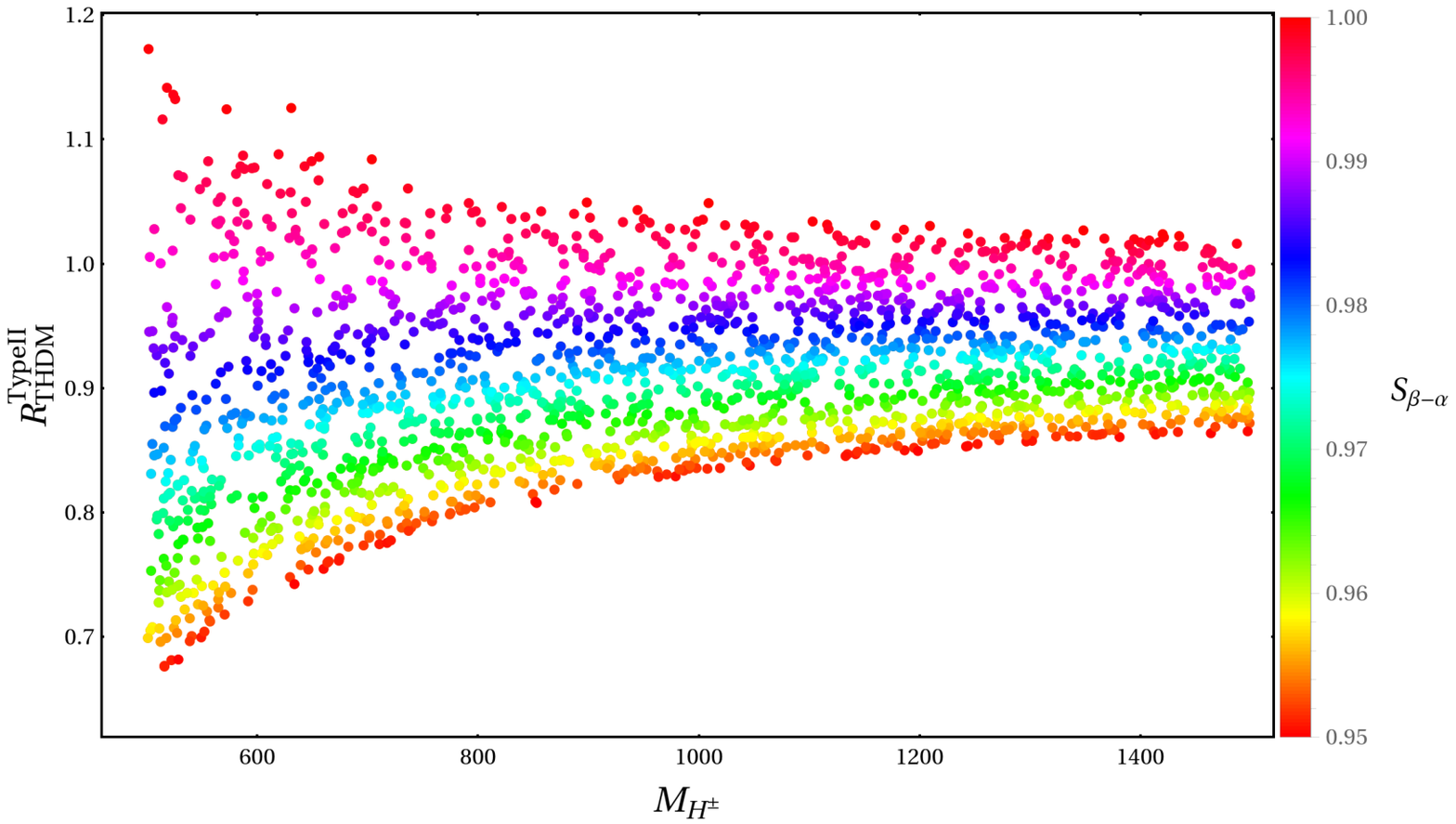}
		\\
		&
		\\
		&
		\\
		\includegraphics[width=8.5cm, height=6cm]
		{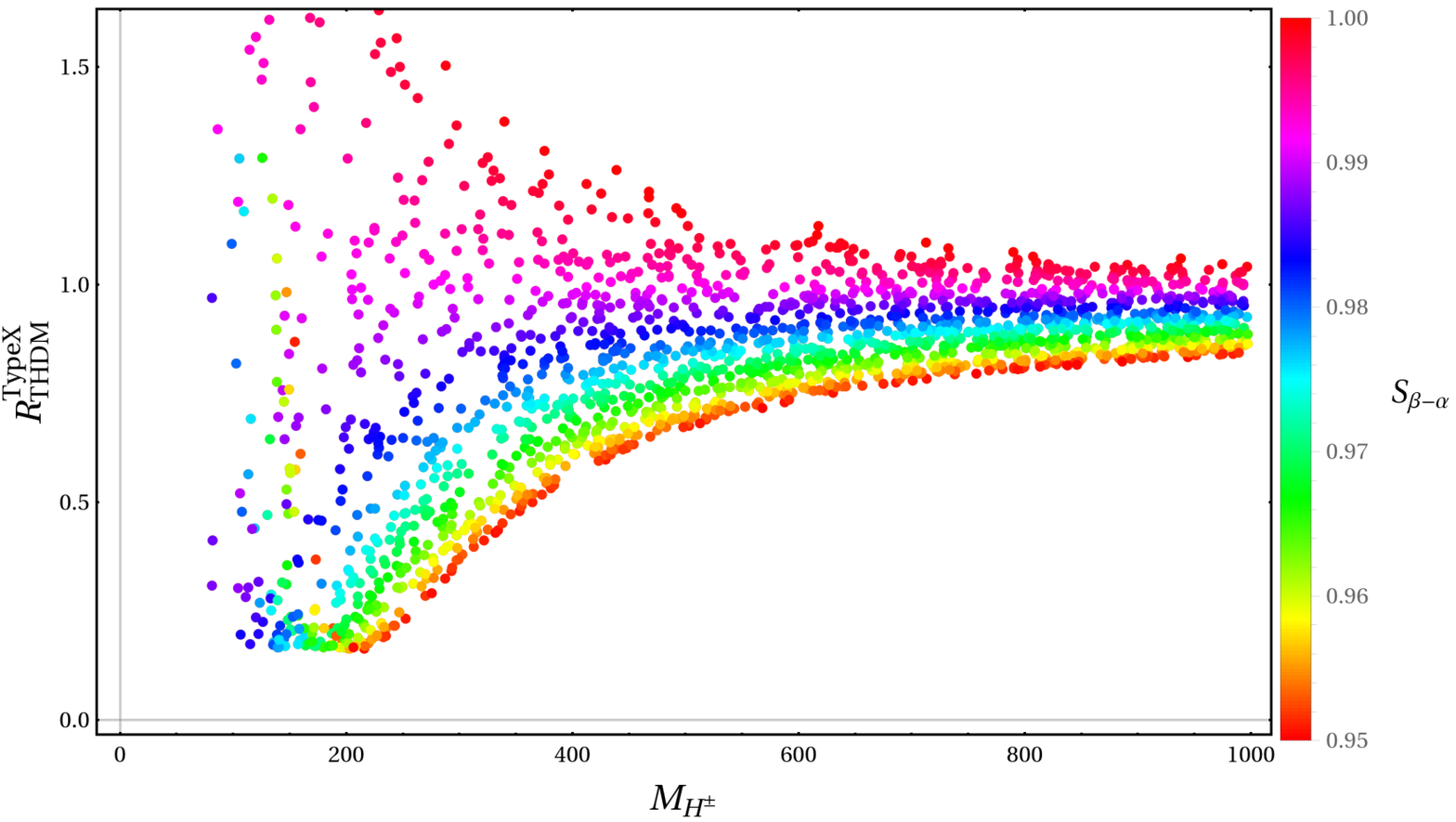}
		& 
		\includegraphics[width=8.5cm, height=6cm]
		{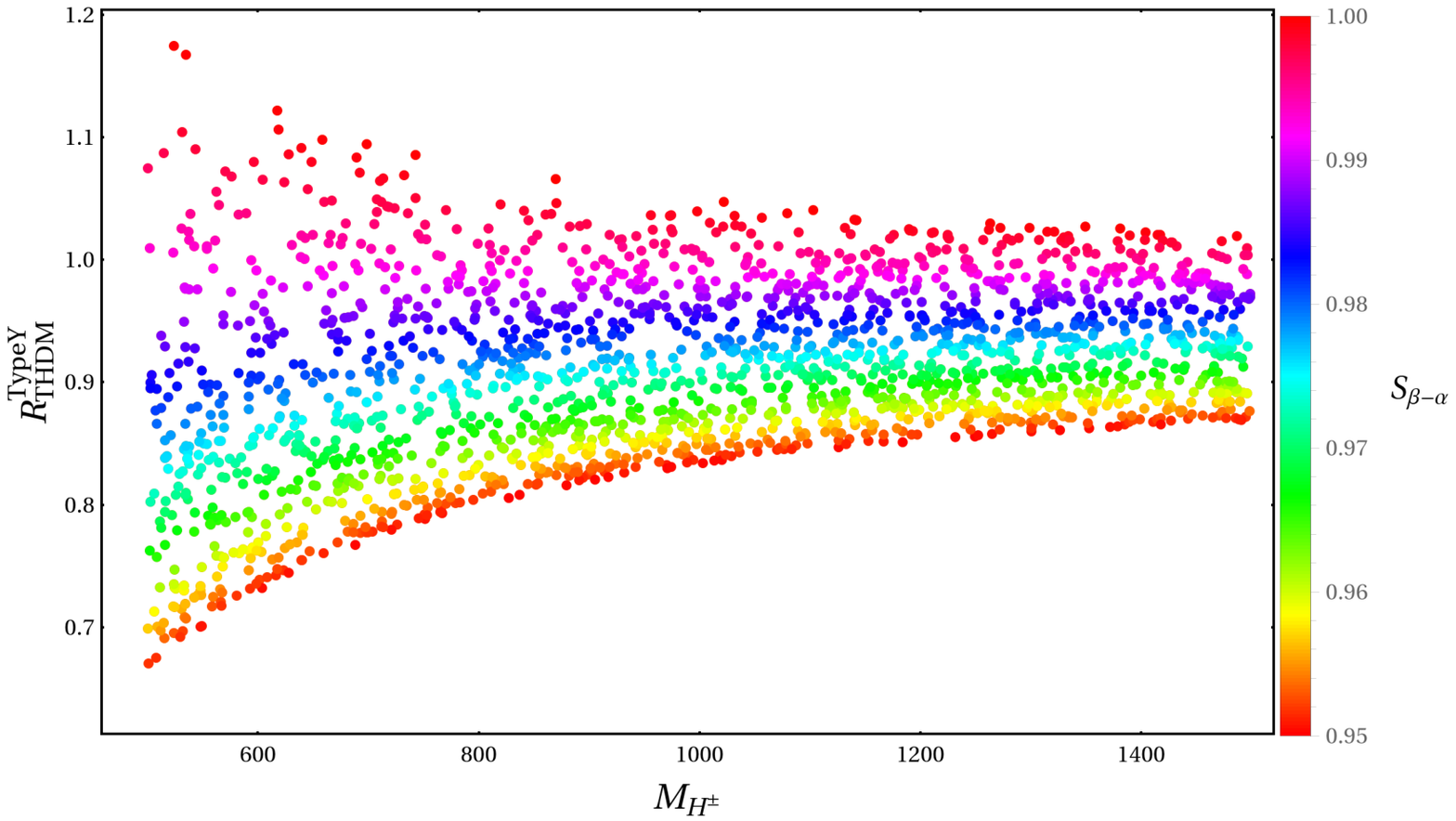}
	\end{array}$
	\caption{\label{RTHDMSAB} The enhancement 
		factor
		for four types of THDM.}
\end{figure}

Fig.~\ref{RTHDMSAB} shows the enhancement factors as functions of $M_{H^{\pm}}$ and $s_{\beta-\alpha}$, where  $t_{\beta}=15$ and $M_{H}=1000$ GeV are 
selected. 
It observes that the enhancement factors of the two THDM types I and X have the same behavior, which also happens in the two THDM types II and Y.  As discussed previously, the slightly different values of the enhancement factors between the two classes of THDM types (I, X) and  (II, Y)  originate from the Yukawa couplings. In the THDM types I and X,  the  enhancement factors have  wide deviations  from 1 in the region  $M_{H^{\pm}}\leq 600$ GeV. In addition, these factors  are more sensitive  with $s_{\beta-\alpha}$ in  smaller values of $M_{H^{\pm}}$. In contrast, large  $M_{H^{\pm}}\geq 800$ GeV corresponds  to the range  $ 0.85 \leq R_{\rm THDM} \leq  1.1$, and converge to the values close to 1 with very heavy $M_{H^{\pm}}$. In the THDM types II and Y, the enhancement factors  have  large deviations from $1$ in all ranges of $M_{H^\pm}$   and  $s_{\beta-\alpha}$. In conclusion, these plots mentioned here show that one may constrain the mixing angle and the singly charged mass at future colliders.

In Table \ref{AFBTHDMIX}, we generate numerical results of the fermion FB asymmetries at several points in the parameter spaces of the THDM types I and X.  The parameters  $t_\beta, s_{\beta - \alpha},  M_{H^\pm}$ and $M_H$ are selected in such a way to cover all possible values in their scanning  ranges mentioned above. The numerical values of these parameters are shown in the first column of Table \ref{AFBTHDMIX}. In the second  (third) column, we report  the values of the FB asymmetries (enclosed with the enhancement factors) of electron and muon, respectively. The results are presented for THDM type I (first line data) and X (second line data). The results in this Table
should compare with the values of the asymmetries in the SM which are given  $\mathcal{A}_{\text{FB}}^{(e)} =0.3673$ and  $\mathcal{A}_{\text{FB}}^{(\mu)} =0.2843$, accordingly. The data indicates that the enhancement factors and the integrated FB asymmetries
are same direction of the derivation to the SM's values. We also find that a bigger derivation of FB 
asymmetry for muon than the one  for electron. It is because we are  not considered the tree diagram 
contributions in both the SM and THDM. 
\begin{table}[ht]
\begin{center}
\begin{tabular}{l@{\hspace{2cm}}c
@{\hspace{1cm}}c@{\hspace{1cm}}}  
\hline \hline 
\\
$\big(t_\beta, s_{\beta - \alpha}, M_{H^\pm}, M_H \big)$
&$\big( R^{(e)}_{\text{THDM}} \,\,,\,\, \mathcal{A}_{\text{FB}}^{(e)} \big)$  
&$\big( R^{(\mu)}_{\text{THDM}} \,\,,\,\, \mathcal{A}_{\text{FB}}^{(\mu)} \big)$ 
\\
\\ \hline
\\
$\big( 5, 0.95, 200, 400 \big)$  
& $(0.9516 \,\,,\,\, 0.3548)$ 
& $(0.8754 \,\,,\,\, 0.3221)$  \\
& $(0.9424 \,\,,\,\, 0.3543)$
& $(0.8135 \,\,,\,\, 0.3415)$  \\
\\ \hline 
\\  
$\big( 5, 0.99, 400, 200 \big)$  
& $(0.9946 \,\,,\,\, 0.3487)$ 
& $(0.9119 \,\,,\,\, 0.3176)$  \\
& $(0.9925 \,\,,\,\, 0.3499)$
& $(0.8276 \,\,,\,\, 0.3483)$  \\
\\ \hline   
\\
$\big( 10, 0.95, 400, 600 \big)$  
& $(0.8439 \,\,,\,\, 0.3338)$ 
& $(0.7809 \,\,,\,\, 0.3017)$  \\
& $(0.8423 \,\,,\,\, 0.3431)$
& $(1.1276 \,\,,\,\, 0.2448)$  \\
\\ \hline   
\\
$\big( 10, 0.99, 600, 400 \big)$  
& $(0.9968 \,\,,\,\, 0.3493)$ 
& $(0.9112 \,\,,\,\, 0.3190)$  \\
& $(0.9972 \,\,,\,\, 0.3512)$
& $(0.8573 \,\,,\,\, 0.3478)$  \\
\\ \hline   
\\
$\big( 15, 0.95, 800, 600 \big)$  
& $(0.8721 \,\,,\,\, 0.3399)$ 
& $(0.8025 \,\,,\,\, 0.3088)$  \\
& $(0.8875 \,\,,\,\, 0.3469)$
& $(1.9386 \,\,,\,\, 0.1499)$  \\
\\ \hline   
\\
$\big( 15, 0.99, 600, 800 \big)$  
& $(0.9851 \,\,,\,\, 0.3473)$ 
& $(0.9007 \,\,,\,\, 0.3172)$  \\
& $(0.9879 \,\,,\,\, 0.3558)$
& $(0.9299 \,\,,\,\, 0.3154)$  \\
\\ \hline   
\\
$\big( 20, 0.95, 500, 1000 \big)$  
& $(0.5310 \,\,,\,\, 0.2421)$ 
& $(0.5197 \,\,,\,\, 0.2097)$  \\
& $(0.5469 \,\,,\,\, 0.2466)$
& $(2.8866 \,\,,\,\, 0.0687)$  \\
\\ \hline   
\\
$\big( 20, 0.99, 1000, 500 \big)$  
& $(0.9857 \,\,,\,\, 0.3474)$ 
& $(0.9008 \,\,,\,\, 0.3174)$  \\
& $(0.9935 \,\,,\,\, 0.3557)$
& $(1.1117 \,\,,\,\, 0.2737)$  \\
\\ \hline   
\hline      
\end{tabular}
\caption{\label{AFBTHDMIX}
The enhancement factor $R^{(l)}_{\text{THDM}}$ and integrated FB asymmetries 
$\mathcal{A}_{\text{FB}}^{(l)}$  with $l = e, \mu$  in the THDM types I (first line) and X 
(second line). The masses are shown in GeV.}
\end{center}
\end{table}

Similarly, Table~\ref{AFBTHDMIIY} shows numerical results relating to the two THDM types II and Y,  with the same notations defined in Table \ref{AFBTHDMIX}. We have the same conclusions as those  derived from the THDMs types I and X. The THDM has only more diagrams with singly charged Higgs exchange in the loop, which give in smaller contributions than the other ones. Therefore, we also have same conclusions for  $A_{\rm FB}^{(l)}$ as in the IDM.
\begin{table}[ht]
\begin{center}
\begin{tabular}{l@{\hspace{2cm}}c
@{\hspace{1cm}}c@{\hspace{1cm}}}  
\hline \hline 
\\
$\big(t_\beta, s_{\beta - \alpha}, M_{H^\pm}, M_H \big)$
&$\big( R^{(e)}_{\text{THDM}} \,\,,\,\, \mathcal{A}_{\text{FB}}^{(e)} \big)$  
&$\big( R^{(\mu)}_{\text{THDM}} \,\,,\,\, \mathcal{A}_{\text{FB}}^{(\mu)} \big)$
\\
\\ \hline
\\
$\big( 5, 0.95, 800, 600 \big)$  
& $(0.9133 \,\,,\,\, 0.3536)$ 
& $(0.7998 \,\,,\,\, 0.3476)$  \\
& $(0.9150 \,\,,\,\, 0.3577)$ 
& $(0.8569 \,\,,\,\, 0.3296)$  \\
\\ \hline  
\\
$\big( 5, 0.99, 600, 800 \big)$  
& $(1.0573 \,\,,\,\, 0.3634)$ 
& $(0.8899 \,\,,\,\, 0.3598)$  \\
& $(1.0576 \,\,,\,\, 0.3621)$ 
& $(0.9643 \,\,,\,\, 0.3356)$  \\
\\ \hline  
\\
$\big( 10, 0.95, 500, 1000 \big)$  
& $(0.7885 \,\,,\,\, 0.3215)$ 
& $(1.0588 \,\,,\,\, 0.2123)$  \\
& $(0.7967 \,\,,\,\, 0.3245)$ 
& $(0.7378 \,\,,\,\, 0.2901)$  \\
\\ \hline  
\\
$\big( 10, 0.99, 1000, 500 \big)$  
& $(0.9903 \,\,,\,\, 0.3532)$ 
& $(0.8389 \,\,,\,\, 0.3478)$  \\
& $(0.9908 \,\,,\,\, 0.3523)$ 
& $(0.9136 \,\,,\,\, 0.3228)$  \\
\\ \hline  
\\
$\big( 15, 0.95, 800, 600 \big)$  
& $(0.8654 \,\,,\,\, 0.3476)$ 
& $(1.9231 \,\,,\,\, 0.1498)$  \\
& $(0.8675 \,\,,\,\, 0.3429)$ 
& $(0.8091 \,\,,\,\, 0.3477)$  \\
\\ \hline  
\\
$\big( 15, 0.99, 600, 800 \big)$  
& $(0.9867 \,\,,\,\, 0.3567)$ 
& $(0.9273 \,\,,\,\, 0.3132)$  \\
& $(0.9841 \,\,,\,\, 0.3539)$ 
& $(0.9002 \,\,,\,\, 0.3297)$  \\
\\ \hline  
\\
$\big( 20, 0.95, 500, 1000 \big)$  
& $(0.5455 \,\,,\,\, 0.2521)$ 
& $(2.8719 \,\,,\,\, 0.0732)$  \\
& $(0.5256 \,\,,\,\, 0.2452)$ 
& $(0.5207 \,\,,\,\, 0.2087)$  \\
\\ \hline  
\\
$\big( 20, 0.99, 1000, 500 \big)$  
& $(0.9849 \,\,,\,\, 0.3537)$ 
& $(1.0996 \,\,,\,\, 0.2667)$  \\
& $(0.9888 \,\,,\,\, 0.3567)$ 
& $(0.9047 \,\,,\,\, 0.3227)$  \\
\\ \hline   
\hline      
\end{tabular}
\caption{\label{AFBTHDMIIY} The enhancement factor $R^{(l)}_{\text{THDM}}$
and integrated FB asymmetries  $\mathcal{A}_{\text{FB}}^{(l)}$ with $l = e, \mu$  in the  THDM types II (first line) and Type Y (second line data), respectively. The masses are presented  in GeV. }
\end{center}
\end{table}
\subsection{THM}   
We arrive at the  phenomenological results  for the last model in this work. 
The theoretical  and experimental constraints on the mentioned scanning parameters in THM 
were shown precisely in Refs. \cite{Chun:2012jw,Chen:2013dh, Arhrib:2011uy, Arhrib:2011vc,Akeroyd:2012ms,Akeroyd:2011zza,Akeroyd:2011ir,Aoki:2011pz,Kanemura:2012rs,Chabab:2014ara,Han:2015hba,Chabab:2015nel,Ghosh:2017pxl, Ashanujjaman:2021txz,Zhou:2022mlz}. Combining all 
of these constraints,  it is  reasonable to scan the mentioned parameters for 
the THM in the  following ranges: $v_{\Delta} =1$ MeV or $5$ MeV,  $0.95 \leq c_{\alpha}\leq 1$, 
$60$ GeV $\leq M_{H^{\pm}}, M_{H^{\pm\pm}}, M_{A^0}, M_H \leq 1000$ GeV.

Since we want to generate scatter plots as functions of $\lambda_1$.  The couplings of the SM-like Higgs boson to singly and doubly charged Higgs bosons will be expressed in terms of this
parameter.  In the scope of this paper, we are concerned the case of  $\frac{v_{\Delta} }{v_{\Phi} } \ll 1$. When $\frac{v_{\Delta} }{v_{\Phi} }  \rightarrow 0$, one has  $s_{\beta^{\pm}}\rightarrow{0}$
as well as using Eq. \eqref{aprTHM}, the  results read
\begin{align}
g^{\rm THM}_{hH^{\pm}H^{\mp}}&=
\lambda_1v_{\Phi}c_{\alpha} 
+ \frac{\lambda_4}{2}v_{\Phi}c_{\alpha}
=\frac{v}{2}(2\lambda_1+\lambda_4)c_{\alpha} =
\lambda_1 \; v\; c_{\alpha} 
+ \frac{2(M_{A^0}^2 - 
M^2_{H^{\pm}})}{v}c_{\alpha},
\\
g^{\rm THM}_{hH^{\pm\pm}H^{\mp\mp}}
&=\lambda_1v_{\Phi}c_{\alpha}
=\lambda_1\; v\; c_{\alpha}.
\end{align}
For our numerical studies, we use 
the following relation: 
\begin{eqnarray}
 M_{A^0}^2 &=& 
 2 M_{H^{\pm}}^2 - M_{H^{\pm\pm}}^2. 
 \end{eqnarray}

\begin{figure}[ht]
 	\centering
 	$
 	\begin{array}{cc}
 		\includegraphics[width=8cm, height=6cm]
 		{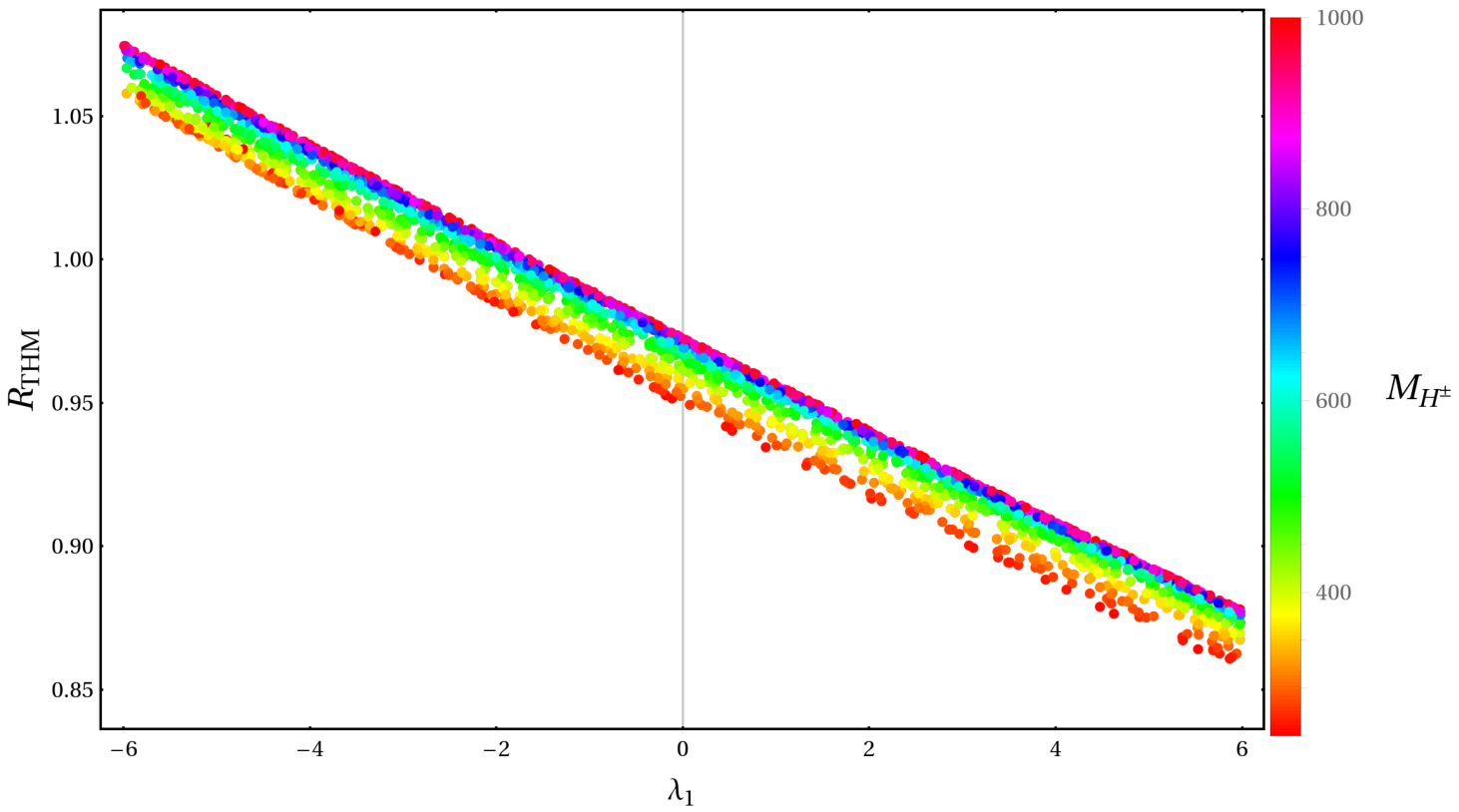}
 		& 
 		\includegraphics[width=8cm, height=6cm]
 		{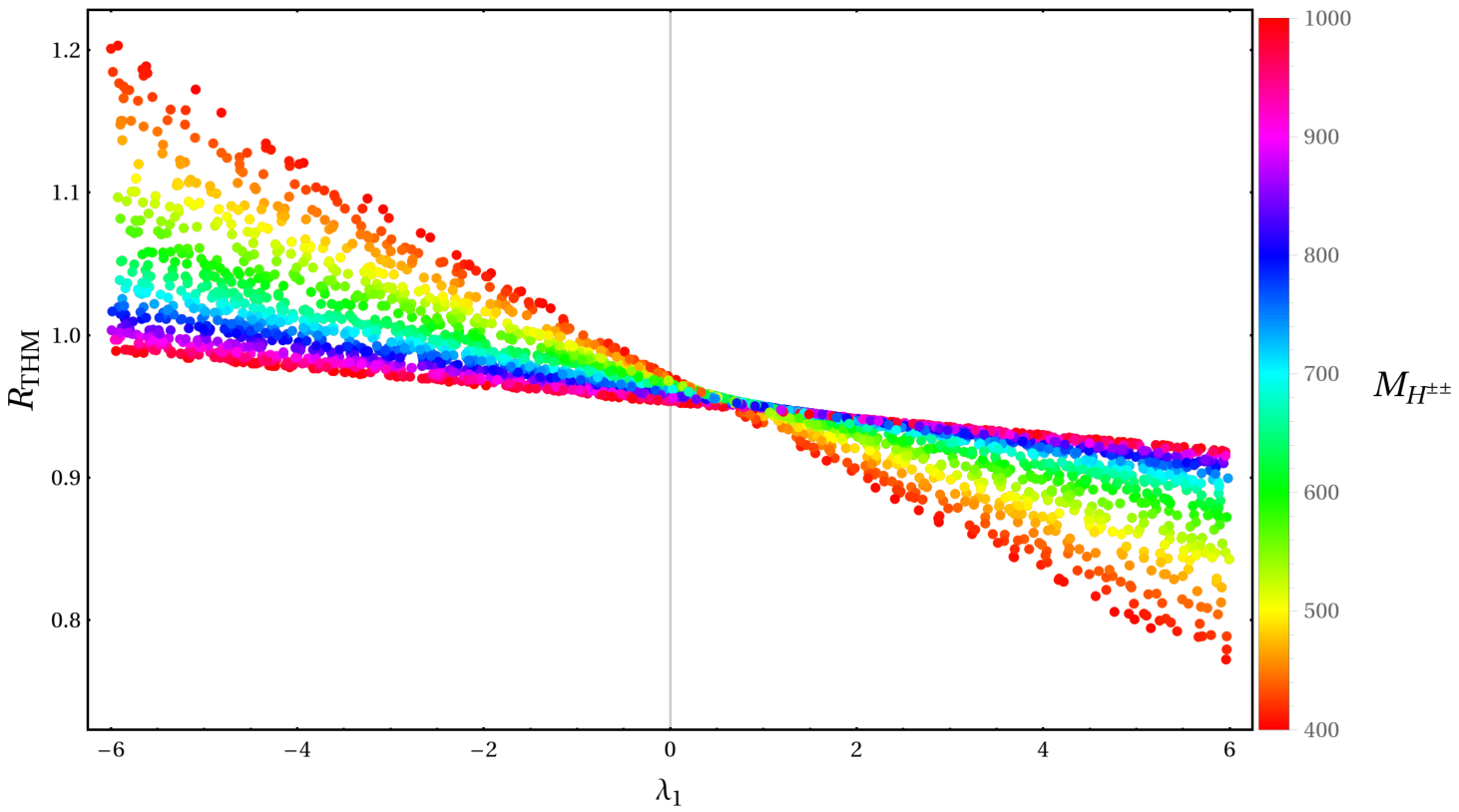}
 	\end{array}$
 	\caption{\label{RTHM} 
 		The enhancement factors are functions of $\lambda_1$ and
 		singly (left panel) and doubly  (right panel)
 		charged Higgs mass. }
 \end{figure}

 The enhancement factor as function of  $\lambda_1$, singly (doubly) charged Higgs boson  (right panel) are plotted in the left (right) panel of Fig.~\ref{RTHM}, where $c_{\alpha} =0.98$ is fixed and the scanning range of $\lambda_1$ is $-6 \leq \lambda_1 \leq 6$. 
  In addition, $M_{H^{\pm\pm}}=600$ ($M_{H^{\pm}}=500$) GeV is fixed in the left (right) panel.
 In the left panel of Fig. \ref{RTHM}, we find that the enhancement factor is smaller (larger) than $1$ for $\lambda_1 \geq -3$ ($\lambda_1 \leq -3$), and depends nearly linearly on $\lambda_1$. While this factor  is not sensitive with $M_{H^{\pm}}$. In the right panel, the enhancement factor is greater (smaller)  than $1$ when $\lambda_1 \leq  -1$ ($\lambda_1 \geq -1$)  for all values of  the  doubly charged Higgs mass. With a fixed value of $m_{H^{\pm\pm}}$, this factor develops linearly with $\lambda_1$ and gives large deviations from 1 for large $|\lambda_1|$.

\begin{figure}[ht]
	\centering
	$
	\begin{array}{cc}
		\includegraphics[width=8cm, height=6cm]
		{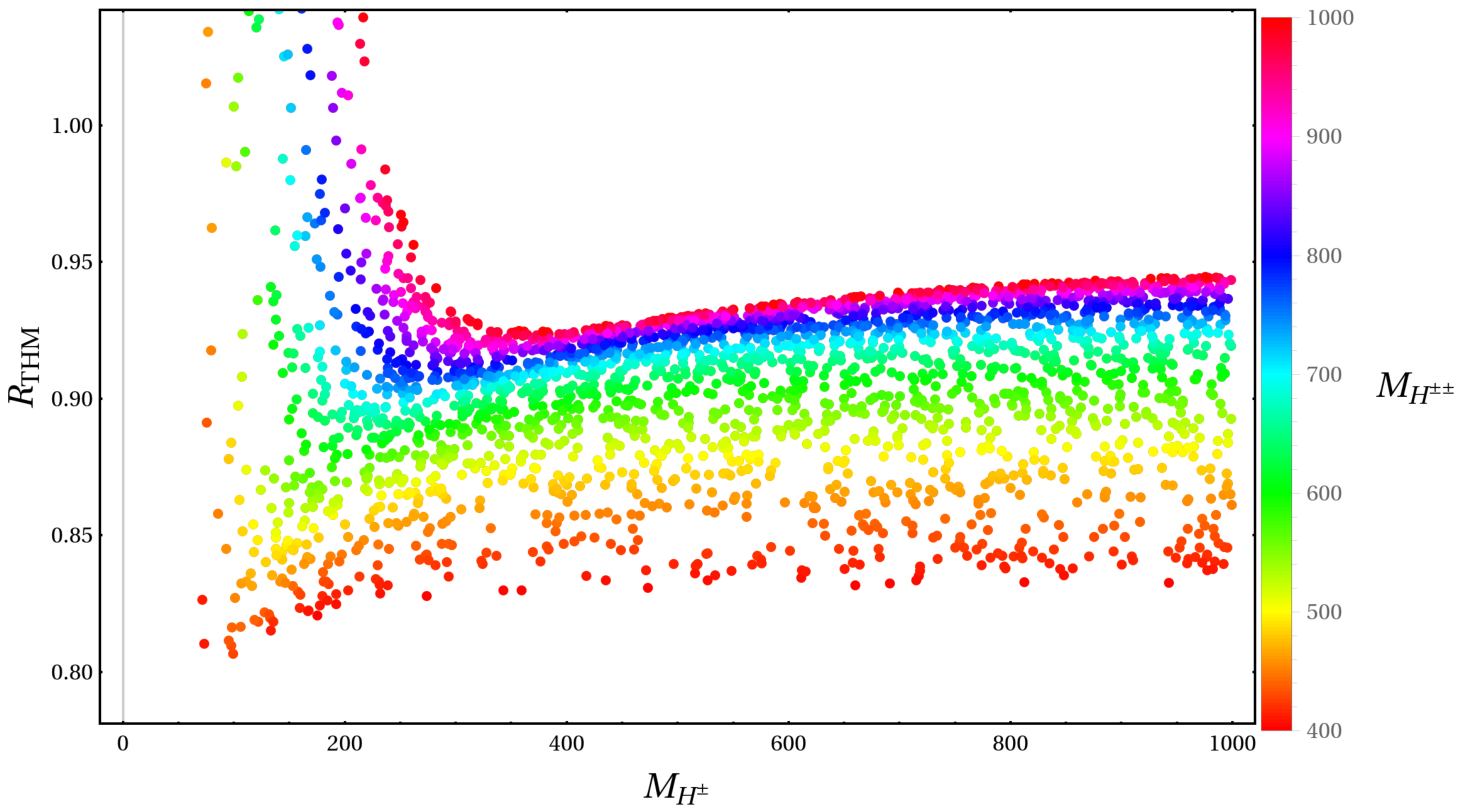}
		& 
		\includegraphics[width=8cm, height=6cm]
		{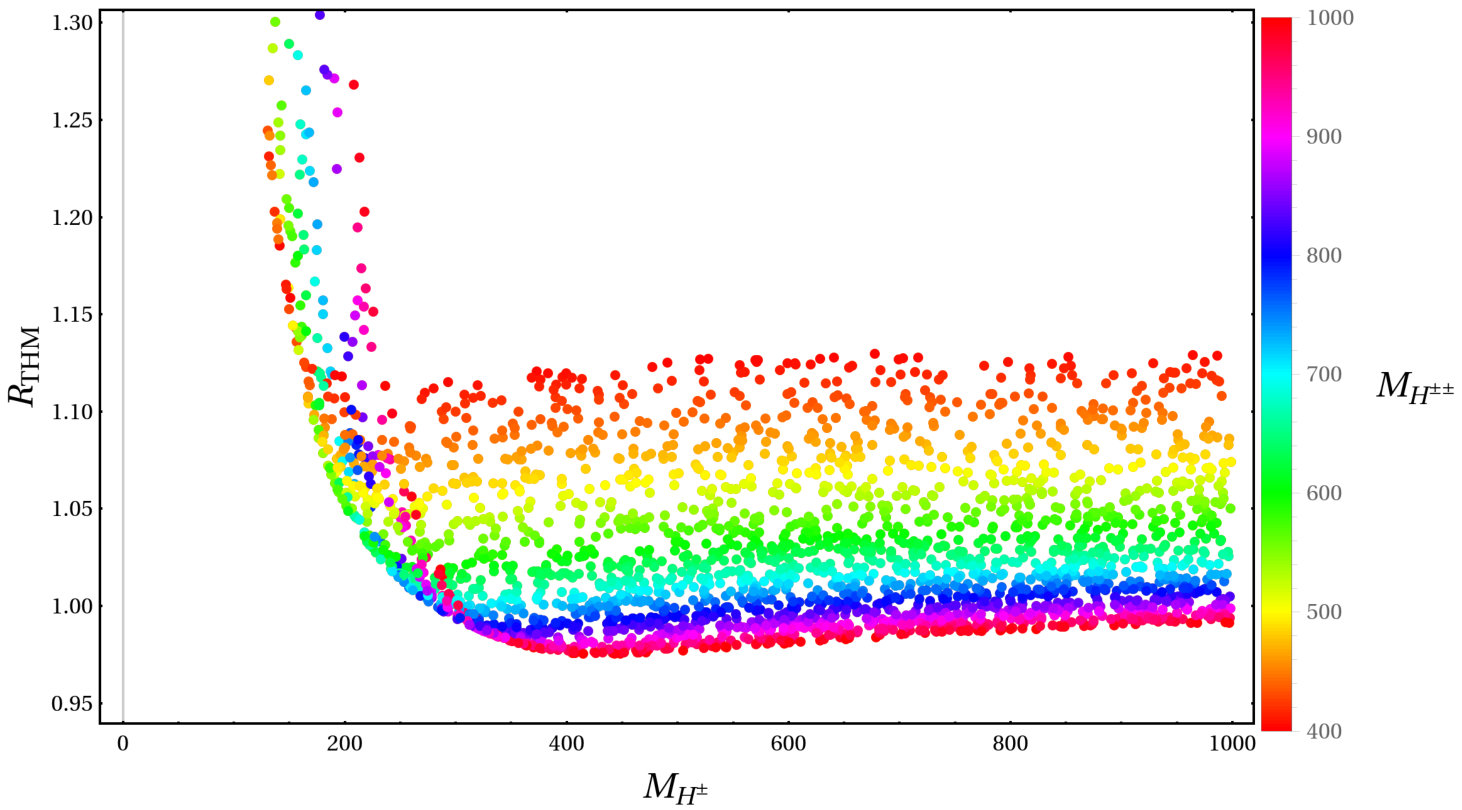}
	\end{array}$
	\caption{\label{RTHMLAB} The enhancement factor
		as a function of $M_{H^{\pm}}$ and $M_{H^{\pm\pm}}$ with two fixed values of
		$\lambda_1= -4$ (4) in the left (right) panel.}
\end{figure}

We next study the enhancement factor as function of $M_{H^{\pm}}$ and $M_{H^{\pm\pm}}$ at the two fixed values of $\lambda_1=\pm 4$  and  $c_{\alpha} =0.98$,  as shown in Fig.~\ref{RTHMLAB}. 
In the  left panel corresponding to  $\lambda_1=-4$,  we find that the enhancement factors have wide deviations from $1$ in the region  $M^{\pm}\leq  200$ GeV. On the other hand,   $ 0.83\leq R_{THM}\leq 0.95$ in the region  $M^{\pm}\geq 200$ GeV. The results show that the enhancement factors are more sensitive with $M_{H^{\pm\pm}}$ in the region of light $M_{H^{\pm}}$. The right panel shows numerical results of  the enhancement factor with $\lambda_1=4$. At small  $M_{H^{\pm}} \leq \sim 200$ GeV,  the enhancement factors are much greater than $1$ and more sensitive with $M_{H^{\pm\pm}}$.  In contrast,  the region with  of $M_{H^{\pm}} \geq \sim 200$ GeV, the enhancement factors develop in a range of $[0.98,1.13]$.  Two scenarios for different $\lambda_1$ can be distinguished from these plots and  could be probed at future colliders.  We also show that the indirect effect of singly and doubly charged Higgs bosons could be probed by measuring these factors in future colliders.

In Table~\ref{AFBTHM}, we present the fermion  FB asymmetries as functions of   singly and doubly charged Higgs masses and $\lambda_1$. The first column list numerical values of   $M_{H^\pm}, M_{H^{\pm \pm}}$, and  $\lambda_1$ with fixed  $v_\Delta=5$ MeV and $c_{\alpha}=0.95, 0.98$. The second (third) column shows numerical values of  the FB asymmetries of  electron (muon). We cover all possible values of $M_{H^\pm}, M_{H^{\pm \pm}}$
in the scanning ranges of  the THM discussed  above. We obtain the same conclusions  with IDM and THDM, except that   the THM predicts  bigger deviations from the SM result than those predicted by  the two models  IDM and THDM. Since, we   have not only singly and doubly charged Higgs in the loop of  $Z^*$-pole diagrams but also  the coupling of $Z^*$ to singly charged Higgs boson depends on $c_{\beta^{\pm}}$. This explains the  bigger deviation of the fermion FB asymmetries in THM in comparison with  other models. These effects could be tested at future colliders. 
\begin{table}[ht]
\begin{center}
\begin{tabular}{l@{\hspace{2cm}}c@{\hspace{2cm}}c}  
\hline \hline 
\\
$\big(M_{H^\pm}, M_{H^{\pm \pm}}, \lambda_1 \big)$
&$\big( R^{(e)}_{\text{THM}} \,\,,\,\, \mathcal{A}_{\text{FB}}^{(e)} \big)$  
&$\big( R^{(\mu)}_{\text{THM}} \,\,,\,\, \mathcal{A}_{\text{FB}}^{(\mu)} \big)$  
\\
\\ \hline
\\
$\big( 100 , 400 , +4 \big)$  
& $(0.7942 \,\,,\,\, 0.3086)$ 
& $(0.8257 \,\,,\,\, 0.2480)$  \\
& $(0.7475 \,\,,\,\, 0.3152)$ 
& $(0.7769 \,\,,\,\, 0.2532)$  \\
\\ \hline  
\\
$\big( 200 , 500 , +4 \big)$  
& $(0.8573 \,\,,\,\, 0.2907)$ 
& $(0.8768 \,\,,\,\, 0.2374)$  \\
& $(0.8104 \,\,,\,\, 0.2955)$ 
& $(0.8278 \,\,,\,\, 0.2414)$  \\
\\ \hline  
\\
$\big( 400 , 600 , +4 \big)$  
& $(0.8997 \,\,,\,\, 0.2644)$ 
& $(0.9104 \,\,,\,\, 0.2182)$  \\
& $(0.8514 \,\,,\,\, 0.2672)$ 
& $(0.8611 \,\,,\,\, 0.2208)$  \\
\\ \hline  
\\
$\big( 600 , 600 , +4 \big)$  
& $(0.9057 \,\,,\,\, 0.2513)$ 
& $(0.9161 \,\,,\,\, 0.2080)$  \\
& $(0.8584 \,\,,\,\, 0.2532)$ 
& $(0.8667 \,\,,\,\, 0.2099)$  \\
\\ \hline 
\\
$\big( 1000 , 1000 , +4 \big)$  
& $(0.9452 \,\,,\,\, 0.2612)$ 
& $(0.9481 \,\,,\,\, 0.2175)$  \\
& $(0.8981 \,\,,\,\, 0.2637)$ 
& $(0.8989 \,\,,\,\, 0.2199)$  \\
\\ \hline  
\\
$\big( 150 , 400 , -4 \big)$  
& $(1.1449 \,\,,\,\, 0.5012)$ 
& $(1.1099 \,\,,\,\, 0.4245)$  \\
& $(1.1106 \,\,,\,\, 0.5166)$ 
& $(1.0712 \,\,,\,\, 0.4395)$  \\
\\ \hline  
\\
$\big( 200 , 600 , -4 \big)$  
& $(1.0489 \,\,,\,\, 0.4778)$ 
& $(1.0240 \,\,,\,\, 0.3989)$  \\
& $(1.0013 \,\,,\,\, 0.4930)$ 
& $(0.9826 \,\,,\,\, 0.4130)$  \\
\\ \hline  
\\
$\big( 400 , 700 , -4 \big)$  
& $(1.0013 \,\,,\,\, 0.3198)$ 
& $(0.9935 \,\,,\,\, 0.2673)$  \\
& $(0.9558 \,\,,\,\, 0.3260)$ 
& $(0.9457 \,\,,\,\, 0.2732)$  \\
\\ \hline  
\\
$\big( 800 , 800 , -4 \big)$  
& $(1.0042 \,\,,\,\, 0.2751)$ 
& $(0.9959 \,\,,\,\, 0.2309)$  \\
& $(0.9577 \,\,,\,\, 0.2784)$ 
& $(0.9473 \,\,,\,\, 0.2342)$  \\
\\ \hline  
\\
$\big( 1000 , 400 , -4 \big)$  
& $(1.1414 \,\,,\,\, 0.2773)$ 
& $(1.0990 \,\,,\,\, 0.2369)$  \\
& $(1.0862 \,\,,\,\, 0.2806)$ 
& $(1.0514 \,\,,\,\, 0.2404)$  \\
\\ \hline   
\hline      
\end{tabular}
\caption{\label{AFBTHM}
The enhancement factor for the decay rates $R^{(l)}_{\text{THM}}$ and integrated FB asymmetries 
$\mathcal{A}_{\text{FB}}^{(l)}$ with $l = e, \mu$ in the THM. The first (second) line shows the numerical values of  fixed $c_\alpha = 0.98, v_\Delta = 5$ MeV ($c_\alpha = 0.95, v_\Delta = 1$ MeV). The masses are shown in GeV.}
\end{center}
\end{table}

\section{ \label{sec_con} Conclusions}   
In this work, we have studied systematically one-loop corrections to the decay  channels of the SM-like Higgs boson  $h\rightarrow l\bar{l}\gamma$ with $l=\nu_{e,\mu, \tau}, e, \mu$, performed  in three HESM frameworks, handling the computations in the  HF gauge. The final formulas of one-loop form factors are expressed in terms of the logarithm and di-logarithmic functions, which have a great advantage for numerical investigation using simple numerical packages. In addition, we indicated that these analytic formulas can be reduced to the previous results of one-loop contributions to the decay channels $h\to \gamma \gamma, Z\gamma$, therefore confirming the consistency between them. Our results were used to investigate the enhancement factors of the decays $h\rightarrow l\bar{l}\gamma$  and the respective fermion FB asymmetries in the three particular HESM, namely the IDM, THDM, and THM. The numerical results show that $R_{\mathrm{IDM}}$ can approach the largest deviation of 10\% from the SM prediction. On the other hand, both $R_{\mathrm{THDM}}$ and $R_{\mathrm{THM}}$ can reach larger deviation values of 30\%. Last but not least, the IDM predicts values of the fermion FB asymmetries to be nearly the same as the SM value. In contrast, the two THDM and HTM predict larger deviations from the SM, namely they can reach 30\% and 50\%, respectively. In conclusion, we have shown that direct impacts of mixing of neutral Higgs bosons and indirect effects of singly and doubly charged Higgs bosons exchanges to the decay channels $h\rightarrow l\bar{l}\gamma$  could be probed at future colliders. 
\\

\noindent
{\bf Acknowledgment:} 
This research is funded by Vietnam National Foundation for Science and
Technology Development (NAFOSTED) under the grant number $103.01$-$2019.346$
\\ 
\appendix 
\section{ \label{app_coupling} The couplings}                       
\subsection*{Inert Doublet Model} 
Deriving all couplings in 
Table~\ref{IDM-coupling} for 
IDM are presented in this Appendix.
We first collect the terms which give
the coupling of SM-like Higgs to 
singly charged Higgs as follows:
\begin{eqnarray}
\mathcal{L}_{\rm IDM} \supset
\lambda_3|\Phi_1|^2|\Phi_2|^2
\supset
\frac{1}{2}\lambda_3v^2AA+\frac{1}{2}
\lambda_3v^2H^{\mp}H^{\pm} 
+ \frac{1}{4}\lambda_3v^2HH 
+ \lambda_3vhH^{\mp}H^{\pm}.
\end{eqnarray}
From the kinetic terms in the IDM Lagrangian, 
one can derive the couplings of SM-like Higgs to 
gauge bosons as:
\begin{eqnarray}
\mathcal{L}_{\rm IDM}
&=&(D_{\mu}\Phi_1)^{\dagger}(D^{\mu}\Phi_1)
+ (D_{\mu}\Phi_2)^{\dagger}(D^{\mu}\Phi_2) 
\\
&\supset& \Big[\partial_{\mu}H^{\mp} 
+ \frac{i}{2}(g_Zc_{2W}Z_{\mu} + g_Zs_{2W}\gamma_{\mu})H^{\mp}+\frac{i}{2}g_2W^-_{\mu}(H+iA)
\Big]\times \nonumber\\
&& \times \Big[
\partial^{\mu}H^{\pm} -\frac{i}{2}(g_Z c_{2W}Z^{\mu} 
+ g_Z s_{2W}\gamma^{\mu})H^{\pm} 
- \frac{i}{2}g_2W^{+, \mu}(H+iA)
\Big] \nonumber\\
&=&\frac{ig_Zc_{2W}}{2}Z^{\mu}
(H^{\mp}\partial_{\mu}H^{\pm}-\partial_{\mu}H^{\mp}H^{\pm})
+ \frac{ig_Zs_{2W}}{2}\gamma^{\mu}(H^{\mp}\partial_{\mu}H^{\pm} - 
\partial_{\mu}H^{\mp}H^{\pm})
\nonumber\\
&=&i \frac{M_Z c_{2W}}{v}\; Z^{\mu}(H^{\mp}\partial_{\mu}H^{\pm} 
-\partial_{\mu}H^{\mp}H^{\pm})+ 
i \frac{2 M_W s_{W}}{v} \; 
A^{\mu}(H^{\mp}\partial_{\mu}H^{\pm} 
-\partial_{\mu}H^{\mp}H^{\pm}). 
\nonumber
\end{eqnarray}
Where $A_{\mu}$ is photon field.  
\subsection*{Two Higgs Doublet Model}  
We next derive all couplings in 
Table~\ref{THDM-coupling} for 
THDM in this Appendix. 
We expand the kinetic terms of 
Higgs Lagrangian in THDM, 
\begin{eqnarray}
\mathcal{L}_{\rm THDM}
&=&(D_{\mu}\Phi_1)^{\dagger}(D^{\mu}\Phi_1)
+(D_{\mu}\Phi_2)^{\dagger}(D^{\mu}\Phi_2) 
\nonumber\\
&\supset&
i \frac{M_Z c_{2W} }{v} 
Z^{\mu}
(H^{\mp}\partial_{\mu}H^{\pm}
-\partial_{\mu}H^{\mp}H^{\pm})
+i\frac{M_Z s_{2W} }{v}
A^{\mu}(H^{\mp}\partial_{\mu}H^{\pm} - 
\partial_{\mu}H^{\mp}H^{\pm})
\nonumber\\
&&+\frac{M_Z^2 s_{\beta-\alpha}}{v}
hZ^{\mu}Z_{\mu}
+
\frac{2M_W^2 s_{\beta-\alpha}}{v}
h W^{\mu}W_{\mu}.
\end{eqnarray}
The Higgs potential gives us the 
coupling for $hH^{\pm}H^{\mp}$, 
\begin{eqnarray}
\mathcal{V}_{\rm THDM} 
&=&\frac{\lambda_1}{2}(\Phi^\dagger_1\Phi_1)^2
+\frac{\lambda_2}{2}(\Phi^\dagger_2\Phi_2)^2 
+\lambda_3(\Phi_1^{\dagger}\Phi_1)(\Phi_2^{\dagger}\Phi_2)
+\lambda_4(\Phi_1^{\dagger}\Phi_2)(\Phi_2^{\dagger}\Phi_1) \notag\\
&& 
+\lambda_5[(\Phi_1^{\dagger}\Phi_2)^2+(\Phi_2^{\dagger}\Phi_1)^2]  \\
&\supset
&  \Big[
\lambda_1\; v \; c_{\beta}\; c_{\alpha} s_{\beta}^2 
+ \lambda_2 \;v\; s_{\beta}s_{\alpha}c_{\beta}^2 
+{\lambda_3v}[s_{\beta}s_{\beta}^2s_{\alpha} 
+ c_{\beta}c_{\beta}^2c_{\alpha}] \nonumber\\
&& \hspace{4cm} 
-(\lambda_4 + \lambda_5)
\;v \;s_{\beta}c_{\beta}
(s_{\beta}c_{\alpha}+c_{\beta}s_{\alpha})
\Big]\times h H^{\pm}H^{\mp}
\nonumber\\
&=&
\dfrac{1}{v}
\Big[
(2\mu^2 -2 M_{H^{\pm}}^2 -2 m_h^2)
s_{\beta-\alpha}
+ 2(\mu^2 -m_{h}^2)\text{cot}_{2\beta}
c_{\beta-\alpha}
\Big]h H^{\pm}H^{\mp}.
\nonumber
\end{eqnarray}

The generalized Yukawa Lagrangian of THDM is
\begin{align}
    -\mathcal{L}_Y=-\bar{Q}_L\frac{M_u}{v_i}\tilde{\Phi}_iu_R-\bar{Q'}_L\frac{M_d}{v_j}{\Phi}_jd_R-\bar{L}_L\frac{M_l}{v_k}{\Phi}_kl_R+h.c.
\end{align}
Specifically, 
we have four Yukawa Lagrangian types 
as follows:
\begin{itemize}
 \item \underline{Type I:}\\
 \begin{align}
-\mathcal{L}_Y^{I}&
=-\bar{Q}_L\frac{M_u}{v_2}\tilde{\Phi}_2u_R-\bar{Q'}_L\frac{M_d}{v_2}{\Phi}_2d_R-\bar{L}_L\frac{M_l}{v_2}{\Phi}_2l_R+h.c \notag\\
&{\supset}\frac{M_l}{\sqrt{2}v}\frac{s_\alpha}{s_\beta}H\bar{e}_ie_i
+\frac{M_l}{\sqrt{2}v}\frac{c_\alpha}{s_\beta}
h\bar{e}_ie_i+i\frac{M_l\cot{\beta}\gamma^5}{\sqrt{2}v}A\bar{e}^i{e}^i \notag\\
&+\frac{M_u}{\sqrt{2}v}\frac{s_\alpha}{s_\beta}
H\bar{u^i}u^i
+\frac{M_u}{\sqrt{2}v}\frac{c_\alpha}{s_\beta}
h\bar{u^i}u^i-i\frac{M_u\cot{\beta}\gamma^5}{\sqrt{2}v}A\bar{u}^i{u}^i \notag\\
&+\frac{M_d}{\sqrt{2}v}\frac{s_\alpha}{s_\beta}
H\bar{d^i}d^i
+\frac{M_d}{\sqrt{2}v}\frac{c_\alpha}{s_\beta}
h\bar{d^i}d^i
+i\frac{M_d\cot{\beta}\gamma^5}{\sqrt{2}v}A\bar{d}^i{d}^i.
\end{align}
\item \underline{Type II:}
\begin{align}
-\mathcal{L}_Y^{II}&=-\bar{Q}_L\frac{M_u}{v_2}\tilde{\Phi}_2u_R-\bar{Q'}_L\frac{M_d}{v_1}{\Phi}_1d_R-\bar{L}_L\frac{M_l}{v_1}{\Phi}_1l_R+h.c \notag\\
&{\supset}\frac{M_l}{\sqrt{2}v}\frac{c_\alpha}{c_\beta}H\bar{e^i}e^i
-\frac{M_l}{\sqrt{2}v}\frac{s_\alpha}{c_\beta}
h\bar{e^i}e^i-i\frac{M_l\gamma^5\tan{\beta}}{\sqrt{2}v}A\bar{e}^i{e}^i \notag\\
&+\frac{M_d}{\sqrt{2}v}\frac{c_\alpha}{c_\beta}H\bar{d^i}d^i
-\frac{M_d}{\sqrt{2}v}\frac{s_\alpha}{c_\beta}
h\bar{d^i}d^i-i\frac{M_d\gamma^5\tan\beta}{\sqrt{2}v}A\bar{d}^id^i \notag\\
&+\frac{M_u}{\sqrt{2}v}\frac{s_\alpha}{s_\beta}H\bar{u^i}u^i
+\frac{M_u}{\sqrt{2}v}\frac{c_\alpha}{s_\beta}
h\bar{u^i}u^i-i\frac{M_u\gamma^5\cot\beta}{\sqrt{2}v}A\bar{u}^i{u}^i. 
\end{align}
\item \underline{Type X:}
\begin{align}
-\mathcal{L}_Y^{X}&=-\bar{Q}_L\frac{M_u}{v_2}\tilde{\Phi}_2u_R-\bar{Q'}_L\frac{M_d}{v_2}{\Phi}_2d_R-\bar{L}_L\frac{M_l}{v_1}{\Phi}_1l_R+h.c \notag\\
&{\supset}\frac{M_l}{\sqrt{2}v}\frac{c_\alpha}{c_\beta}H\bar{e^i}e^i
-\frac{M_l}{\sqrt{2}v}\frac{s_\alpha}{c_\beta}
h\bar{e^i}e^i-i\frac{M_l\gamma^5\tan\beta}{\sqrt{2}v}A\bar{e}^i{e}^i \notag\\
&+\frac{M_d}{\sqrt{2}v}\frac{s_\alpha}{s_\beta}H\bar{d^i}d^i
+\frac{M_d}{\sqrt{2}v}\frac{c_\alpha}{s_\beta}
h\bar{d^i}d^i+i\frac{M_d\gamma^5\cot\beta}{\sqrt{2}v}A\bar{d}^i{d}^i \notag\\
&+\frac{M_u}{\sqrt{2}v}\frac{s_\alpha}{s_\beta}H\bar{u^i}u^i
+\frac{M_u}{\sqrt{2}v}\frac{c_\alpha}{s_\beta}h\bar{u^i}u^i-i\frac{M_u\gamma^5\cot\beta}{\sqrt{2}v}A\bar{u}^i{u}^i.  
\end{align}
\item \underline{Type Y:}
\begin{align}
-\mathcal{L}_Y^{Y}&=-\bar{Q}_L\frac{M_u}{v_2}\tilde{\Phi}_2u_R-\bar{Q'}_L\frac{M_d}{v_1}{\Phi}_1d_R-\bar{L}_L\frac{M_l}{v_2}{\Phi}_2l_R+h.c \notag\\
&{\supset}\frac{M_l}{\sqrt{2}v}\frac{s_\alpha}{s_\beta}H\bar{e}_ie_i
+\frac{M_l}{\sqrt{2}v}\frac{c_\alpha}{s_\beta}h\bar{e}_ie_i+i\frac{M_l\gamma^5\cot\beta}{\sqrt{2}v}A\bar{e}^i{e}^i \notag\\
&+\frac{M_u}{\sqrt{2}v}\frac{s_\alpha}{s_\beta}H\bar{u}^i\; u^i
+\frac{M_u}{\sqrt{2}v}\frac{c_\alpha}{s_\beta}h\bar{u}^i u^i-i\frac{M_u\gamma^5\cot\beta}{\sqrt{2}v}A\bar{u}^i{u}^i \notag\\
&+\frac{M_d}{\sqrt{2}v}\frac{c_\alpha}{c_\beta}H\bar{d^i}d^i
-\frac{M_d}{\sqrt{2}v}\frac{s_\alpha}{c_\beta}h\bar{d^i}d^i-i\frac{M_d\gamma^5\tan\beta}{\sqrt{2}v}A\bar{d}^i{d}^i.
\end{align}
\end{itemize}
\subsection*{Triplet Higgs Model}   
We expand the kinematic term in the 
Lagrangian
of Triplet Higgs Model,
\begin{align}
\mathcal{L}_{\rm THM}
&= {\rm Tr}[(D^\mu{\Delta})^\dagger(D_\mu{\Delta})]
+(D^\mu{\Phi})^\dagger(D_\mu{\Phi}) 
\nonumber\\
&\supset
ie \; 
A^{\mu}(H^{\pm}{\partial_{\mu} }H^{\mp}
-H^{\mp}{\partial_{\mu} }H^{\pm})
\\
&+ i\frac{M_Z}{ \sqrt{v^2+2v_{\Delta}^2} }
(c^2_{W}-s_W^2-c^2_{\beta^{\pm}})
Z^{\mu} (H^{\pm}{\partial_{\mu}}H^{\mp} 
- H^{\mp}{\partial_{\mu}}H^{\pm}) \notag\\
&
+ 
\frac{2M_W^2}{v} ( c_{\alpha}\; c_{\beta^{\pm} }  
+ \sqrt{2} s_{\alpha} \; s_{\beta^{\pm} } )
W^{\pm, \mu}W^{\mp}_{\mu}h 
+ 
\frac{2M_Z^2}{\sqrt{v^2+2v_{\Delta}^2 }}
(c_{\beta^0} c_{\alpha} 
+ 2 s_{\beta^0} s_{\alpha} )
Z^{\mu}Z_{\mu}h \notag\\
&
+
i\frac{2M_Z}{  \sqrt{v^2+2v_{\Delta}^2} 
}(c^2_{W}-s_W^2)
Z^{\mu} (H^{\pm\pm}{\partial_{\mu} }H^{\mp\mp}
-H^{\mp\mp}{\partial_{\mu} }H^{\pm\pm}) \notag\\
&
+i(2e )
A^{\mu}
(H^{\pm\pm}{\partial_{\mu} }H^{\mp\mp}
-H^{\mp\mp}{\partial_{\mu} }H^{\pm\pm}). 
\nonumber
\end{align}
The Higgs potential gives us 
the couplings of $hH^{\pm\pm}H^{\mp\mp},
hH^{\pm}H^{\mp}$ as follows:
\begin{align}
\mathcal{V}_{\rm THM}& =
\frac{\lambda}{4}(\Phi^\dagger\Phi)^2 
+ {\lambda_1}(\Phi^\dagger\Phi)
{\rm Tr}(\Delta^\dagger\Delta) 
+ (\lambda_2+\lambda_3) 
{\rm Tr}(\Delta^\dagger\Delta)^2 
+ \lambda_4\Phi^\dagger\Delta
\Delta^\dagger\Phi \notag\\
&{\supset}
\Bigg[\frac{\lambda}{2}s^2_{\beta^{\pm}}v_{\Phi}c_{\alpha}+\lambda_1(s^2_{\beta^{\pm}}v_{\Delta}s_{\alpha}+c^2_{\beta^{\pm}}v_{\Phi}c_{\alpha})+2(\lambda_2+\lambda_3)c^2_{\beta^{\pm}}v_{\Delta}s_{\alpha} \notag\\
& \hspace{3cm}
+\frac{\lambda_4}{2} 
(-\sqrt{2}s_{\beta^{\pm}}c_{\beta^{\pm}}(v_{\Phi}s_{\alpha}+v_{\Delta}c_{\alpha})+v_{\Phi}c_{\alpha}c^2_{\beta^{\pm}})\Bigg]hH^{\pm}H^{\mp} \notag\\
&+ 
\Big[\lambda_1v_{\Phi}c_{\alpha}+2(\lambda_2+\lambda_3)v_{\Delta}s_{\alpha}
\Big] hH^{\pm\pm}H^{\mp\mp}.
\end{align}
When $\frac{v_{\Delta} }{v_{\Phi} } \rightarrow{0}$, one has  
$s_{\beta^{\pm}}\rightarrow{0}$ and the resulting reads
\begin{align}
g^{\rm THM}_{hH^{\pm}H^{\mp}}&=
\lambda_1v_{\Phi}c_{\alpha} 
+ \frac{\lambda_4}{2}v_{\Phi}c_{\alpha}
=\frac{v}{2}(2\lambda_1+\lambda_4)c_{\alpha} 
= \lambda_1 \; v\; c_{\alpha} 
+ \frac{2(M_{A^0}^2 
- M^2_{H^{\pm}})}{v}c_{\alpha},
\\
g^{\rm THM}_{hH^{\pm\pm}H^{\mp\mp}}
&=\lambda_1v_{\Phi}c_{\alpha}
= \lambda_1\; v\; c_{\alpha}.
\end{align}
Here $\lambda_1 \rightarrow \frac{4 M_{H^{\pm}}^2 
- 3M_{A^0}^2 + m_h^2 }{v^2}$ 
and $2 M_{H^{\pm}}^2 = M_{H^{\pm\pm}}^2+M_{A^0}^2$. 
The Yukawa Lagrangian
is to generate masses of neutrinos
\begin{align}
-\mathcal{L}_{\rm Yukawa}
&=-\mathcal{L}_{\rm Yukawa}^{\rm SM}
+ y_{\nu}l^T_LC(i\sigma_2\Delta)l_L
+ {\rm h.c}
\notag\\
&=y_{\nu}
\left(
\begin{array}{cc}
\nu^i_L & e^i_L 
\end{array}
\right)
C
\left(
\begin{array}{cc}
0 & 1 \\
-1 & 0
\end{array}
\right)
\left(
\begin{array}{cc}
\frac{\delta^{\pm}}{\sqrt{2}} 
& \delta^{\pm\pm} \\
\frac{1}{\sqrt{2}}(v_\Delta 
+\eta_\Delta+i\chi_\Delta) 
& -\frac{\delta^{\pm}}{\sqrt{2}}
\end{array}
\right)
\left(
\begin{array}{c}
\nu^i_L  \\
e^i_L 
\end{array}
\right) 
+
{\rm h.c}
\notag\\
&=\frac{y_{\nu}v_{\Delta}}{\sqrt{2}}(\nu_L^iC{\nu}_L^i+\nu_L^iC^{\dagger}{\nu}_L^i)+\frac{y_{\nu}s_{\alpha}}{\sqrt{2}}h(\nu_L^iC{\nu}^i_L+\nu_L^iC^{\dagger}{\nu}^i_L)+\frac{y_{\nu}c_{\alpha}}{\sqrt{2}}H(\nu_L^iC\nu^i_L+\nu_L^iC^{\dagger}\nu^i_L) \notag\\
&+i\frac{y_{\nu}s_{\beta^0}}{\sqrt{2}}G^0(\nu_L^iC\nu^i_L-\nu_L^iC^{\dagger}\nu^i_L) 
+ i\frac{y_{\nu}c_{\beta^0}}{\sqrt{2}} 
A(\nu_L^iC\nu^i_L-\nu_L^iC^{\dagger}\nu^i_L) \notag\\
&-\frac{y_{\nu}s_{\beta^{\pm} }}{\sqrt{2}}G^{\pm}(\nu_L^iCe_L^i+e_L^iC\nu_L^i) 
- \frac{y_{\nu}c_{\beta^{\pm} } } {\sqrt{2}}H^{\pm}(\nu_L^iCe_L^i+e_L^iC\nu_L^i) 
\notag\\
&
- \frac{y_{\nu}s_{\beta^{\pm} }}{\sqrt{2}}G^{\mp}(\nu_L^iC^{\dagger}e_L^i+e_L^iC^{\dagger}\nu_L^i)
-\frac{y_{\nu}c_{\beta^{\pm} }}{\sqrt{2}}H^{\mp}(\nu_L^iC^{\dagger}e_L^i+e_L^iC^{\dagger}\nu_L^i)
\nonumber\\
&- y_{\nu}(\delta^{\pm\pm}e_L^iCe^i_L 
+ \delta^{\mp\mp}e_L^iC^{\dagger}e_L^i).
\end{align}
Here we replace 
$\delta^{\pm\pm} = H^{\pm\pm}$ for later. 
We then collect all couplings related to
the channels under consideration in THM as 
shown in Table~\ref{THM-coupling}.  
\section{ \label{app_loga} Expressions for  
scalar one-loop integrals}               
Expressions for 
all scalar one-loop integrals 
in terms of logarithm and 
di-logarithm functions are presented 
in this appendix.
The $\epsilon$-expansions are 
performed with the help of 
{\tt Package-X}. We have used the  
shorthand notation for kinematic 
variables 
as follows: 
$\tau_i = 4 M_i^2 / m_h^2$ 
and $\eta_i (\zeta_i) = 
4 M_i^2 / q_{23} (q_{13})$, 
$\lambda_i = 4 M_i^2 / q_{12}$ 
where $M_i$ can be one of masses
$\{m_f, M_Z, M_W \}$. It is 
stress that internal masses 
are taken the form of 
$M_i^2 \rightarrow 
M_i^2 -i \rho$
for $\rho\rightarrow 0$ where 
Feynman's $i\rho$
prescription is taken into 
account. Expanding scalar 
one-loop integrals, there are 
exist the following parameters 
$1/\epsilon$ and $\mu^2$. The
first parameter explains for
$UV$-divergent part of Feynman
loop integrals. The later parameter
is to renormalization 
scale introducing to   
help to track of the 
correct dimension of the integrals 
in space-time dimension $d$. Besides that
Spence functions ${\rm Li}_2(x)$, 
$\epsilon$-expansions
for scalar box-integrals relate
to Beenakker-Denner continued 
dilogarithm functions $\mathcal{L}i_2(x,y)$
as defineed in \cite{Beenakker:1988jr}.
\begin{itemize}   
\item \underline{
$B_{0}(m_h^2,M_i^2,M_i^2) 
- B_{0}(q_{23},0,M_i^2)$}:

Scalar one-loop two-point functions
appear in our work can be expanded
in terms of logarithm functions
as follows:
\begin{eqnarray}
B_{0}(m_h^2,M_i^2,M_i^2)
&=&
2
+
\dfrac{1}{\epsilon}
-
\log (M_i^2)
+
\sqrt{1-\tau_i} 
\log 
\Bigg[
1
+
\dfrac{2 
(\sqrt{1-\tau_i}-1)}{\tau_i}
\Bigg], 
\\
B_{0}(q_{23},0,M_i^2)
&=&
2
+
\dfrac{1}{\epsilon}
-
\log (M_i^2)
-
\Big(
\dfrac{\eta_i}{4} - 1
\Big) 
\log 
\Bigg[
\dfrac{\eta_i}{\eta_i-4}
\Bigg]. 
\end{eqnarray}
Since internal masses takes
$M_i^2 -i \rho$ for $\rho\rightarrow 0$, 
the logarithm functions in above are well 
defined in complex plane. 
Within the dimensional regularization,
the ultraviolet divergences will be 
cancelled once 
subtract two scalar two-points 
integrals $B_0$. In detail, we 
obtain the finite term as follows:
\begin{eqnarray}
B_{0}(m_h^2,M_i^2,M_i^2) 
- 
B_{0}(q_{23},0,M_i^2)
&=&  \\
&&\hspace{-4cm}
=
\Big(
\dfrac{\eta_i}{4} - 1
\Big) 
\log 
\Bigg[
\dfrac{\eta_i}{\eta_i-4}
\Bigg]
+
\sqrt{1-\tau_i} 
\log 
\Bigg[
1
+
\dfrac{2 (\sqrt{1-\tau_i}-1)}{\tau_i}
\Bigg]. 
\nonumber
\end{eqnarray}
\item \underline{
$C_{0}(0,m_h^2,q_{23}(q_{13}),
0,M_i^2,M_i^2)$}:

Expansion for scalar 
one-loop three-point functions
in this configuration 
is presented as
\begin{eqnarray}
&&\hspace{-1.0cm}
C_{0}(0,m_h^2,q_{23},0,M_i^2,M_i^2)
=
\n \\
&=&
-\dfrac{\eta_i \tau_i}{4 M_i^2 (\eta_i-\tau_i)}
\Bigg\{
\text{Li}_2
\Bigg[
\dfrac{\eta_i}{\tau_i}
+
i \rho 
\dfrac{
(\eta_i-4) (\eta_i-\tau_i)
}{\eta_i^3 \tau_i^2}
\Bigg]
\\
&&
+\text{Li}_2
\Bigg[
\dfrac{ \eta_i (\tau_i-4) +4 \tau_i}
{
\eta_i \tau_i
+
2 (\eta_i - \tau_i) (\sqrt{1-\tau_i}-1)}
-i \rho 
\Big\{
\big(\eta_i - \tau_i + 8 \big) 
\eta_i^{-1} \tau_i^{-1}
-4 \big(
\eta_i^{-2}
+ 
\tau_i^{-2}
\big)
\Big\}
\Bigg]
\n \\
&&
+\text{Li}_2
\Bigg[
\dfrac{\eta_i^2}
{
\eta_i^2
+4 (\tau_i - \eta_i)
}
+ i \rho 
\big(4 \eta_i^{-1}-1\big)
\Bigg]
-
\text{Li}_2
\Bigg[
\dfrac{\eta_i^2 (\tau_i-4)+4 \eta_i \tau_i}
{
\tau_i (\eta_i^2-4 \eta_i+4 \tau_i)}
\Bigg]
\n \\
&&
-\text{Li}_2
\Bigg[
\dfrac{\eta_i^2-4 \eta_i+4 \tau_i}{\eta_i \tau_i}
+
i \rho 
\dfrac{
(\eta_i-4) (\eta_i-\tau_i) 
(\eta_i^2-4 \eta_i+4 \tau_i)}{\eta_i^5 \tau_i^2}
\Bigg]
\n \\
&&
-\text{Li}_2
\Bigg[
\dfrac{\eta_i \tau_i
}{
\eta_i \tau_i
+
2 (\tau_i - \eta_i) 
(\sqrt{1-\tau_i}+1)}
+
i \rho 
(\tau_i^{-1}-\eta_i^{-1}) 
\Bigg]
\n \\
&&
+
\text{Li}_2
\Bigg[
\dfrac{\eta_i (\tau_i-4) +4 \tau_i
}{
\eta_i \tau_i
+
2 (\tau_i - \eta_i) (\sqrt{1-\tau_i}+1) }
\Bigg]
-
\text{Li}_2
\Bigg[
\dfrac{\eta_i \tau_i
}{
\eta_i \tau_i
+
2 (\eta_i - \tau_i) 
(\sqrt{1-\tau_i}-1)
}
\Bigg]
\Bigg\}.
\n
\end{eqnarray}
Another $C_0$ function is obtained
by
\begin{eqnarray}
C_{0}(0,m_h^2,q_{13},0,M_i^2,M_i^2)
&=&
C_{0}(q_{13},m_h^2,0,0,M_i^2,M_i^2).
\end{eqnarray}
\item \underline{
$C_{0}(0,q_{12},
0,0,M_i^2,M_i^2)$}: 

For this kinematic configuration,
$C_0$ function is expressed as:
\begin{eqnarray}
C_{0}(0,q_{12},0,0,M_i^2,M_i^2)
&=&
-\dfrac{\lambda_i}{24 M_i^2}
\Bigg\{
\pi ^2
-6 \, \text{Li}_2
\Bigg[
\dfrac{\lambda_i}{\lambda_i
+2 (\sqrt{1-\lambda_i}-1)}
\Bigg]
\\
&& \hspace{-5cm}
+6 \, \text{Li}_2
\Bigg[
\dfrac{\lambda_i-4}{\lambda_i
-2 (\sqrt{1-\lambda_i}+1)}
\Bigg]
-6 \, \text{Li}_2
\Bigg[
1-\dfrac{4}{\lambda_i}
\Bigg]
-6 \, \text{Li}_2
\Bigg[
\dfrac{4-\lambda_i
}{
2 (\sqrt{1-\lambda_i}+1)}
\Bigg]
\nonumber \\
&&
\hspace{-5cm}
+6 \, \text{Li}_2
\Bigg[
\dfrac{\lambda_i} 
{2 (\sqrt{1-\lambda_i}+1)}
\Bigg]
+ 3 \log ^2
\Bigg[
\dfrac{1}{2} (\sqrt{1-\lambda_i}+1)
\Bigg]
-3 \log ^2
\Bigg[
\dfrac{\lambda_i
+ 2 (\sqrt{1-\lambda_i}-1) }
{ 2 (\sqrt{1-\lambda_i}+1)}
\Bigg]
\Bigg\}. 
\nonumber
\end{eqnarray}

\item  \underline{
$C_{0}(0,0,q_{23}(q_{13}),0,0,M_i^2)$
}:

In further, one has 
\begin{eqnarray}
&&\hspace{-1.0cm}
C_{0}(0,0,q_{23},0,0,M_i^2)
=
\n \\
&=&
\dfrac{\eta_i}{8 M_i^2}
\Bigg\{
\log \Big(
\dfrac{\eta_i}{\eta_i-4}
\Big) 
\Bigg[
\dfrac{2}{\epsilon}
+
\log \Big(\dfrac{\eta_i}
{\eta_i-4}\Big)
+2 \log \Big(\dfrac{\mu ^2}{M_i^2}\Big)
\Bigg]
-2 \, \text{Li}_2
\Bigg[
\dfrac{4}{4-\eta_i}
\Bigg]
\Bigg\}. 
\end{eqnarray}
%
\item \underline{
$C_{0}(0,0,q_{23}(q_{13}),
0,M_i^2,M_i^2)$}:

In additional, we have 
\begin{eqnarray}
C_{0}(0,0,q_{23},0,M_i^2,M_i^2)
&=&
\dfrac{\eta_i}{8 M_i^2}
\Bigg\{
2 \, \text{Li}_2
\Bigg[
\dfrac{4}{4-\eta_i}
\Bigg]
+
\log ^2
\Bigg[
\dfrac{\eta_i}{\eta_i-4}
\Bigg]
\Bigg\}.
\end{eqnarray}
%
\item 
\underline{
$C_{0}(q_{12},0,m_h^2,
M_i^2,M_i^2,M_i^2)$}:

In this kinematic configuration, 
scalar one-loop three-point functions
are expanded as
\begin{eqnarray}
&&\hspace{-1.0cm}
C_{0}(q_{12},0,m_h^2,M_i^2,M_i^2,M_i^2)
=
\n \\
&=&
\dfrac{\lambda_i \tau_i}
{
8 M_i^2 (\tau_i - \lambda_i)}
\Bigg\{
\log ^2
\Bigg[
1 +
\dfrac{2 (\sqrt{1-\lambda_i}-1)}{\lambda_i}
\Bigg]
-\log ^2
\Bigg[
1
+
\dfrac{
2 (\sqrt{1-\tau_i}-1)}{\tau_i}
\Bigg]
\Bigg\}.
\end{eqnarray}
\item 
\underline{
$D_{0}(0,q_{13},
m_h^2,q_{23},0,0,0,0,M_i^2,M_i^2)$}:

Scalar one-loop box integrals for
thid kinematic configuration are 
derived as 
\begin{eqnarray}
&&\hspace{-1.0cm}
D_{0}(0,q_{13},m_h^2,q_{23},0,0,0,0,M_i^2,M_i^2)
=
\n \\
&=&
\dfrac{\eta_i \zeta_i}{8 M_i^4
(4-\eta_i-\zeta_i)}
\Bigg\{
2 \log \Big(\dfrac{\eta_i}{\eta_i-4}\Big)
\Bigg[
\dfrac{1}{\epsilon }+\log
\Big(\dfrac{\mu ^2}{M_i^2}\Big)
+2 \log
\Big(\dfrac{\zeta_i}{\zeta_i-4}\Big)
\Bigg]
\\
&&
+\log
\Big(\dfrac{\zeta_i}{\zeta_i-4}\Big) 
\Bigg[
\dfrac{2}{\epsilon }+2 
\log \Big(\dfrac{\mu ^2}{M_i^2}\Big)
+\log \Big(\dfrac{\zeta_i}{\zeta_i-4}\Big)
\Bigg]
+\log ^2\Big(\dfrac{\eta_i}{\eta_i-4}
\Big)
\n \\
&&
+2 \, \mathcal{L}i_2
\Bigg[
\dfrac{\eta_i-4}{\eta_i}
-i\dfrac{\rho}{\eta_i}
,
1-\dfrac{2(\sqrt{1-\tau_i}+1)}{\tau_i}
-i\rho 
\dfrac{
\sqrt{1-\tau_i}}{\tau_i}
\Bigg]
\n \\
&&
+2 \, \mathcal{L}i_2
\Bigg[
\dfrac{\eta_i
-4}{\eta_i}
- i\dfrac{\rho}{\eta_i}
,
1 +
\dfrac{ 2(\sqrt{1-\tau_i}-1)}{\tau_i}
+i\rho \times 
\dfrac{
\sqrt{1-\tau_i}}{\tau_i}
\Bigg]
\n \\
&&
+4 \, \mathcal{L}i_2
\Bigg[
\dfrac{\eta_i
}{\eta_i-4}
+i
\dfrac{\rho }{\eta_i}
,
\dfrac{\zeta_i}{\zeta_i-4}
+ i\dfrac{\rho}{\zeta_i}
\Bigg]
-4
\Bigg[
\text{Li}_2
\Big(\dfrac{4}{4-\eta_i}\Big)
+
\text{Li}_2
\Big(\dfrac{4}{4-\zeta_i}\Big)
\Bigg]
\n \\
&&
-2 \, \mathcal{L}i_2
\Bigg[
1-\dfrac{2(\sqrt{1-\tau_i}+1)
}{
\tau_i}
-i\rho \dfrac{
\sqrt{1-\tau_i}}{\tau_i}
,
\dfrac{\zeta_i}{\zeta_i-4}
+
i
\dfrac{\rho}{\zeta_i}
\Bigg]
\n \\
&&
-2 \, \mathcal{L}i_2
\Bigg[
1
+
\dfrac{2(\sqrt{1-\tau_i}-1)}{\tau_i}
+i\rho 
\dfrac{\sqrt{1-\tau_i}}{\tau_i}
,
\dfrac{\zeta_i}{\zeta_i-4}
+i
\dfrac{\rho}{\zeta_i}
\Bigg]
\n \\
&&
+\Bigg[
\log \Big(\dfrac{\eta_i}{\eta_i-4}\Big)
+\log \Big(\dfrac{\zeta_i}{\zeta_i-4}\Big)\Bigg] 
\log \Bigg[
\dfrac{\eta_i
(\zeta_i-4)}{(\eta_i-4)
\zeta_i}
+i\rho 
\Big(
\dfrac{\zeta_i - \eta_i}{\zeta_i \eta_i}
\Big)
\Bigg]
\Bigg\}.
\n
\end{eqnarray}
\item
\underline{
$D_{0}(0,q_{12},
0,q_{23}(q_{13}),0,m_h^2,0,
M_i^2,M_i^2,M_i^2)$}:

Furthermore, we have another 
scalar box integrals for the 
following kinematic configuration
which can be get the form of
\begin{eqnarray}
&&\hspace{-1.0cm}
D_{0}(0,q_{12},0,q_{23},0,m_h^2,0,M_i^2,M_i^2,M_i^2)
=
\\
&=&
-\dfrac{\lambda_i \tau_i^{1/2}}{4 M_i^4 
\Delta(\tau_i, \lambda_i, \eta_i)
}
\Bigg\{
\mathcal{L}i_2
\Big[
1-4\eta_i^{-1}
- i \rho 
\times
\eta_i^{-1},\Delta^{+}(\tau_i, \lambda_i, \eta_i)
+i \rho 
\times
\Sigma(\tau_i, \lambda_i, \eta_i)
\Big]
\n \\
&&
-
\mathcal{L}i_2
\Big[
1-4\eta_i^{-1}
-i\rho \times
\eta_i^{-1},\Delta^{-}(\tau_i, \lambda_i, \eta_i)
-i\rho \times
\Sigma(\tau_i, \lambda_i, \eta_i)
\Big]
\n \\
&&
+
\mathcal{L}i_2
\Bigg[
\dfrac{\lambda_i}{\lambda_i-2 
\Lambda^{+}}
+i\rho \times
\dfrac{
\Big[(\Lambda^{+} + 1) 
\lambda_i - 2
\Lambda^{+}\Big]}{\lambda_i
(\lambda_i - 2 \Lambda^{+})}
,
\Delta^{+}(\tau_i, \lambda_i, \eta_i)
+i\rho \times
\Sigma(\tau_i, \lambda_i, \eta_i)
\Bigg]
\n \\
&&
-
\mathcal{L}i_2
\Bigg[
\dfrac{\lambda_i}{\lambda_i-2 
\Lambda^{+}}
+i\rho \times
\dfrac{
\Big[(\Lambda^{+}+1) 
\lambda_i-2
\Lambda^{+}\Big]}{\lambda_i
(\lambda_i - 2 \Lambda^{+})}
,
\Delta^{-}
(\tau_i, \lambda_i, \eta_i)
-i\rho \times
\Sigma(\tau_i, \lambda_i, \eta_i)
\Bigg]
\n \\
&&
+
\mathcal{L}i_2
\Bigg[
\dfrac{\lambda_i}{\lambda_i+2 \Lambda^{-}}
-i\rho 
\dfrac{
\Big[
(\Lambda^{-}-1)
-
2 \lambda_i^{-1}
\Lambda^{-} 
\Big]
}
{\lambda_i+2 \Lambda^{-}}
,
\Delta^{+}(\tau_i, \lambda_i, \eta_i)
+i\rho \times
\Sigma(\tau_i, \lambda_i, \eta_i)
\Bigg]
\n \\
&&
-
\mathcal{L}i_2
\Bigg[
\dfrac{\lambda_i}{\lambda_i+2 \Lambda^{-}}
-i\rho 
\dfrac{
\Big[
(\Lambda^{-}-1)
-
2 \lambda_i^{-1}
\Lambda^{-} \Big]}
{\lambda_i+2 \Lambda^{-}}
,
\Delta^{-}(\tau_i, \lambda_i, \eta_i)
- i\rho \times
\Sigma(\tau_i, \lambda_i, \eta_i)
\Bigg]
\n \\
&&
-
\mathcal{L}i_2
\Big[
1-4\tau_i^{-1}
-i\rho \times
\tau_i^{-1}, 
\Delta^{+}(\tau_i, \lambda_i, \eta_i)
+i\rho \times
\Sigma(\tau_i, \lambda_i, \eta_i)
\Big]
\n \\
&&
+
\mathcal{L}i_2
\Big[
1-4\tau_i^{-1}
-i\rho \times
\tau_i^{-1},
\Delta^{-}(\tau_i, \lambda_i, \eta_i)
-i\rho \times
\Sigma(\tau_i, \lambda_i, \eta_i)
\Big]
\n \\
&&
+
\text{Li}_2
\Bigg[
\dfrac{
\tau_i(\eta_i-4)
+
\Delta(\tau_i, \lambda_i, \eta_i)
\eta_i\sqrt{\tau_i}
}{2 \lambda_i
(\eta_i-\tau_i)}
+i\rho \times
\Gamma^{-}(\tau_i, \lambda_i, \eta_i)
\Bigg]
\n \\
&&   
+
\text{Li}_2
\Bigg[
\dfrac{
\tau_i(\eta_i-4)
+
\Delta(\tau_i, \lambda_i, \eta_i)
\eta_i\sqrt{\tau_i}
}{2 \lambda_i
(\eta_i-\tau_i)}
+i\rho \times
\Gamma^{+}(\tau_i, \lambda_i, \eta_i)
\Bigg]
\n \\
&&   
-
\text{Li}_2
\Bigg[
\dfrac{
\tau_i(\eta_i-4)
-
\Delta(\tau_i, \lambda_i, \eta_i)
\eta_i\sqrt{\tau_i}
}{2 \lambda_i
(\eta_i-\tau_i)}
+i\rho \times
\Gamma^{+}(\tau_i, \lambda_i, \eta_i)
\Bigg]
\n \\
&&
-
\text{Li}_2
\Bigg[
\dfrac{
\tau_i(\eta_i-4)
-
\Delta(\tau_i, \lambda_i, \eta_i)
\eta_i\sqrt{\tau_i}
}{2 \lambda_i
(\eta_i-\tau_i)}
+i\rho \times
\Gamma^{-}(\tau_i, \lambda_i, \eta_i)
\Bigg]
\Bigg\}.
\n
\end{eqnarray}
Where related kinematic variables
$\Delta^{\pm}, \Lambda^{\pm}, 
\Gamma^{\pm}, \Delta$ 
and $\Sigma$ are defined 
in terms of dimensionless 
parameters $\tau_i, \lambda_i, 
\eta_i$ as follows:
\begin{eqnarray}
\Delta(\tau_i, \lambda_i, \eta_i)
&=&
\Big[
\tau_i
+
8 \eta_i^{-2}
\Big(
2 \lambda_i \eta_i
-2 \lambda_i \tau_i
- \tau_i \eta_i
+2 \tau_i
\Big)
\Big]^{1/2},
\\
\Delta^{\pm}(\tau_i, \lambda_i, \eta_i)
&=&
\dfrac{2 \tau_i (2-\lambda_i)
+\eta_i 
\Big[
2 \lambda_i-\tau_i
\pm \sqrt{\tau_i}
\Delta(\tau_i, \lambda_i, \eta_i)
\Big]
}{2
\lambda_i (\eta_i-\tau_i)}, 
\\
\Sigma(\tau_i, \lambda_i, \eta_i)
&=&
\dfrac{\lambda_i (\eta_i-\tau_i)}{
M_i^2 \eta_i \sqrt{\tau_i}
\Delta(\tau_i, \lambda_i, \eta_i)}, 
\\
\Lambda^{\pm}
&=&
\sqrt{1-\lambda_i} \pm 1
\end{eqnarray}
and
\begin{eqnarray}
\Gamma^{\pm}(\tau_i, \lambda_i, \eta_i)
&=&
\dfrac{1}{2 M_i^2 \eta_i \lambda_i
   (\eta_i-\tau_i) \sqrt{\tau_i} \Delta(\tau_i, \lambda_i, \eta_i)}
\times
\\
&& \hspace{-1.5cm} 
\times
\Bigg\{
- \eta_i^2 \Bigg[
2 \lambda_i^2 \pm 2 M_i^2
   \sqrt{\tau_i} \Delta(\tau_i, \lambda_i, \eta_i)
   \lambda_i \mp M_i^2 \tau_i^{3/2}
   \Big(\sqrt{\tau_i}+\Delta(\tau_i, \lambda_i, \eta_i)\Big)
\Bigg]
\n \\
&&\hspace{-1.5cm}
   +2 \tau_i \eta_i
   \Bigg[
   2
   \lambda_i^2 \pm M_i^2 \lambda_i \Big(\sqrt{\tau_i} \Delta(\tau_i, \lambda_i, \eta_i)+8\Big) 
   \mp 2 M_i^2 \sqrt{\tau_i} \Big(
   2 \sqrt{\tau_i} + \Delta(\tau_i, \lambda_i, \eta_i)
   \Big)
   \Bigg]
\n \\
&&\hspace{-1.5cm}
-2 \tau_i^2 \Bigg[
\lambda_i^2
\pm 8 M_i^2
   (\lambda_i-1)
   \Bigg]
\Bigg\}. 
\n
\end{eqnarray}
\section{ \label{eq_UV}                  
Check for the calculations}              
We verify that the  ultraviolet divergences
for $V^*$-pole contributions come from 
the following term
\begin{eqnarray}
 B_{0}(m_h^2,M_i^2,M_i^2) 
- B_{0}(q_{23},0,M_i^2).
\end{eqnarray}
This term gives 
a UV-finite result. 
In case of $F^{\text{Non-pole, Z}}_{1,L}$, 
we recognize that appearance of ultraviolet 
divergences by the $\epsilon^{-1}$ terms in 
such Passarino-Veltman functions 
\begin{eqnarray}
C_{0}(0,q_{13}(q_{23}),0,0,0,M_Z^2)
\; {\rm and} \; 
D_{0}(0,q_{13},m_h^2,q_{23},0,0,0,0,M_Z^2,M_Z^2)
\nonumber
\end{eqnarray}
as presented in next subsections.
Therefore, we consider $\epsilon^{-1}$ terms 
and proof that divergences in the form 
factor are analytically eliminated as follows
\begin{eqnarray}
F^{\text{Non-pole, Z}}_{1,L,\text{UV-term}}
&\supset&
q_{23} 
C_{0}(0,0,q_{23},0,0,M_Z^2)
+ 
q_{13} 
C_{0}(0,q_{13},0,0,0,M_Z^2)
\\
&&
+
\Bigg[
M_Z^2 (q_{13} + q_{23})
-
q_{13} \, q_{23}
\Bigg]
D_{0}(0,q_{13},m_h^2,
q_{23},0,0,0,0,M_Z^2,M_Z^2)
\nonumber\\
&=&
\dfrac{4 M_Z^2}{\eta_Z} 
C_{0}(0,0,q_{23},0,0,M_Z^2)
+ 
\dfrac{4 M_Z^2}{\zeta_Z} 
C_{0}(0,q_{13},0,0,0,M_Z^2)
\n \\
&&
+
4 M_Z^4
\Bigg[
\dfrac{\eta_Z 
+ \zeta_Z - 4}{\eta_Z \zeta_Z}
\Bigg]
D_{0}(0,q_{13},m_h^2,q_{23},
0,0,0,0,M_Z^2,M_Z^2).
\nonumber
\end{eqnarray}
Kinematic variables $\eta_Z$ and 
$\zeta_Z$ are 
defined as previous section. Expanding 
one-loop scalar integrals
in terms of logarithm and di-logarithm
functions, one has
\begin{eqnarray}
F^{\text{Non-pole, Z}}_{1,L,\text{UV-term}}
&\supset&
\dfrac{4 M_Z^2}{\eta_Z} 
\Bigg[
\dfrac{\eta_Z}{8 M_Z^2}
\log \Big(
\dfrac{\eta_Z}{\eta_Z-4}
\Big) 
\Bigg]
\Bigg[
\dfrac{2}{\epsilon}
+2 \log \Big(\dfrac{\mu ^2}{M_Z^2}\Big)
\Bigg]
\n \\
&&
\hspace{2.7cm}
+ 
\dfrac{4 M_Z^2}{\zeta_Z} 
\Bigg[
\dfrac{\zeta_Z}{8 M_Z^2}
\log \Big(
\dfrac{\zeta_Z}{\zeta_Z-4}
\Big) 
\Bigg]
\Bigg[
\dfrac{2}{\epsilon}
+ 2 
\log \Big(\dfrac{\mu^2}{M_Z^2}\Big)
\Bigg]
\n \\
&&
+
4 M_Z^4
\Bigg[
\dfrac{\eta_Z  
+ \zeta_Z - 4}{\eta_Z \zeta_Z}
\Bigg] \times
\nonumber\\
&& 
\hspace{1cm}
\times 
\Bigg\{
\dfrac{\eta_Z \zeta_Z}{8 M_Z^4
(4-\eta_Z-\zeta_Z)}
\Bigg[
2 \log \Big(\dfrac{\eta_Z}{\eta_Z-4}\Big)
\Bigg(
\dfrac{1}{\epsilon }
+\log
\Big(\dfrac{\mu^2}{M_Z^2}\Big)
\Bigg)
\n \\
&&\hspace{4cm}
+
\log
\Big(\dfrac{\zeta_Z}{\zeta_Z-4}\Big) 
\Bigg(
\dfrac{2}{\epsilon }
+2 \log \Big(\dfrac{\mu ^2}{M_Z^2}\Big)
\Bigg)
\Bigg]
\Bigg\}
\n \\
&=&
\log \Big(
\dfrac{\eta_Z}{\eta_Z-4}
\Big) 
\Bigg[
\dfrac{1}{\epsilon }
+\log
\Big(\dfrac{\mu ^2}{M_Z^2}\Big)
\Bigg]
+ 
\log \Big(
\dfrac{\zeta_Z}{\zeta_Z-4}
\Big) 
\Bigg[
\dfrac{1}{\epsilon }
+\log
\Big(\dfrac{\mu ^2}{M_Z^2}\Big)
\Bigg]
\n \\
&&
-
\dfrac{1}{2}
\Bigg\{
2 
\log 
\Big(\dfrac{\eta_Z}{\eta_Z-4}\Big)
\Bigg[
\dfrac{1}{\epsilon }
+\log
\Big(\dfrac{\mu ^2}{M_Z^2}\Big)
\Bigg]
\nonumber\\
&& 
\hspace{5cm}
+
\log
\Big(\dfrac{\zeta_Z}{\zeta_Z-4}\Big) 
\Bigg[
\dfrac{2}{\epsilon }
+ 2 \log
\Big(\dfrac{\mu ^2}{M_Z^2}\Big)
\Bigg]
\Bigg\}
\n \\
&=&0.
\n
\end{eqnarray}
The final results are independent
of $\mu^2$ and $1/\epsilon$. 
\end{itemize}
\section{ \label{app_Ik}The basic integrals $I_1, I_2$}
The basic integrals $I_1$ and $I_2$ 
are taken in ~\cite{Djouadi:2005gi}
and presented in term of $f$, $g$
functions as follows:
\begin{align}
I_1(\tau,\lambda) 
&= \frac{\tau\lambda}{2(\tau-\lambda)} 
+ \frac{\tau^2\lambda^2}{2(\tau-\lambda)^2}
\Big[f(\tau)-f(\lambda) \Big] 
+ \frac{\tau^2\lambda}{(\tau-\lambda)^2}
\Big[ g(\tau)-g(\lambda) \Big],\\
I_2(\tau, \lambda)
&= 
-\frac{\tau\lambda}{2(\tau-\lambda)}
\Big[ f(\tau)-f(\lambda) \Big].
\end{align}
Two complex functions $f,\; g$ can 
be expressed as follows:
\begin{align}
f(\tau) &=
\left\{ \begin{array}{rcl}
\arcsin^{2}\sqrt{\tau} & \mbox{for}
& \tau\leq 1, \\
& \\
-\frac{1}{4}
\left[\log\dfrac{1+\sqrt{1-\tau^{-1}}}
{1-\sqrt{{1-\tau^{-1}}}}-i\pi\right]^{2} 
& \mbox{for} & \tau > 1,
\end{array}\right.
\end{align}
and
\begin{align}
g(\tau) &=
\left\{ \begin{array}{rcl}
\sqrt{\tau^{-1}-1}\arcsin\sqrt{\tau} & 
\mbox{for}
& \tau\geq 1, \\
& \\
\dfrac{\sqrt{1-\tau^{-1}}}{2}
\left[\log\dfrac{1+\sqrt{1-\tau^{-1}}}
{1-\sqrt{{1-\tau^{-1}}}}-i\pi\right] & 
\mbox{for} & \tau< 1.
\end{array}\right.
\end{align}
\section{ \label{app_Phase}Appendix $E$: Phase-space 
$1\rightarrow 3$}
In order to generate the forward-backward
asymmetries, we are working on rest
frame of Higgs boson. In this frame, involved
kinematic variables are taken
\begin{eqnarray}
q_{12}
&=&
m_h^2 - 2 m_h E_\gamma,
\n \\
q_{13}
&=&
m_l^2 +
2 E_\gamma
\left(
E_1
-
|\vec{p}_1| \cos \theta_l
\right).
\end{eqnarray}
Where $E_1$ is energy of lepton and 
three momentum of lepton is 
taken as $|\vec{p}_1| = \sqrt{E_1^2 - m_l^2}$. 
The energy $E_1$ is then calculated as 
follows: 
\begin{eqnarray}
E_1
&=&\dfrac{
\dfrac{m_h (m_h - E_\gamma)
(m_h - 2 E_\gamma)}{2}
+ 
\dfrac{E_\gamma \cos \theta_l}{2}
\sqrt{
m_h^2 (m_h - 2 E_\gamma)^2
+ 4 m_l^2 
\Big[
E^2_\gamma \cos^2 \theta_l 
- (E_\gamma - m_h)^2
\Big]  }
}
{
(E_\gamma - m_h)^2 
- E^2_\gamma \cos^2 \theta_l
}. 
\nonumber\\
\end{eqnarray}
As a result, we present the 
differential decay rate with 
respect to $E_\gamma$ 
and $\cos \theta$ as follows
\begin{eqnarray}
\dfrac{d \Gamma}{d q_{12} q_{13}}
&=&
\dfrac{d \Gamma}{d E_\gamma \cos \theta_l}
\times
\Big|
\dfrac{\partial q_{12}}{\partial E_\gamma}
\dfrac{\partial q_{13}}{\partial \cos \theta_l}
-
\dfrac{\partial q_{12}}{\partial \cos \theta_l}
\dfrac{\partial q_{13}}{\partial E_\gamma}
\Big|^{-1}.
\end{eqnarray}
We then perform the 
above integrand over 
$E_{\gamma}^{\text{cut}} 
\le E_\gamma \le 
\dfrac{m_h}{2}$ 
and $-1 \le \cos \theta_l \le +1$ or 
\begin{eqnarray}
\Gamma 
=\int\limits_{E_{\gamma}^{\text{cut}}}^{m_h/2} 
dE_{\gamma}
\int\limits_{-1}^{1} d\cos \theta_l
\dfrac{d \Gamma}{d E_\gamma \cos \theta_l}
\times
\Big|
\dfrac{\partial q_{12}}{\partial E_\gamma}
\dfrac{\partial q_{13}}{\partial \cos \theta_l}
-
\dfrac{\partial q_{12}}{\partial \cos \theta_l}
\dfrac{\partial q_{13}}{\partial E_\gamma}
\Big|^{-1}.
\end{eqnarray}

\end{document}